\documentclass[twocolumn,pra,groupedaddress.floatfix%,nofootinbib
]{revtex4-2}%uncomment nofootinbib to have footnotes on same page
\pdfoutput=1
\usepackage{amsfonts}
\usepackage{color}
\usepackage{amsmath}
\usepackage{amssymb}
\usepackage{graphicx}
\usepackage{caption,subcaption}
\usepackage{anyfontsize}
\RequirePackage{fix-cm}
\definecolor{darkblue}{rgb}{0,0,0.5}
\providecommand{\U}[1]{\protect\rule{.1in}{.1in}}

\newenvironment{proof}[1][Proof]{\noindent\textbf{#1.} }{\ \rule{0.5em}{0.5em}}
\newenvironment{spmatrix}{
    \left(
    \begin{smallmatrix}
    }
    {
    \end{smallmatrix}\right)}

\DeclareMathOperator*{\argmax}{arg\,max}

\newcommand{\dif}{\mathop{}\!\mathrm{d}}

\renewcommand{\d}{\mathrm{d}}
 
\newcommand{\atermM}{P(k|0;M)}
\newcommand{\btermM}{P(k|\pi;M)}
\newcommand{\ns}{{\bar N}_{\rm S}}
\let\originalleft\left
\let\originalright\right
\renewcommand{\left}{\mathopen{}\mathclose\bgroup\originalleft}
\renewcommand{\right}{\aftergroup\egroup\originalright}
\newcommand{\mb}{\mathbf}
\newcommand{\mbb}{\mathbb}
\newcommand{\us}{\underset}
\newcommand{\ssd}[1]{\scriptstyle #1\displaystyle} %Stands for sicriptstyle display - to be used for shrinking expressions in displayed Eq.s.
\newcommand{\mat}[2]{\left(\begin{array}{#1}#2\end{array}\right)}
\newcommand{\X}{\mb{X}}

\begin{document}

\title{Transceiver designs to attain the entanglement assisted communications capacity}%{Structured transceiver design to attain the many-fold enhancement in communication capacity afforded by pre-shared entanglement}% Infinite-fold enhancement in communications capacity using pre-shared entanglement}

\author{Ali Cox}
\author{Quntao Zhuang}
\author{Christos N. Gagatsos}
\author{Boulat Bash}
\author{Saikat Guha}
\affiliation{College of Optical Sciences, University of Arizona, Tucson AZ 85721}
\affiliation{Department of Electrical Engineering, University of Arizona, Tucson AZ 85721}

\begin{abstract}
Pre-shared entanglement can significantly boost communication rates in the high thermal noise and low-brightness transmitter regime. In this regime, for a lossy-bosonic channel with additive thermal noise, the ratio between the entanglement-assisted capacity and the Holevo capacity---the maximum reliable-communications rate permitted by quantum mechanics without any pre-shared entanglement---scales as $\log(1/{\bar N}_{\rm S})$, where the mean transmitted photon number per mode, ${\bar N}_{\rm S} \ll 1$. Thus, pre-shared entanglement, e.g., distributed by the quantum internet or a satellite-assisted quantum link, promises to significantly improve low-power radio-frequency communications. In this paper, we propose a pair of structured quantum transceiver designs that leverage continuous-variable pre-shared entanglement generated, e.g., from a downconversion source, binary phase modulation, and non-Gaussian joint detection over a codeword block, to achieve this scaling law of capacity enhancement. Further, we describe a modification to the aforesaid receiver using a front-end that uses sum-frequency generation sandwiched with dynamically-programmable in-line two-mode squeezers, and a receiver back-end that takes full advantage of the output of the receiver's front-end by employing a non-destructive multimode vacuum-or-not measurement to achieve the entanglement-assisted classical communications capacity. %Finally, the implication of this result to the breaking of the well-known {\em square-root law} for covert communications, with pre-shared entanglement assistance, is discussed.

\end{abstract}
%\date{December 17, 2019}
\maketitle

\section{Introduction}

An emerging focus has been drawn to architecting the {\em quantum internet}~\cite{Kim08, Weh18}: a global network built using quantum repeaters~\cite{Guh15, Mur16} and satellites~\cite{Pir21} to distribute entanglement at high rates to multiple distant users on-demand~\cite{Pir19,Pan19,Das18}. There are several well-known applications of shared {\em entanglement}, a new information currency: distributed quantum computing~\cite{Met16}, communications with physics-based security~\cite{E91}, provably-secure access to quantum computers on the cloud~\cite{Chi05}, and entanglement-enhanced distributed sensing~\cite{Zhu17a,Pro18,Guo19,Xia19}. In this paper, we design a system for another impactful application of shared entanglement: substantially improving classical communication rates in certain regimes.

Transmission of electromagnetic (EM) waves in linear media, such as optical fiber, atmosphere, and vacuum, can be described as propagation of a set of mutually-orthogonal spatio-temporal-polarization modes over the single-mode lossy bosonic channel with additive thermal-noise ${\cal N}_\eta^{{\bar N}_{\rm B}}$, described by the Heisenberg evolution ${\hat a}_{\rm R} = \sqrt{\eta}\, {\hat a}_{\rm S} + \sqrt{1-\eta}\,{\hat a}_{\rm E}$, where $\eta \in (0, 1]$ is the modal (power) transmissivity, and the environment ${\hat a}_{\rm E}$ is excited in a zero-mean thermal state of mean photon number per mode ${\bar N}_{\rm B}$. Alice encodes classical information by modulating the state of the ${\hat a}_{\rm S}$ modes, with the constraint of ${\bar N}_{\rm S}$ mean photons transmitted per mode. The quantum limit of the classical communications capacity, known as the {\em Holevo capacity}~\cite{Hol98, Sch97}, in bits per transmitted mode, is given by:
\begin{equation}
C(\eta, {\bar N}_{\rm S}, {\bar N}_{\rm B}) = g({\bar N}_{\rm S}^\prime) - g((1-\eta){\bar N}_{\rm B}),
\label{eq:UltHol}
\end{equation}
where ${\bar N}_{\rm S}^\prime \equiv \eta {\bar N}_{\rm S} + (1-\eta){\bar N}_{\rm B}$ is the mean photon number per the ${\hat a}_{\rm R}$ mode at the channel's output received by Bob, and $g(x) \equiv (1+x)\log(1+x) - x\log(x)$ is the von Neumann entropy of a zero-mean single-mode thermal state with mean photon number $x$~\cite{Gio04,Gio14}. In this paper, we denote base-$2$ and natural logarithms by $\log$ and $\ln$, respectively.

If Alice and Bob pre-share an unlimited amount of entanglement as a resource for transmitting classical data over ${\cal N}_\eta^{{\bar N}_{\rm B}}$ with a transmit photon number constraint of ${\bar N}_{\rm S}$ photons per mode, the capacity (in bits per mode) increases to:~\cite{BSST,Gio03,Hol01,Hol02,Hol03,Hol04}:
\begin{equation}
C_{\rm E}(\eta, {\bar N}_{\rm S}, {\bar N}_{\rm B}) = g({\bar N}_{\rm S}) + g({\bar N}_{\rm S}^\prime) - g(A_+) - g(A_-),
\label{eq:CE}
\end{equation}
where $C_{\rm E}$ is the {\em entanglement assisted classical capacity} of the quantum channel ${\cal N}_\eta^{{\bar N}_{\rm B}}$, and
$
A_{\pm} = \frac12(D-1 \pm ({\bar N}_{\rm S}^\prime - {\bar N}_{\rm S})), 
$
with
$
D = \sqrt{({\bar N}_{\rm S} + {\bar N}_{\rm S}^\prime + 1)^2 - 4\eta {\bar N}_{\rm S}({\bar N}_{\rm S}+1)}.
$

\begin{figure}[!h]
\centering
\includegraphics[width=.9\columnwidth]{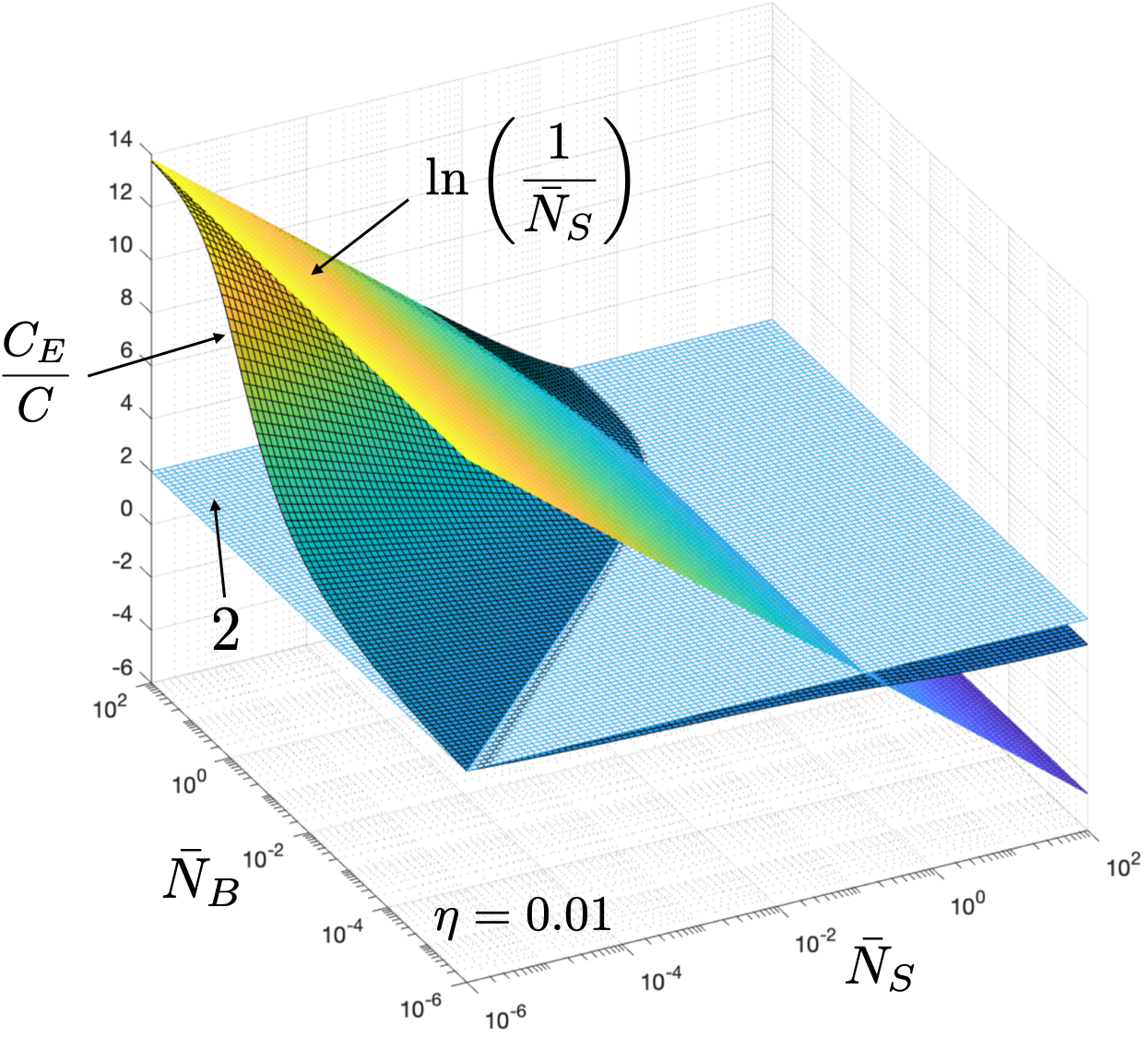}
\caption{The ratio $C_{\rm E}/C$ is plotted as a function of ${\bar N}_{\rm S}$ and ${\bar N}_{\rm B}$ for channel transmissivity, $\eta = 0.01$. Also shown is a plot of $\ln(1/{\bar N}_{\rm S})$, which is the scaling of $C_{\rm E}/C$ as ${\bar N}_{\rm S} \to 0$ and ${\bar N}_{\rm B} \to \infty$.}
\label{fig:capacity_ratio}
\end{figure}

In the regime of a low-brightness transmitter (${\bar N}_{\rm S} \ll 1$) and high thermal noise (${\bar N}_{\rm B} \gg 1$),
\begin{equation}
{C_{\rm E}}/{C} \approx \ln\left({1}/{{\bar N}_{\rm S}}\right),
\label{eq:scaling}
\end{equation}
which tends to infinity as ${\bar N}_{\rm S} \to 0$~\cite{Shi19}. Physically, this means that for a fixed number of channel uses, i.e., transmitted modes, a receiver that has access to a quantum system entangled with the transmitted modes can extract many more message bits reliably per mode, in the low-signal-brightness high-thermal-noise regime, compared with a receiver that has no such access. The practical implications are potentially revolutionary in radio-frequency (RF) communications, since the condition ${\bar N}_{\rm B} \gg 1$ is naturally satisfied at a long center wavelength. The ${\bar N}_S \ll 1$ regime is of particular interest to covert communications, where the transmitter tries to hide the presence of the communication attempt~\cite{Gag20}. Pre-shared entanglement between Alice and Bob---distributed, e.g., via the quantum internet---allows for an order of magnitude or more enhancement in classical communications rate, depending on the operational regime of loss, noise, and transmit power (see Fig.~\ref{fig:capacity_ratio}).

Despite this large capacity advantage attainable with pre-shared entanglement having been known for decades, a structured transmitter-receiver design to harness this enhancement has been elusive. Continuous-variable (CV) superdense coding yields a factor-of-two capacity advantage in the noiseless case, but does not provide any advantage in the noisy regime~\cite{Soh03}. Phase-only encoding on many copies of pre-shared two-mode squeezed vacuum (TMSV) states, $|\psi\rangle_{\rm SI} = \sum_{j=0}^\infty \sqrt{{\bar N}_{\rm S}^j/(1+{\bar N}_{\rm S})^{j+1}} |j\rangle_{\rm S}|j\rangle_{\rm I}$, attains $C_{\rm E}$ in the ${\bar N}_{\rm S} \ll 1$, ${\bar N}_{\rm B} \gg 1$ regime~\cite{Shi19}, but with a receiver measurement whose structured optical design is unknown. Receivers based on optical parametric amplification (OPA)~\cite{Guh09} and sum-frequency-generation (SFG)~\cite{Zhu17} yield at most a factor-of-$2$ improvement over $C$, as shown in~\cite{Shi19} and Appendix~\ref{app:OPA} of this paper.

In this paper, we take an important step in solving this long-standing problem. We use insights from the SFG receiver proposed for the quantum-illumination radar~\cite{Zhu17,Zhu21} to map the task of jointly detecting long code blocks consisting of binary phase-correlated signal-idler mode pairs onto a task of jointly detecting code blocks of coherent states, which in turn can be done using the linear-optical `Green Machine' (GM) joint-detection receiver which has been shown to attain superadditive communication capacity with phase modulation of coherent states~\cite{Guh11b}~\footnote{Jet Propulsion Laboratory developed a decoding algorithm for the first-order length-$n$ Reed Muller codes that employed the fast Hadamard transform in a specialized circuit that used $(n\log n)/2$ symmetric butterfly circuits, for sending images from Mars to the Earth as part of the Mariner 1969 Mission. This circuit came to be known as the {\em Green Machine} named after its JPL inventor. Guha developed an optical version of the Green Machine decoding circuit, replacing the butterfly elements by 50-50 beamsplitters, which he showed achieved superadditive communication capacity with Hadamard-coded coherent-state BPSK modulation, i.e., communication capacity in bits transmissible reliably per BPSK symbol that is fundamentally higher than that is physically permissible with any receiver that detects each BPSK modulated pulse one at a time~\cite{Guh11b}. This paper's joint detection receiver for entanglement assisted communications leverages insights from that optical Green Machine.}. In section \ref{sec:JDRdesign}, we propose two structured designs, one improving upon the other, each consisting of a transmitter, modulation format, and receiver. We also show later in section \ref{sec:JDRdesign}  that both designs achieve the $\ln(1/{\bar N}_{\rm S})$ scaling in entanglement-assisted capacity gain over the Holevo capacity in Eq.~\eqref{eq:scaling}. Preliminary results on the first of these designs and the associated capacity scaling were published in a conference proceedings article by a subset of the co-authors~\cite{Guha2020EAJDR}. These proposed receivers also set the stage for designing structured receivers, which with binary-phase modulated pre-shared TMSV entanglement, can achieve the entanglement-assisted capacity, $C_E$. We give a possible design of a $C_E$-achieving transceiver based on an adaptive multi-mode non-destructive vacuum-or-not measurement in Section~\ref{future}. To close the gap to $C_E$, this new receiver reveals that not only is it necessary to choose an optimal classical JDR on the blocks of coherent states once the received correlated states are mapped to the coherent states, but it is also necessary to modify the front-end of the receiver to avoid losing information in the mapping process.

\section{Quantum limits of unassisted classical communications}\label{sec:classcap_review}
The front end of the receiver designs proposed in this paper convert classical information encoded in the direction of correlation between the received signal and the stored idler modes of a two-mode zero-mean Gaussian state into the mean field of a single-mode displaced thermal state. %The commented out part that follows is a well-written explanation of the transceiver but doesn't belong in this section.
\iffalse Before transmission through the channel, the two modes that are to ultimately undergo this conversion are initialized as a two-mode squeezed vacuum (a maximally entangled Gaussian state) whose `idler' part is sent to and stored at the receiver, and whose `signal' part is modulated  at the transmitter by a binary phase-shift corresponding to a single bit of the message intended for transmission. When the signal is transmitted to the receiver through the single-mode lossy bosonic channel $\mathcal{N}^{\bar{N}_B}_\eta$, the entanglement between the signal and idler modes is broken by the bright thermal noise so that the two-mode state to be processed by the receiver is merely classically correlated. Nevertheless, the amount of classical correlation, and hence the information available to the receiver, is proportional to the amount of initial correlation between the signal and idler, so an initially entangled state, whose correlation strength exceeds the maximum amount possessable by a classical state, leads to a higher classical correlation at the receiver. \fi 
At the back-end, the displaced thermal states produced by the receiver front-end are passed to joint-detection receivers (JDRs) designed for classical communication, i.e., using a coherent state modulation without entanglement assistance~\cite{Guh11a,Rengas20}. Therefore in this section we review the problem setup and JDR methods for coherent-state-encoded classical communications over a lossy bosonic channel without pre-shared entanglement. In particular, section \ref{sec:Holevocapacity} is a review of the definition of the unassisted Holevo capacity for arbitrary modulation formats, among which section \ref{sec:capacityBPSKOOK} covers binary modulation formats, section \ref{sec:classicalcommJDR} reviews joint-detection receiver designs for BPSK modulation that exhibit superadditivity to approach the Holevo capacity, and finally section \ref{sec:receiverforHolevo} discusses code and receiver design ideas to achieve the ultimate unassisted Holevo capacity. Some results in this section were derived in an unpublished memo from 2011~\cite{Guh11memo}. In subsection~\ref{sec:classicalcommJDR} we also present a new JDR design for classical communications, which we use later as a module to build a JDR for entanglement-assisted communications.
\begin{figure}
    \includegraphics[width=\linewidth]{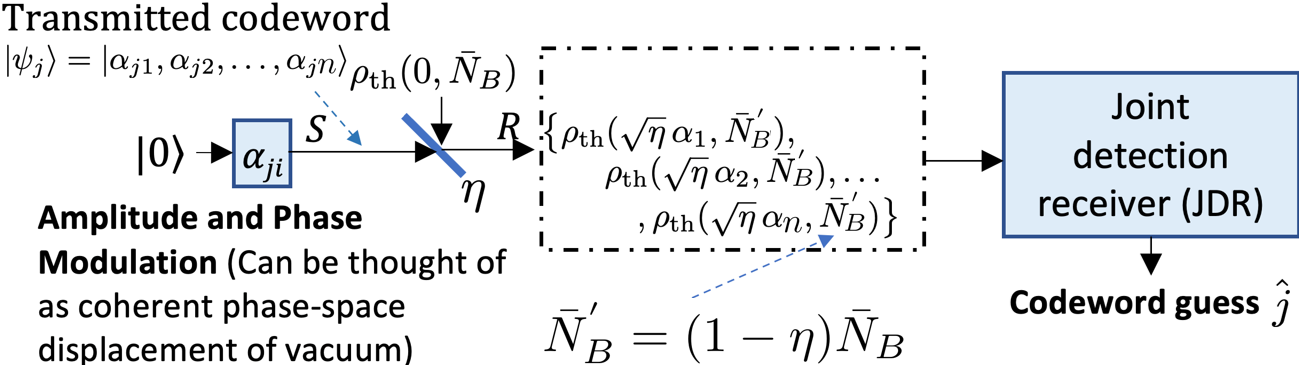}
    \caption{Schematic of classical optical communications over a lossy, noisy channel. Each transmitted symbol of a length-$L$ code word is a single bosonic mode excited in a coherent state $|\alpha_{ji}\rangle$, $1 \le i \le L$, $1 \le j \le 2^{nR}$, which results in a displaced thermal state at the channel output: $\hat{\rho}_{\rm th}(\sqrt{\eta}\alpha_{ji}, (1-\eta){\bar N}_B)$ of mean field $\sqrt{\eta}\alpha_{ji}$ and mean thermal photon number $(1-\eta){\bar N}_B$. The total mean photon number of each received mode is ${\bar N}_S^\prime = \eta {\bar N}_S+(1-\eta){\bar N}_B$. To achieve the {\em Holevo capacity}, the ultimate limit to the reliable communication rate, the receiver must perform a joint quantum measurement on a long code word block. Such a receiver is called a joint-detection receiver (JDR).}
    \label{fig:classicalcomms}
\end{figure}

\subsection{The ultimate Holevo capacity}\label{sec:Holevocapacity}
Modulation and encoding using coherent states (or ideal laser light pulses) suffices to attain the ultimate (Holevo) capacity of classical communications over the lossy-noisy bosonic channel ${\cal N}_\eta^{{\bar N}_B}$~\cite{Gio14}. If the modulation is unrestricted, i.e., the coherent-state amplitude $\alpha_i$ of the $i$-th transmitted mode can be chosen over the entire complex plane, subject to a mean photon number constraint ${\bar N}_S = \langle \alpha_i \rangle$, the Holevo capacity is given by Eq.~\eqref{eq:UltHol} \iffalse: 
\begin{equation}
    C(\eta,\bar{N}_S,\bar{N}_B)=g(\eta {\bar N}_S+(1-\eta){\bar N}_B)-g((1-\eta){\bar N}_B),
    \label{eq:UltHol}
\end{equation}
where $g(x)=(x+1)\log_2(x+1)-x\log_2(x)$.\fi To achieve communications at a rate approaching this capacity, the transmitter picks a code book consisting of $2^{nR}$ length-$n$ (i.e., $n$-mode) coherent state code words $|\psi_j\rangle = |\alpha_{j1}, \alpha_{j2}, \ldots, \alpha_{jn}\rangle$, where each symbol of each code word $\alpha_{ji}$, $1 \le j \le 2^{nR}$, $1 \le i \le n$ is chosen i.i.d. from a Gaussian distribution, $p(\alpha) = (1/\pi {\bar N}_S)e^{-|\alpha|^2/{\bar N}_S}$, $\alpha \in \mathbb{C}$. If a coherent state $|\alpha\rangle$, $\alpha \in {\mathbb C}$ is transmitted at the input of the channel, the output state ${\hat \rho}_{\rm th}(\sqrt{\eta}\alpha, {\bar N}^{'}_{\rm B})$ is a single-mode displaced thermal state with {\em mean field amplitude} $\sqrt{\eta}\alpha$, and {\em thermal-noise mean photon number}, ${\bar N}^{'}_{\rm B} = (1-\eta){\bar N}_B$. Its density operator is given by:
\begin{equation}
{\hat \rho}_{\rm th}(\sqrt{\eta}\alpha, {\bar N}_{\rm B}^{'}) = \int_{\mathbb C} \frac{1}{\pi {\bar N}_{\rm B}^{'}}e^{-|\beta - \sqrt{\eta}\alpha|^2/{\bar N}_{\rm B}^{'}}|\beta\rangle \langle \beta | d^2\beta,
\end{equation} 
where $|\beta\rangle, \beta \in {\mathbb C}$ is a coherent state. The receiver performs a joint quantum measurement on the $n$-mode received code word, each of whose $n$ symbols are displaced thermal states as above, to make a guess $\hat j$ on which code word was transmitted. Fig.~\ref{fig:classicalcomms} depicts this encoding-transmission-decoding process schematically. If the rate of the code is chosen to satisfy $R < C(\eta,\bar{N}_S,\bar{N}_B)$ (bits per transmitted mode), then if the transmitter picks a random code book as described above and the receiver uses the {\em square root measurement}~\cite{Sch97} (a structured optical realization of which is not known), the  receiver's probability of error in picking the correct transmitted code word $P_e^{(L)} \to 0$, as $n \to \infty$~\cite{Hol98}. The Holevo capacity expression in Eq.~\eqref{eq:UltHol} can be derived as follows:
\begin{eqnarray}
    C(\eta, \bar{N}_S, \bar{N}_B)
    =\max_{p(\alpha), \alpha \in {\mathbb C}}\, \left\{ S\left({\bar{\hat{\rho}}}_R\right) - {\bar S}\right\},
    \label{eq:HSW_Bosonic}
\end{eqnarray}
where $S(\hat{\rho}) = -{\rm Tr}(\hat{\rho} \log \hat{\rho})$ is the von Neumann entropy of the state $\hat{\rho}$, ${\bar{\hat{\rho}}}_R = \int p(\alpha) {\hat \rho}_{\rm th}(\sqrt{\eta}\alpha, {\bar N}_{\rm T}) d^2\alpha$ is the average state of the output mode, ${\bar S} = \int p(\alpha) S\left({\hat \rho}_{\rm th}(\sqrt{\eta}\alpha, {\bar N}_{\rm T})\right) \, d^2\alpha$ is the average entropy of the possible output states, and the maximum is taken over all probability distribution functions $p(\alpha)$, with $\alpha \in {\mathbb C}$, satisfying the transmit mean-photon number constraint $\int p(\alpha) |\alpha|^2 d^2\alpha = {\bar N}_S$. It can be shown that no other quantum states used at the transmitter can achieve a capacity higher than that achieved using coherent states with a Gaussian prior: the Holevo capacity as given in Eq.~\eqref{eq:UltHol}~\cite{Gio14}. The unrestricted-modulation Holevo capacity, in the lossless ($\eta = 1$), noiseless (${\bar N}_B = 0$) case is given by $C(1, {\bar N}_S, 0) = g({\bar N}_S)$, which in the small-${\bar N}_S$ limit, can be expanded as~\cite{Guh11memo}:
\begin{equation}
C(1, {\bar N}_S, 0) = -{\bar N}_S \ln {\bar N}_S + {\bar N}_S + \frac{{\bar N}_S^2}{2} + {\text{h.o.t.}},
\label{eq:capacityHolevoscaling}
\end{equation}
where the unit used is {\em nats} per mode, with $\ln 2$ nats equalling 1 bit, and h.o.t. refers to ``higher order terms". 

\subsection{Holevo capacity with binary modulation constellations}\label{sec:capacityBPSKOOK}
If the encoding is restricted to the binary phase-shift keying (BPSK) alphabet, i.e., coherent states $|\pm\alpha\rangle$ for each transmitted mode, $|\alpha|^2 = {\bar N}_S$, the maximum possible communication rate, in bits per transmitted mode, is the Holevo capacity of the BPSK alphabet,
\begin{eqnarray}
    \chi_\text{BPSK}(\eta, \bar{N}_S, \bar{N}_B)
    =\max_{p \in [0,1]}\, \left\{ S\left({\bar{\hat{\rho}}}_R\right) - {\bar S}\right\},
    \label{eq:BPSKHol_thermal}
\end{eqnarray}
where ${\bar{\hat{\rho}}}_R = p\hat{\rho}_{\rm th}(\sqrt{\eta}\alpha,{\bar N}_T) + (1-p)\hat{\rho}_{\rm th}(-\sqrt{\eta}\alpha,{\bar N}_T))$ is the average state of the output mode, and ${\bar S} = pS(\hat{\rho}_{\rm th}(\sqrt{\eta}\alpha,{\bar N}_T)) + (1-p)S(\hat{\rho}_{\rm th}(-\sqrt{\eta}\alpha,{\bar N}_T))$ is the average entropy of the two possible output states. To achieve a rate approaching this capacity, a random code book should be constructed as explained in Section~\ref{sec:Holevocapacity}, but with each symbol of each code word $\alpha_{ji}$ chosen i.i.d. with equal priors, i.e., $p = 1/2$, from the two possible amplitudes $\{\pm\alpha\}$. In the lossless ($\eta = 1$), noiseless case (${\bar N}_B=0$), the BPSK Holevo capacity is given by:
\begin{equation}
    \chi_\text{BPSK}(1, \bar{N}_S, 0)
    =h_2\left([1 + e^{-2\bar{N}_S}]/2\right),
    \label{eq:BPSKHol}
\end{equation}
where $h_2(x) = -x\log x - (1-x)\log(1-x)$, $x \in (0, 1)$, is the binary entropy function. Expanding the expression in the small-${\bar N}_S$ limit, we get~\cite{Guh11memo}:
\begin{equation}
\chi_\text{BPSK}(1, {\bar N}_S, 0) = -{\bar N}_S \ln {\bar N}_S + {\bar N}_S + {{\bar N}_S^2 \ln {\bar N}_S} + {\text{h.o.t.}},
\label{eq:capacityBPSKHolevoscaling}
\end{equation}
in units of nats per transmitted mode.

If the encoding is restricted to an on-off-keyed (OOK) modulation alphabet, i.e., coherent state $|\alpha\rangle$ or vacuum $|0\rangle$ for each transmitted mode, the maximum communication rate, the OOK Holevo capacity is given by:
\begin{align}
    &\chi_\text{OOK}(\eta,\bar{N}_S,{\bar N}_B) = \max_{p \in [0,1]}\, \left\{ S\left({\bar{\hat{\rho}}}_R\right) - {\bar S}\right\},
    \end{align}
where ${\bar{\hat{\rho}}}_R = p\hat{\rho}_{\rm th}(\sqrt{\eta}\alpha,{\bar N}_T) + (1-p)\hat{\rho}_{\rm th}(0,{\bar N}_T))$ is the average state of the output mode, ${\bar S} = pS(\hat{\rho}_{\rm th}(\sqrt{\eta}\alpha,{\bar N}_T)) + (1-p)S(\hat{\rho}_{\rm th}(0,{\bar N}_T))$ is the average output entropy per mode, ${\bar N}_S = p|\alpha|^2$ is the mean transmitted photon number per mode, and $p$ is the prior probability with which the ``on" symbol ($|\alpha\rangle$) is transmitted. For the lossless, noiseless case,
\begin{align}
&\chi_\text{OOK}(1,\bar{N}_S,0) = \nonumber\\
    &\max_{0<p<1}h_2\left(\frac{1}{2}\left(1-\sqrt{(1-2 p)^2+4e^{-\frac{\bar{N}_S}{p}}p(1-p)}\right)\right),
    \label{eq:OOKHol}
\end{align}
where the optimal on-prior, $p \sim \sqrt{{\bar N}_S/2}$ decreases as ${\bar N}_S$ decreases~\cite{Guh11memo}, resulting in the mean photon number of the (infrequent) transmitted on-pulse given by, ${\bar N}_S / p \sim \sqrt{{\bar N}_S}$, in the ${\bar N}_S \ll 1$ regime. Expanding $\chi_\text{OOK}(1,\bar{N}_S,0)$ in the small-${\bar N}_S$ regime, we get:
\begin{equation}
\chi_\text{OOK}(1, {\bar N}_S, 0) = -{\bar N}_S \ln {\bar N}_S + {\bar N}_S + \sqrt{2}{{\bar N}_S^{3/2} \ln {\bar N}_S} + {\text{h.o.t.}},
\end{equation}
in units of nats per transmitted mode~\cite{Guh11memo}.

\begin{figure}
    \flushleft
    \includegraphics[width=.9\linewidth]{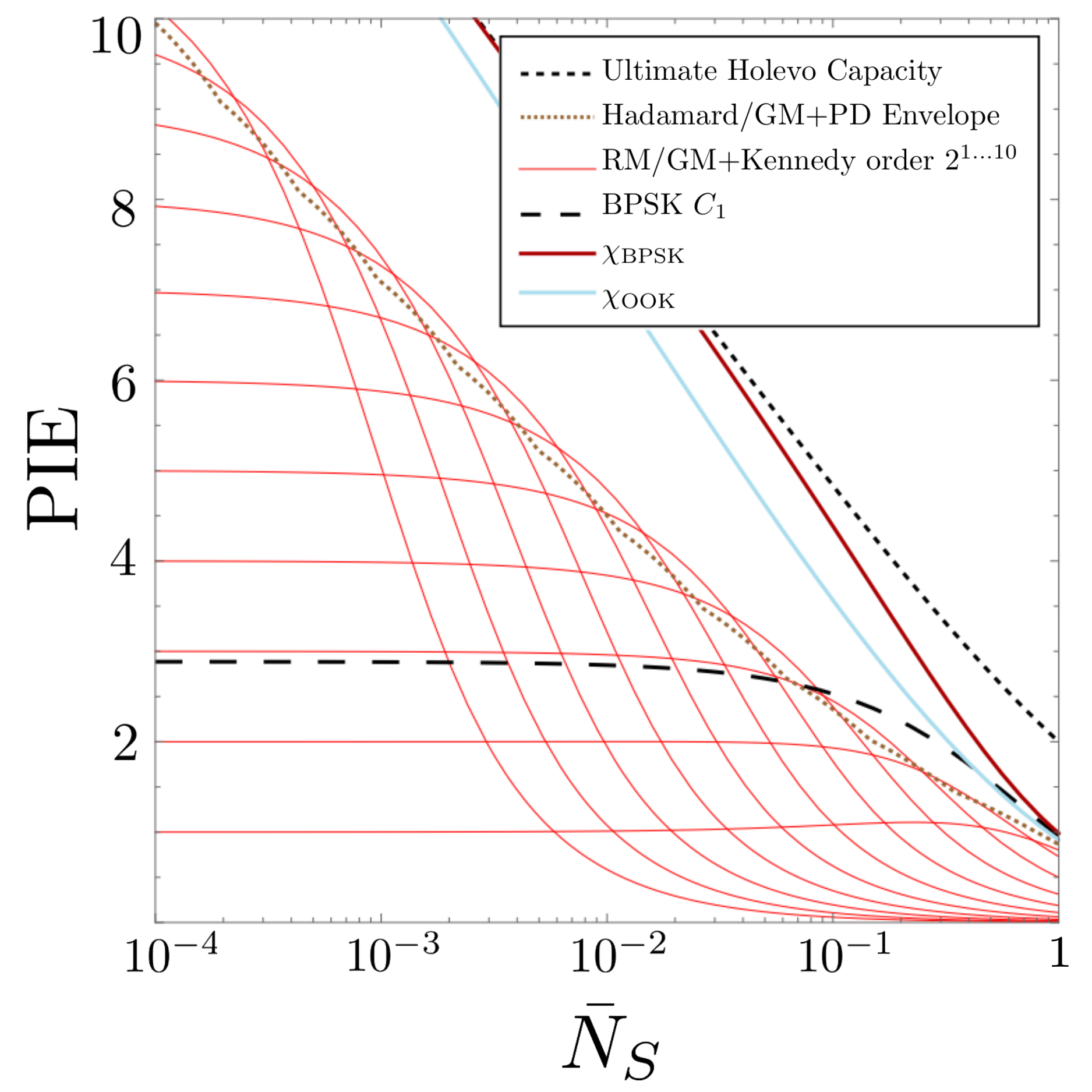}
    \caption{Photon information efficiency (bits per photon) plotted as a function of the mean transmitted photon number per mode ${\bar N}_S$, for the lossless ($\eta = 1$), noiseless (${\bar N}_B=0$) bosonic channel. These plots can be also interpreted as the bits per photon attained for the pure-loss bosonic channel ($\eta \in (0, 1]$), with the x-axis re-labeled $\eta {\bar N}_S$. A detailed description of the assumptions on the modulation and receiver choices for the various plots, appear in the main text.}
    \label{fig:classcaps}
\end{figure}
In Fig.~\ref{fig:classcaps}, we plot the Photon Information Efficiency (PIE)---the bits-per-mode capacity divided by the mean photon number transmitted per mode ${\bar N}_S$---for the ultimate Holevo limit, i.e., $C(1, {\bar N}_S, 0)/{\bar N}_S$ (black fine-dashed line), the BPSK Holevo capacity, i.e., $\chi_\text{BPSK}(1, {\bar N}_S, 0)/{\bar N}_S$ (dark-red solid line), and the OOK Holevo capacity, i.e., $\chi_\text{OOK}(1, {\bar N}_S, 0)/{\bar N}_S$ (light blue solid line). Since each of these PIEs scale as $-\ln{\bar N}_S + 1$ (in nats per photon) up to the second-order term in the small-${\bar N}_S$ limit, it is not surprising that each of these three Holevo-capacity PIE plots converge to one another as ${\bar N}_S \to 0$. Intuitively, this implies that when ${\bar N}_S \ll 1$, the BPSK constellation (and even the OOK constellation) suffices to closely achieve the ultimate Holevo capacity. This is because in that regime, the density operator of the average state for the equi-prior coherent-state BPSK alphabet closely approximates that of the optimal Gaussian-prior coherent-state alphabet. The PIE plots in Fig. \ref{fig:classcaps} can be also interpreted as the bits per photon attained for the pure-loss bosonic channel ($\eta \in (0, 1]$), with the x-axis re-labeled $\eta {\bar N}_S$. This is because a coherent state $|\alpha\rangle$ transmitted through the pure-loss channel ${\cal N}_\eta^{{\bar N}_B}$, ${\bar N}_B=0$, results in coherent state $|\sqrt{\eta}\alpha\rangle$ at the channel's output. 

With the BPSK alphabet $\left\{|\alpha\rangle, |-\alpha\rangle\right\}$, and assuming $\eta = 1$, ${\bar N}_B = 0$, if the receiver uses the optimal symbol-by-symbol detection scheme---the Helstrom measurement (realized by the Dolinar receiver~\cite{Hel76,Dol73}) that achieves the minimum probability of error of discriminating between the BPSK coherent states $\left\{|\alpha\rangle, |-\alpha\rangle\right\}$ given by $q = \frac12[1-\sqrt{1-e^{-4{\bar N}_S}}]$---the receiver induces a binary symmetric channel (BSC) for the detection of each received mode. This results in a (Shannon) capacity $C_1(1, {\bar N}_S, 0) = 1 - h_2(q)$ bits per mode. It is simple to see that $\lim_{{\bar N}_S \to 0} [C_1(1, {\bar N}_S, 0)]/{\bar N}_S = 2/\ln 2 \approx 2.89$ bits per photon~\cite{Tak14,Chu17}. This PIE is plotted by the black coarse-dashed line in Fig. \ref{fig:classcaps}. Closing the PIE gap between this BPSK $C_1$ and the BPSK Holevo limit $\chi_\text{BPSK}$ requires joint detection receivers (JDRs) that act collectively on long BPSK-modulated code words. The fact that a receiver acting collectively on multiple modulated symbols can achieve a fundamentally higher per-symbol communication rate is often termed {\em superaddivity} of capacity.

\subsection{Joint detection receivers for superadditive capacity with BPSK modulation}\label{sec:classicalcommJDR}
\begin{figure}
    \includegraphics[width=\linewidth]{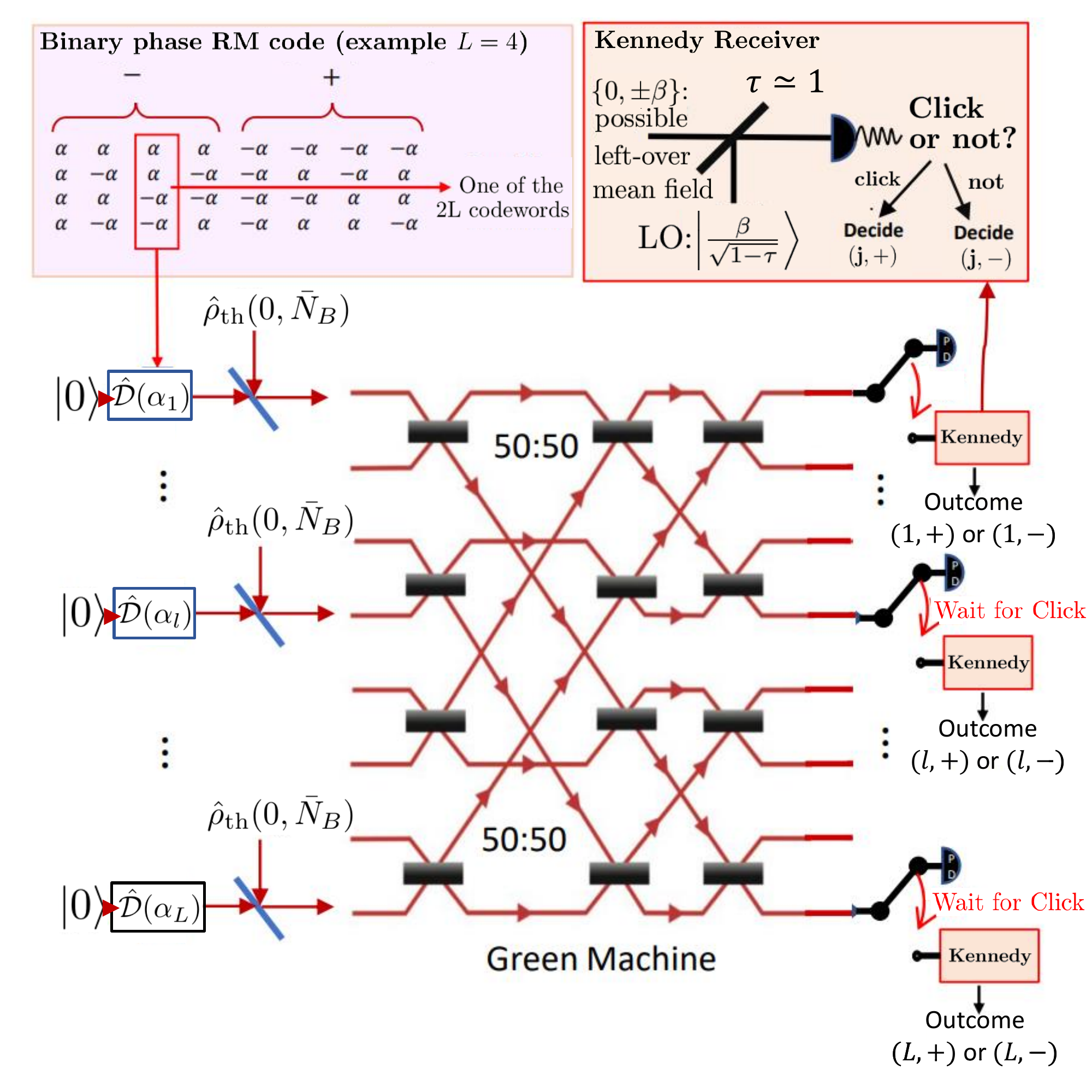}
    \caption{Schematic of a JDR for BPSK-modulated Hadamard and the 1st order Reed Muller (RM) codes. The Green Machine (GM) receiver front-end transforms $L$ Hadamard code words into $L$-ary pulse-position modulation (PPM) with a single pulse of mean field $\sqrt{L\eta}\alpha$ at one output mode. Single-photon detectors at each of the $L$ outputs of the GM are used to identify the pulse-containing mode. If a RM code is used, the GM transforms the $2L$ RM code words into binary-phase-coded PPM. Single photon detectors are employed at each GM output to identify the pulse-containing mode. An electro-optic switch directs the remaining pulse (in the pulse-containing mode), after the arrival of the first photon click, to a Kennedy receiver~\cite{Ken73} to identify the binary phase of that pulse.}
    \label{fig:GMreceiver}
\end{figure}
Two JDR designs for BPSK modulation were proposed~\cite{Guh11b,Guh11a}, depicted collectively in Fig.~\ref{fig:GMreceiver}, employed the Hadamard code~\cite{Guh11b} and the first-order Reed Muller (RM) code~\cite{Guh11a} respectively, with the receiver employing a linear-optical front end known as the Green Machine (GM). There are $L$ Hadamard code words each $L$-mode long, whereas there are $2L$ RM code words of the same length. Assuming $\bar{N}_B=0$,The GM, a linear-optical circuit made out of $(L \log L)/2$ 50-50 beamsplitters sketched in Fig.~\ref{fig:GMreceiver}, turns the $L$-mode BPSK Hadamard code book of mean field $\sqrt{\eta{\bar N}_S}$ per mode, into an $L$-ary pulse-position modulation (PPM) with a $\sqrt{L \eta{\bar N}_S}$ mean field displaced thermal state pulse ${\hat{\rho}}_\text{th}(\sqrt{L}\eta{\bar N}_S,\bar{N}_B^{'})$ in one of $L$ output modes of the GM (and a zero-mean thermal state of mean photon number $\bar{N}_B^{'}=(1-\eta)N_B$ in the remaining $L-1$ modes). Using shot-noise-limited single photon detectors (PD in legend of Fig. \ref{fig:GMreceiver}) at each of the $L$ output modes induces an $L$-input $L+1$-output symmetric erasure channel, with the erasure probability $p_e = e^{-L {\bar N}_S}$. The capacity achieved by this JDR, when $\eta=1$, in bits per mode, is the Shannon capacity of this channel, $(1-p_e)\log(L)$, divided by $L$. The envelope of the PIE attained by this BPSK code-JDR combination, for $\bar{N}_B=0$, $L = 2, 2^2, \ldots, 2^{10}$, is plotted by the light brown dotted line in Fig.~\ref{fig:classcaps}. This PIE exceeds $C_1$, and hence exhibits super-additive capacity~\cite{Guh11b}. 

If the RM code is used, at the output of the GM one obtains a binary-phase-modulated PPM, i.e., one of the $L$ modes carrying one of two displaced thermal states $\hat{\rho}_\text{th}(\pm \sqrt{L\eta{\bar N}_S},\bar{N}_B^{'})$ and the rest carrying thermal states $\hat{\rho}_\text{th}(0,\bar{N}_B^{'})$. Building upon the receiver design in~\cite{Guh11a}, we propose and analyze the following receiver for the BPSK RM code. Each of the $L$ outputs of the GM is detected by a single photon detector. The first detected click in any of those detectors is used to identify the `pulse-containing' mode. If no click occurs, it is deemed an erasure. If a click is detected, the remainder of the (displaced thermal state) pulse in this identified pulse-containing mode has mean field 0, $\beta$ or $-\beta$ where $\beta \in \left[0, \sqrt{L{\bar N}_S}\right)$ is calculated based on the actual arrival time of the first photon click. This remaining pulse is fed into a Kennedy receiver~\cite{Ken73}, which applies a coherent displacement of $\beta$ prior to detecting it, to identify the binary phase of the pulse in that mode. A click of the Kennedy receiver after nulling is more likely to occur if the pulse prior to the nulling was $\beta$, hence the receiver decides on $(+,j)$ if a click occurs, where $j$ is the index of the pulse-carrying mode. The notation $(\pm,l)$ for $1\leq l\leq L$ labels the RM code words such that $l$ is the index of associated Hadamard code word and $\pm$ determines the parity of the code word (since each Reed-Muller code is either a Hadamard code word or the negative of one, as depicted in the top left box of Fig. \ref{fig:GMreceiver}). The Shannon capacity of the $2L$-input $2L+1$ output channel thereby induced, divided by $L$, is the bits-per-mode capacity attained by the above BPSK code-JDR pair, whose PIE we plot for $\eta=1$ and $\bar{N}_B=0$ using the interweaving red curves in Fig.~\ref{fig:classcaps}, for $L = 2, 2^2, \ldots, 2^{10}$~\footnote{This capacity calculation of the coherent-state RM-code along with GM plus Kennedy-received based JDR emerges as a special case (i.e., with the thermal noise mean photon numbers set to zero) of the entanglement-assisted capacity calculation of the TMSV-assisted RM binary phase code modulated JDR2 that we describe in detail in Section~\ref{sec:capacityJDR2}.}. Even though the envelope of these PIE plots is higher than that of the Hadamard code GM-JDR combination, their scaling is inferior to that of the BPSK Holevo limit. 

\subsection{Code and receiver designs to achieve the ultimate Holevo capacity}\label{sec:receiverforHolevo}
In Ref.~\cite{Chu17}, it was proven that no matter which coherent-state (amplitude and/or phase) modulation format is employed, and no matter what error correction code is used, if the JDR is built with passive linear optics, coherent-state local oscillators, shot-noise-limited photon detectors, and electro-optic feedforward, in the small-${\bar N}_S$ limit, assuming $\eta = 1$ and ${\bar N}_B=0$, the capacity achieved by such a linear-optical JDR must scale as:
\begin{equation}
C_{\text{LO-JDR}}({\bar N}_S) = - {\bar N}_S \ln {\bar N}_S - \ln \ln (1/{\bar N}_S) + {\text{h.o.t.}},
\label{eq:capacityLOJDR}
\end{equation}
in units of nats per transmitted mode. The difference between the second term in Eq.~\eqref{eq:capacityLOJDR}, i.e., $- \ln \ln (1/{\bar N}_S)$, and the second terms in Eqs.~\eqref{eq:capacityHolevoscaling},~\eqref{eq:capacityBPSKHolevoscaling},~\eqref{eq:OOKHol}, i.e., $+{\bar N}_S$ captures the scaling gap between the PIE plots of the GM-based JDRs and the Holevo-limit PIEs in Fig.~\ref{fig:classcaps}.

In order to close this gap to the Holevo limit, one would need the JDR to employ quantum transformations that go beyond the linear-optical photon-detection based JDRs considered in~\cite{Chu17}. Referring back to Fig.~\ref{fig:classicalcomms} and its caption, when the code length $L$ is large, and the code rate $R < C$, where $C$ is the Holevo capacity (of the unrestricted, or of the specific modulation format in question), the quantum states of the $2^{LR}$ coherent-state code words---either of the random code as described in Section~\ref{sec:Holevocapacity} or a structured good code---are close to being mutually orthogonal. For this reason, there are multiple JDR measurements that can tell them apart with an error rate that goes to $0$ as $L \to \infty$, thereby achieving reliable communications at a rate approaching $C$. 

Examples of such JDR prescriptions are as follows:
\begin{enumerate}
\item {\textbf{Square root measurement}}---The {\em square-root measurement} (also known as the `pretty-good measurement') acts on the $L$-mode code word of a random code book~\cite{Hol98,Sch97} either in the optical domain, or in a matter-qubit processor, after faithfully transducing the optical quantum state of the received code word into a qubit register. 

\item {\textbf{Sequential decoding}}---This receiver acts on the received $L$-mode code word of a random code book, with a coherent displacement (realized using an array of highly-transmissive beamsplitters and strong coherent-state local oscillators) matched to the negative amplitude of the $j$-th code word, followed by applying an $L$-mode `vacuum-or-not' (VON)---a non-destructive binary projective measurement described by POVM elements $\left\{|0\rangle^{\otimes L}\,{}^{\otimes L}\langle 0|, {\hat I} - |0\rangle^{\otimes L}\,{}^{\otimes L}\langle 0|\right\}$---stopping with decision ${\hat j} = j$ if it gets the `vacuum' outcome, and if not, undoes the displacement on the $L$-mode post-measurement output state, increments $j \to j+1$ and repeats the process~\cite{Wil12}.

\item {\textbf{Polar code with successive cancellation}}---Encodes the information using a classical-quantum Polar code~\cite{Wil13,Wil13a}, and implements---all-optically or in a matter-qubit domain---the {\em successive cancellation decoder} to decode the $R$ bits successively.

\item {\textbf{Unambiguous state discrimination}}---Implements the $2^{LR} + 1$ outcome {\em unambiguous state discrimination} (USD) measurement on the received $L$-mode code word of a random code, which unambiguously identifies the transmitted code word or produces an erasure outcome~\cite{Tak13}.

\item {\textbf{Belief propagation with quantum messages}}---Couples the coherent state in each received mode (of a $N$-ary constellation) of the $L$-mode code word to the state of $\log N$ qubits of a quantum computer, while preserving the relative inner-products of the quantum states of the modulation constellation, followed by acting upon that $L\log N$ qubit register with a quantum belief-propagation decoder~\cite{Rengas20,Delaney21}.
\end{enumerate}

\section{Entanglement enhanced classical communications}\label{sec:EAcap_review}
In parallel with section \ref{sec:classcap_review}, we now describe the task of classical communication over a lossy noisy bosonic channel with the added resource of pre-shared entanglement. In subsection \ref{sec:EAHolevocapacity} we review how the Holevo capacity is defined when the receiver has access to a mode entangled with the transmitted signal for each transmitted symbol. Then subsection \ref{sec:EA_binarymodulation} restricts to the case of binary modulation and TMSV entanglement, and subsection \ref{sec:EAreceiverprinciples} describes the working principle for the transceivers introduced in detail in section \ref{sec:JDRdesign}.

\subsection{Entanglement-assisted capacity of the lossy thermal-noise bosonic channel}\label{sec:EAHolevocapacity}
If the transmitter Alice and the receiver Bob pre-share an unlimited amount of entanglement, with ${\bar N}_{\rm S}$ still denoting the mean number of photons transmitted over the channel ${\cal N}_\eta^{{\bar N}_{\rm B}}$ per mode, the capacity for sending classical data (in bits per mode) increases above the Holevo capacity $C(\eta, {\bar N}_{\rm S}, {\bar N}_{\rm B})$ in Eq.~\eqref{eq:UltHol} to~\cite{BSST,Gio03,Hol01,Hol02,Hol03,Hol04} Eq.~\eqref{eq:CE}, \iffalse:
\begin{equation}
C_{\rm E}(\eta, {\bar N}_{\rm S}, {\bar N}_{\rm B}) = g({\bar N}_{\rm S}) + g({\bar N}_{\rm S}^\prime) - g(A_+) - g(A_-),
\label{eq:CE}
\end{equation}
where $C_{\rm E}$ is the {\em entanglement assisted classical capacity} of the quantum channel ${\cal N}_\eta^{{\bar N}_{\rm B}}$, and
$
A_{\pm} = \frac12(D-1 \pm ({\bar N}_{\rm S}^\prime - {\bar N}_{\rm S})), 
$
with
$
D = \sqrt{({\bar N}_{\rm S} + {\bar N}_{\rm S}^\prime + 1)^2 - 4\eta {\bar N}_{\rm S}({\bar N}_{\rm S}+1)}.
$\fi
scaling like Eq.~\eqref{eq:scaling} in the regime of a low-brightness transmitter (${\bar N}_{\rm S} \ll 1$) and high thermal noise (${\bar N}_{\rm B} \gg 1$), \iffalse
\begin{equation}
{C_{\rm E}}/{C} \approx \ln\left({1}/{{\bar N}_{\rm S}}\right),
\label{eq:scaling_CE}
\end{equation}\fi
which tends to infinity as ${\bar N}_{\rm S} \to 0$~\cite{Shi19}. The goal of this paper is to explore transmitter-receiver designs that can achieve this $\ln\left({1}/{{\bar N}_{\rm S}}\right)$ enhancement in capacity over the Holevo limit.

Intuitively, the scaling ${C_{\rm E}}/{C} \sim \ln\left({1}/{{\bar N}_{\rm S}}\right)$ follows from the dominant term in the expression for $C_{\rm E}$ as ${\bar N}_{\rm S}\to 0$, being $-{\bar N}_{\rm S}\log {\bar N}_{\rm S}$ for any constant ${\bar N}_{\rm B}>0$, while the Taylor series expansion of $C$ at ${\bar N}_{\rm S}=0$ yields $C={\bar N}_{\rm S}\log\left(1+((1-\eta){\bar N}_{\rm B})^{-1}\right)+o({\bar N}_{\rm S})$.
Formally, one can use L'H\^{o}pital's rule to obtain the following limit:
\begin{align}
\label{eq:CEbyClogNs_limit}\lim_{{\bar N}_{\rm S}\to0}\frac{C_{\rm E}}{C\ln \left(\frac{1}{{\bar N}_{\rm S}}\right)}&=\frac{1}{(1+(1-\eta){\bar N}_{\rm B})\ln\left(1+\frac{1}{(1-\eta){\bar N}_{\rm B}}\right)},
\end{align}
which yields the scaling.  Note that the right hand side (RHS) of \eqref{eq:CEbyClogNs_limit} is zero when ${\bar N}_{\rm B}=0$. This is consistent with the known fact that the ratio $C_{\rm E}/C\leq2$ in the noiseless (${\bar N}_{\rm B}=0$) regime.

The plot of $C_{\rm E}/C$ as a function of ${\bar N}_{\rm S}$ and ${\bar N}_{\rm B}$, shown in Fig. ~\ref{fig:capacity_ratio} for $\eta = 0.01$ yields further insight. At optical frequencies, at 300K, the Planck-Law-limited thermal mean photon number per mode ${\bar N}_{\rm B}$ ranges between $10^{-5}$ to $10^{-6}$. At such small ${\bar N}_{\rm B}$, despite the scaling in \eqref{eq:scaling}, the actual capacity ratio is essentially at or below $2$ ($C_{\rm E}/C$ is at most $2$ when ${\bar N}_{\rm B}=0$) over the range of ${\bar N}_{\rm S}$: $10^{-6}$ to $10^2$; and hence the entanglement enhancement is significant only for the meaninglessly-small values of ${\bar N}_{\rm S}$. However, at ${\bar N}_{\rm B} = 100$ corresponding to the thermal noise at microwave wavelengths, $\eta = 10^{-3}$, and $\ns = 10^{-3}$, reasonably achieved with a spontaneous-parametric downconversion (SPDC) based entanglement source employed by our proposed transceiver, $C_{\rm E}/C \approx 7$. This is a substantial improvement over the highest capacity achievable without leveraging pre-shared entanglement.

It was shown in \cite{Gio03} that, in order to achieve $C_{\rm E}$ in Eq.~\eqref{eq:CE}, it suffices for Alice and Bob to pre-share many copies of the following entangled state:
\begin{equation}
|\psi\rangle_{\rm SI} = \sum_{n=0}^\infty \sqrt{\frac{{\bar N}_S^n}{(1+{\bar N}_S)^{1+n}}} \, |n\rangle_{\rm S} |n\rangle_{\rm I},
\label{eq:TMSVexpression}
\end{equation}
known as the {\em two-mode squeezed vacuum} (TMSV). Let us consider a pulsed SPDC source that produces entangled signal and idler pulses, each of duration $T$ seconds and optical bandwidth $W$ (typically $1$-$2$ THz) around their respective center frequencies. The quantum description of the pair of signal-idler entangled pulses is $|\psi\rangle_{\rm SI}^{\otimes M}$, where $M \approx WT$ is the number of mutually-orthogonal temporal modes in each of the signal and idler pulses. 

\subsection{Entanglement-assisted capacity with TMSV entanglement and binary modulation constellations}\label{sec:EA_binarymodulation}

In Ref.~\cite{Zhu17a}, it was shown that not only does pre-sharing TMSV entangled states suffice to attain $C_{\rm E}$, but that phase modulation of the signal mode (at Alice's end) and transmitting that to Bob suffices to attain $C_E$ in the regime of low signal-to-noise (SNR) ratio (${\bar N}_S \ll 1$ and ${\bar N}_B \gg 1$). This low-SNR is the regime where the scaling ${C_{\rm E}}/{C} \sim \ln\left({1}/{{\bar N}_{\rm S}}\right)$ holds true, and hence of interest here. 

\begin{figure}[h]
\centering
\begin{subfigure}{0.95\linewidth}
    \includegraphics[width=\linewidth]{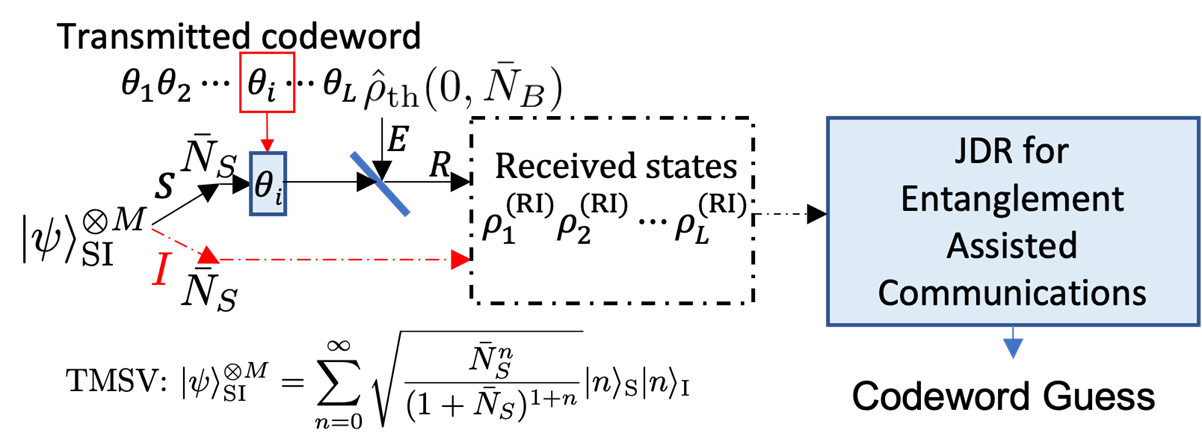}
    \caption{Transmission of a $L$-length phase code word, where each phase is modulated on to an $M$-temporal-mode signal pulse on Alice's end of the pre-shared SPDC-generated entanglement. The mean photon number per mode of each received mode $R$, $\bar{N}_S^\prime = \eta \bar{N}_S + (1-\eta) \bar{N}_B$. Bob's JDR acts on $2ML$ modes.}
    \label{fig:seqeachannel}
\end{subfigure}
\begin{subfigure}{0.95\linewidth}
    \includegraphics[width=\linewidth]{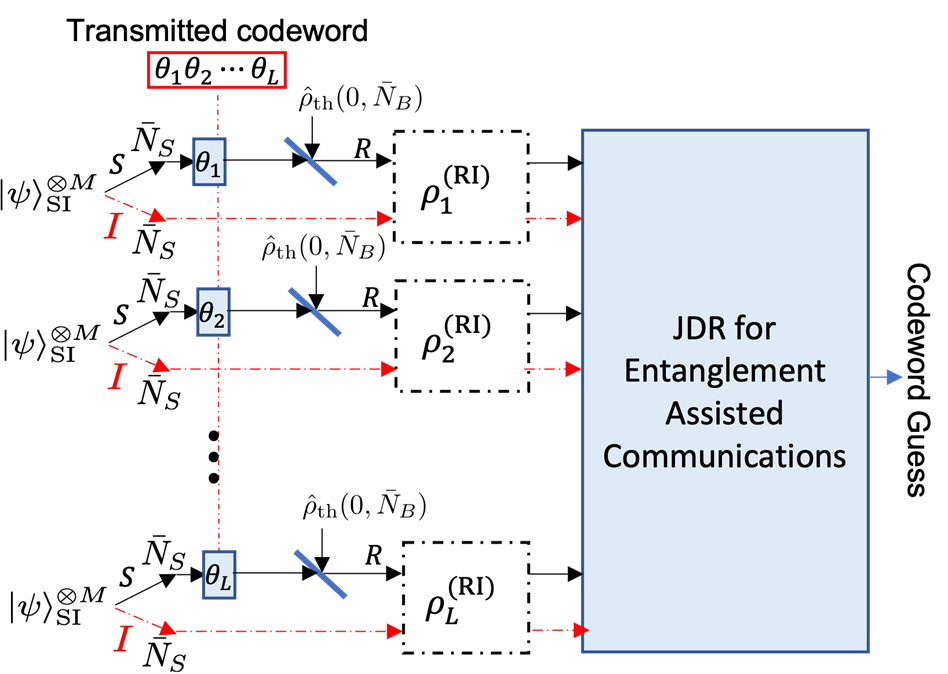}
    \caption{An equivalent parallel representation of the channel, explicitly showing that Bob's received code word comprises of $L$ `symbols' $(\hat{\rho}_1^{{\rm RI}}, \ldots, \hat{\rho}_L^{{\rm RI}})$, where each symbol is uncorrelated and un-entangled with the others. The $M$ received-idler mode pairs within each symbol hold phase-sensitive cross-correlations, i.e., $\langle {\hat a}_R {\hat a}_I \rangle \ne 0$, which encode information about the phase modulated by Alice on the $M$ modes of that transmitted symbol.}
    \label{fig:seqeachannel}
\end{subfigure}
\caption{Diagram of entanglement-assisted communication of classical data over a lossy, noisy quantum channel. The S, I, E and R modes are labels, respectively, for: the initial signal mode (to be sent over the channel ${\cal N}_\eta^{{\bar N}_B}$ to the receiver), the idler mode retained at the receiver (entangled with the corresponding S mode), the environment mode in a thermal state of mean photon number $\bar{N}_B$, and the received mode at the front end of the receiver.}
\label{fig:eacomms}
\end{figure}
Pursuant to phase modulation of the signal modes of pre-shared TMSV states generated using an SPDC source, a schematic system diagram for entanglement-assisted communications is depicted in Fig.~\ref{fig:eacomms}. Alice maps her message to a code book $\mathcal{C}$ (with $|\mathcal{C}|\leq2^{Lr}$), each of whose code words are $L$ phases, e.g., $(\theta_1, \ldots, \theta_L)$. Alice modulates an SPDC-generated signal pulse, i.e., $M$ temporal modes, whose entangled idler-mode counterparts are assumed pre-shared with Bob, with a single phase symbol of that code word. So, transmission of one code word consumes $ML$ uses of the single-mode lossy thermal-noise bosonic channel ${\cal N}_\eta^{{\bar N}_B}$. The receiver Bob acts on $2ML$ modes---$ML$ modes of Alice's code word, corrupted by channel noise, that he receives, and the corresponding $ML$ pre-shared idler modes---with a joint detection receiver (JDR), to produce a guess of the transmitted code word. Bob's received code word comprises of $L$ `symbols' $(\hat{\rho}_1^{{\rm RI}}, \ldots, \hat{\rho}_L^{{\rm RI}})$, where each symbol is uncorrelated and un-entangled with the other symbols. Further, it is instructive to note that even though the signal-idler modes were entangled, there is no entanglement left amid the received-idler mode pairs within each symbol, due to the entanglement-breaking lossy-noisy channel ${\cal N}_\eta^{{\bar N}_B}$. However, the received-idler mode pairs within each symbol hold phase-sensitive cross-correlations, i.e., $\langle {\hat a}_R {\hat a}_I \rangle$, which encode information about the phase value modulated by Alice on that symbol at the transmitter. If Bob is able to faithfully recover the transmitted message, let the reliable-communications rate achieved by the above scheme be $r$ modes per transmitted symbol. Then since each transmitted symbol comprises of $M$ temporal modes, the rate in bits per mode, $R = r/M$. 

\begin{figure}
    \includegraphics[width=\linewidth]{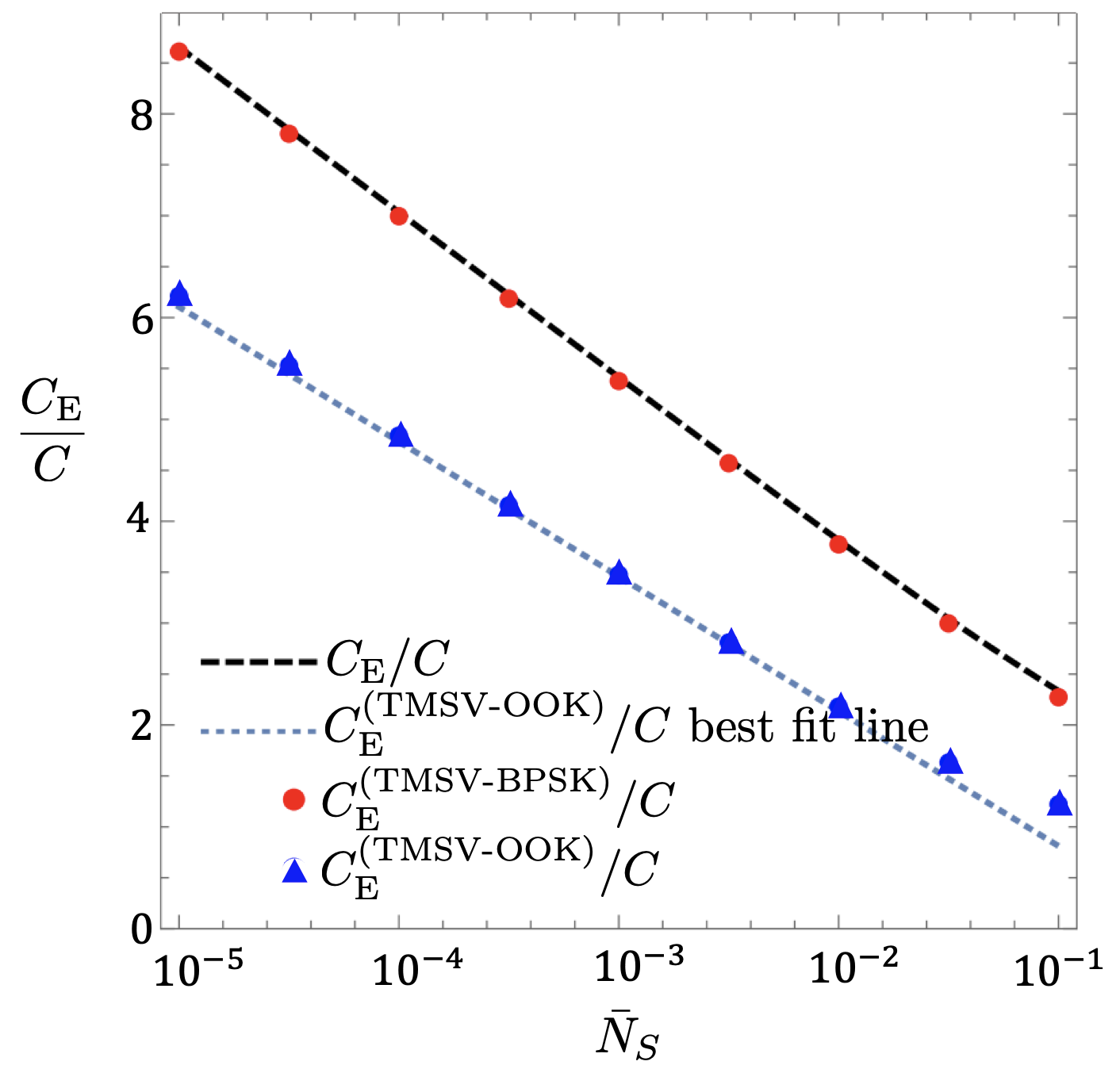}
    \caption{Ratios of the entanglement-assisted capacities for specific binary modulation formats and the ultimate classical (Holevo) capacity without pre-shared entanglement $C$, plotted as a function of $\bar{N}_S$ for $\eta=0.1$ and $\bar{N}_B=1$. $M=1$ is assumed for all the $C_E$ plots.}
    \label{fig:eacaps}
\end{figure}
In Fig. \ref{fig:eacaps}, we further show that restricting the aforesaid phase modulation to binary phase shift keying (BPSK), i.e., each of the $L$ symbols in Alice's code words comprising only of two phases $\left\{0, \pi\right\}$, suffices to attain $C_E$ in the low SNR limit. This observation is reminiscent of the coherent state BPSK modulation closely approaching the Holevo limit for unassisted classical communications, as shown in Fig.~\ref{fig:classcaps}. We further evaluate the performance of an on-off keying (OOK) modulation format, wherein Alice's transmitted symbols are either an `on' pulse, which is an $M$-temporal-mode signal pulse of her pre-shared SPDC-entanglement, or an `off' pulse, which is Alice staying silent for an entire pulse duration. To evaluate the entanglement-assisted communication rates using the BPSK and the OOK modulation formats, $C_{\rm E}^{({\text{TMSV-BPSK}})}$ and $C_{\rm E}^{({\text{TMSV-OOK}})}$ respectively, we calculate the (unassisted) Holevo capacities of the modulated received-idler code words at Bob's end. In other words, for BPSK modulation, the maximum entanglement-assisted rate is:
\begin{equation}
C_{\rm E}^{({\text{TMSV-BPSK}})} = \max_{p \in [0, 1]}\frac{\left\{S({\bar{\hat{\rho}}}_{RI}) - {\bar S}\right\}}{M} \, {\text{bits per mode}},
\label{eq:CE_TMSV_BPSK}
\end{equation}
with ${\bar{\hat{\rho}}}_{RI} = p \left({\hat{\rho}}_{RI}^{(0)}\right)^{\otimes M} + (1-p) \left({\hat{\rho}}_{RI}^{(\pi)}\right)^{\otimes M}$, where ${\hat{\rho}}_{RI}^{(\theta)}$ is a two-mode Gaussian state obtained by applying a single-mode phase $e^{i \theta {\hat a}_S^\dagger {\hat a}_S}$ on the signal mode of a two-mode TMSV state of mean photon number per mode ${\bar N}_S$, followed by the transmission of the signal mode through the channel ${\cal N}_\eta^{{\bar N}_B}$. Further, ${\bar S} = pMS\left({\hat{\rho}}_{RI}^{(0)}\right) + (1-p)MS\left({\hat{\rho}}_{RI}^{(\pi)}\right)$. For on-off-keying (OOK) modulation,
\begin{equation}
C_{\rm E}^{({\text{TMSV-OOK}})} = \max_{p \in [0, 1]}\frac{\left\{S({\bar{\hat{\rho}}}_{RI}) - {\bar S}\right\}}{M} \, {\text{bits per mode}},
\label{eq:CE_TMSV_OOK}
\end{equation}
with ${\bar{\hat{\rho}}}_{RI} = p \left({\hat{\rho}}_{RI}^{({\rm on})}\right)^{\otimes M} + (1-p) \left({\hat{\rho}}_{RI}^{({\rm off})}\right)^{\otimes M}$, where ${\hat{\rho}}_{RI}^{({\rm on})}$ is a two-mode Gaussian state obtained by the transmission of the signal mode of a two-mode TMSV state of mean photon number per mode ${\bar N}_S/p$ through the channel ${\cal N}_\eta^{{\bar N}_B}$, and ${\hat{\rho}}_{RI}^{({\rm off})} = \hat{\rho}_{\rm th}(0,{\bar N}_T) \otimes \hat{\rho}_{\rm th}(0,{\bar N}_S/p)$ is a product of two zero-mean thermal states, with ${\bar N}_T = (1-\eta){\bar N}_B$. Further, the average output entropy, ${\bar S} = pMS\left({\hat{\rho}}_{RI}^{({\rm on})}\right) + (1-p)MS\left({\hat{\rho}}_{RI}^{({\rm off})}\right)$. 

The optimal prior $p$ for the BPSK modulation is $1/2$, whereas the optimal on-prior $p$ for the OOK modulation (evaluated numerically) comes out to be much smaller than $1/2$ for ${\bar N}_S \ll 1$, similar to what was seen to be the case for coherent-state OOK-based classical communications, discussed in Section~\ref{sec:capacityBPSKOOK}. There is however an important difference between coherent-state OOK modulation for unassisted communications and TMSV-OOK modulation for entanglement-assisted communications. In the latter, even for the `off' symbols where Alice stays silent (transmits vacuum), she must still consume pre-shared entanglement, i.e., discard the signal pulses during off-symbol transmissions. Since the capacity calculations are done with a fixed mean transmitted photon number per mode ${\bar N}_S$, and since the off symbols---despite consuming pre-shared entanglement---do not transmit any photons over the channel, the pre-shared entangled TMSV states need to have ${\bar N}_S/p$ mean photon number per mode. Since the amount of entanglement (in ebits/mode) in a TMSV of mean photon number per mode ${\bar N}$ is given by $g({\bar N})$, which monotonically increases with $\bar N$, the OOK modulation format for entanglement assisted communications consumes more pre-shared entanglement compared to the BPSK method, and achieves a capacity inferior to that of BPSK, as shown in Fig.~\ref{fig:eacaps}. The advantage the OOK modulation format enjoys however is that no phase modulation is necessary at Alice's end. Further, the sparse on-symbol transmission allows Alice to use a pulse-position modulation (PPM) modulation-code over the OOK alphabet, which allows her to use good outer (classical) error correction codes to achieve the Shannon capacity of the transmitter-JDR pair in question, as described later in Section~\ref{sec:PPM}. 

All the plots in Fig. \ref{fig:eacaps}, except for that of $C_{\rm E}/C$, were generated numerically: by numerical evaluation of the capacity expressions in Eqs.~\eqref{eq:CE_TMSV_BPSK} and~\eqref{eq:CE_TMSV_OOK}, and a Mathematica-generated best-fit line for $C_{\rm E}^{({\text{TMSV-OOK}})}$ in the semi-logarithmic scale, to assess the scaling of the capacity ratio $C_{\rm E}^{({\text{TMSV-OOK}})}/C$. It was found (numerically) that $C_{\rm E}^{({\text{TMSV-OOK}})}/C \sim \ln(1/{\bar N}_S)$ in the ${\bar N}_S \ll 1$, ${\bar N}_B \gg 1$ regime, confirming that even an OOK modulation of pre-shared TMSV states can achieve the optimal scaling of the ratio to the unassisted Holevo capacity $C$. 

Each of the plots in Fig. \ref{fig:eacaps} assume $M=1$, i.e., each individual temporal mode of the SPDC-generated entanglement is individually modulated. This is hard to achieve in practice as doing this will require THz-class electro-optic modulators (EOMs). For the performance evaluation of the joint-detection receivers we present in Section~\ref{sec:JDRdesign}, we will use more reasonable values of modes-per-symbol $M$, e.g., $M = 10^5$ used for the rate plots in Fig.~\ref{fig:cap_ratio_M100000}, which for an SPDC optical bandwidth (modes per second) $W = 10^{12}$ Hz, translates to a much more reasonable EOM modulation bandwidth (modulated symbols per second) of $10^7$, i.e., $10$ MHz.

\subsection{Receiver design principles}\label{sec:EAreceiverprinciples}
Although the entanglement-assisted capacity of ${\cal N}_\eta^{{\bar N}_B}$ can be saturated by encoding information in the phase of the signal modes of pre-shared signal-idler SPDC mode pairs, it is not immediately clear how to design a structured receiver to extract this information on the receiver end. Initial guesses at designing a receiver could be to employ either an optical parametric amplifier (OPA) or the phase-conjugate receiver (PCR) proposed in~\cite{Guh09} in the context of building receivers for a quantum-illumination radar, to convert the information-bearing phase-sensitive cross correlations $\langle {\hat a}_R {\hat a}_I \rangle$ into a photon-number signature. Such a receiver acts jointly upon each received noisy-modulated symbol ($M$ temporal modes) and the corresponding $M$-mode idler pulse, either with an OPA or a PCR~\cite{Shi19}, followed by performing a maximum-likelihood determination of the phase encoded in that symbol. However, the bits-per-mode capacity attained by such a measurement that acts on one modulated {\em symbol} at a time as they are received, cannot exceed twice the ultimate classical Holevo capacity $C$ in the low $\bar{N}_S$, high $\bar{N}_B$ regime (see Appendix~\ref{app:OPA} for proof).

\begin{figure}
\centering
\includegraphics[width=.95\linewidth]{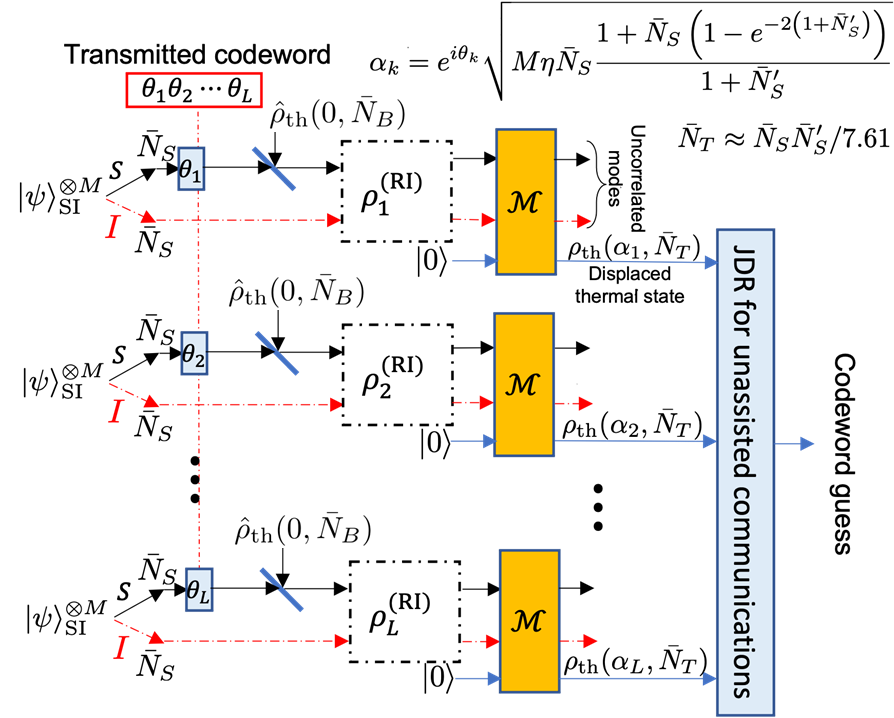}
\caption{The big picture view of the receivers proposed in this paper. Joint detection on multiple signal-idler mode pairs is made possible by using a quantum map (denoted $\mathcal{M}$) that depletes the cross-correlation between the received-idler mode pair, converting it to the coherent displacement of an ancilla mode.}
\label{fig:bigpicture}
\end{figure}
In order to design a receiver to achieve $C_E$, or for that matter to exceed $2C$ bits per mode, we must design a joint detection receiver (JDR)---along the lines of JDRs for superadditive classical communications, as described in Sections~\ref{sec:classicalcommJDR} and~\ref{sec:receiverforHolevo}, and illustrated in Fig. \ref{fig:eacomms}---that acts on multiple modulated symbols collectively. The challenge is to use a code word block of signal-idler mode pairs to produce a strong detectable signature (e.g., photon number at the output of the receiver) to pinpoint which code word was received. The way in which this is done in the receiver design proposed here is to first convert phase-sensitive cross-correlations in the return-idler mode pairs into the coherent displacement of a single bosonic mode using a non-linear sum-frequency generation (SFG) module, and then use a coherent state JDR as described in Sections~\ref{sec:classicalcommJDR} and~\ref{sec:receiverforHolevo} on the (code word of) coherently-displaced modes, as depicted in Fig. \ref{fig:bigpicture}. The next Section describes how this works, in further detail.

\section{Joint detection receiver design for entanglement-assisted communications}\label{sec:JDRdesign}

\subsection{Structure of the transmitter and receiver}

\begin{figure}
\centering
% \begin{minipage}[b][%0.55\paperheight
% ][t]{.7\columnwidth}
	\includegraphics[width=\columnwidth]{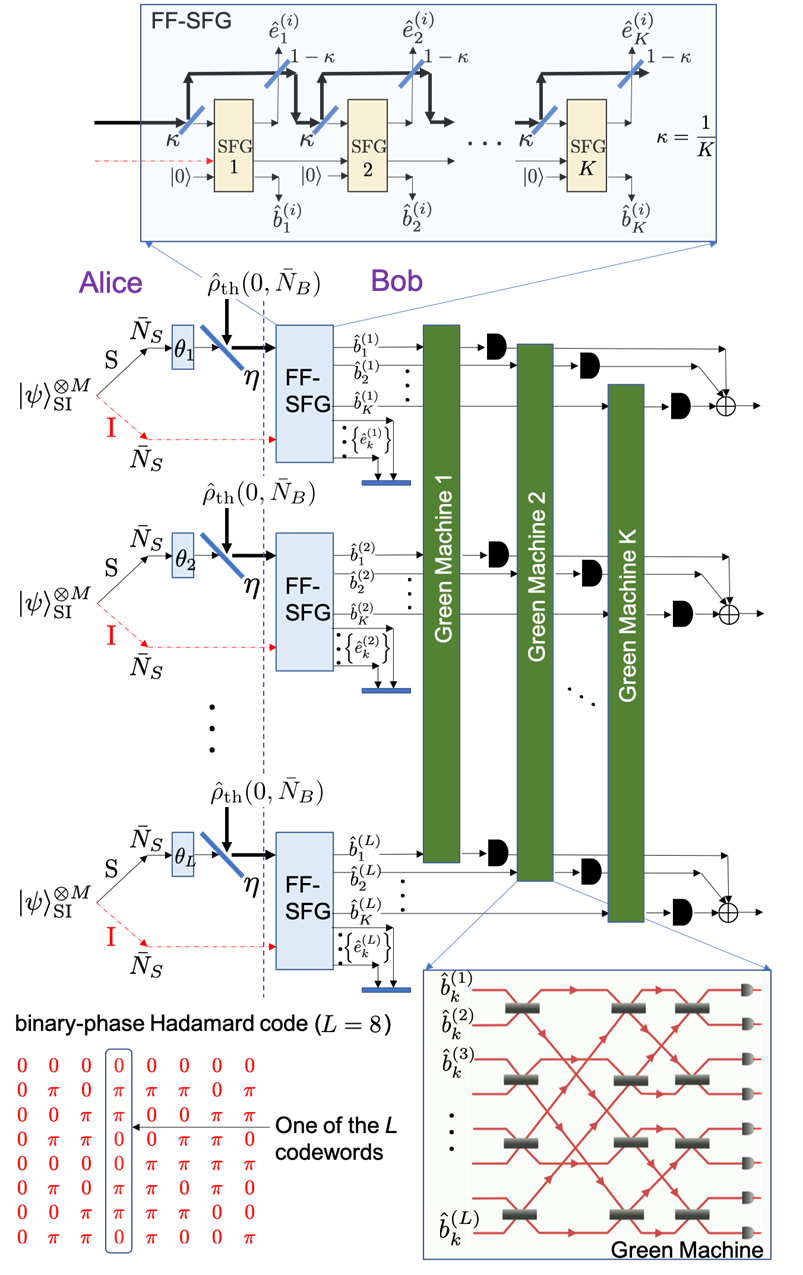}
	\caption{Our first transceiver design JDR1, which uses the BPSK Hadamard code, and applies a Green-Machine at the end of each SFG cycle to the $k^\text{th}$ set of sum-frequency modes $\{b_k^{(i)}\}_i$ and classically adds the outcomes of the post-green machine photon detectors. Thin and thick lines depict weak (e.g., ${\bar N}_{\rm S} \ll 1$) and strong (e.g., ${\bar N}_{\rm B} \gg 1$) mean photon number signals respectively. In an actual realization, only one $L$-mode Green Machine is needed as the sum-frequency modes ${\hat b}_k^{(i)}$, $1 \le k \le K$, for each $i \in \left\{1, \ldots, L\right\}$, appear in a temporal sequence.}
\label{fig:JDR}
% \end{minipage}
% \caption{Two SFG joint-detection receiver designs for EA-communications. Alice modulates $L$ blocks of $M$-fold repeated signal modes of pre-shared two-mode squeezed vacuum (TMSV) states, which are received by Bob after transmission through $ML$ uses of the single-mode lossy-noisy Bosonic channel ${\cal N}_\eta^{{\bar N}_{\rm B}}$. The receivers act on the received $ML$ modes, and $ML$ idler modes held by Bob, entangled with Alice's transmitted modes.} 
% \label{fig:JDRs}
\end{figure}
Consider the transmitter-receiver structure in Fig. ~\ref{fig:JDR}. Alice uses a binary phase shift keying (BPSK) modulation with a Hadamard code of order $L$, on the signal (S) modes of TMSV states $|\psi\rangle_{\rm SI}$ whose idler (I) modes are pre-shared with Bob. Let us assume $L$ is an integer power of $2$ such that a Hadamard code exists. The $l$-th `pulse' (comprising $M$ orthogonal temporal modes) of the signal output of a pulsed spontaneous parametric downconversion (SPDC) source, an $M$-fold tensor product TMSV $|\psi\rangle_{\rm SI}^{\otimes M}$, is modulated by Alice, with the binary phase $\theta_l \in \left\{0, \pi\right\}$. The transmission of an entire Hadamard code word thus consumes $L$ SPDC signal pulses, modulated with phases $\theta_l, 1 \le l \le L$, over $ML$ uses of the single-mode channel ${\cal N}_\eta^{{\bar N}_{\rm B}}$. The corresponding idler modes are assumed losslessly pre-shared with Bob, e.g., using a fault-tolerant quantum internet. Given that in our regime of interest: $(1-\eta){\bar N}_{\rm B} > \eta$, the channel ${\cal N}_\eta^{{\bar N}_{\rm B}}$ is entanglement-breaking. Alice's phase modulation of the signal mode ${\hat a}_{\rm S}$ of a TMSV state, followed by its transmission through ${\cal N}_\eta^{{\bar N}_{\rm B}}$, results in an output mode ${\hat a}_{\rm R}$ received by Bob with a (large) mean photon number ${\bar N}_{\rm S}^\prime = \eta {\bar N}_{\rm S} + (1-\eta){\bar N}_{\rm B}$. Mode ${\hat a}_{\rm R}$ and the (weak) idler mode of the TMSV ${\hat a}_{\rm I}$ with mean photon number ${\bar N}_{\rm S}$ held by Bob, are individually in zero-mean thermal states. However, their joint quantum state is a classically-correlated zero-mean Gaussian state (no longer {\em entangled}), with a phase sensitive cross correlation $\langle {\hat a}_{\rm R} {\hat a}_{\rm I} \rangle = \pm \sqrt{\eta {\bar N}_{\rm S} ({\bar N}_{\rm S}+1)}$, where the sign depends on the phase ($0$ or $\pi$) modulated by Alice. Note that the amount of cross correlation in the received state is proportional to the amount of cross correlation in the initially generated state which, being entangled, is correlated by an amount even beyond the maximum allowed by classical physics. So the term `entanglement-assisted' applies in spite of the entanglement-breaking channel.

The receiver employs the SFG, a non-linear optical process that runs SPDC in reverse per the Hamiltonian ${\hat H}_{\rm SFG} = \hbar g \sum_{m=1}^M \left({\hat b}^\dagger {\hat a}_{{\rm S}_m} {\hat a}_{{\rm I}_m} + {\hat b} {\hat a}_{{\rm S}_m}^\dagger {\hat a}_{{\rm I}_m}^\dagger\right)$, with $\hbar$ the reduced Planck constant, and $g$ the non-linear interaction strength. Signal-idler photon pairs from the $M$ input mode pairs $\left\{{\hat a}_{{\rm S}_m}, {\hat a}_{{\rm I}_m}\right\}$, $1 \le m \le M$, are up-converted to a sum-frequency mode $\hat b$. The phase-sensitive cross-correlation at the input of the SFG, $\langle {\hat a}_{{\rm S}_m}{\hat a}_{{\rm I}_m} \rangle$, manifests as the mean field amplitude of a thermal state of $\hat b$~\cite{Zhu17}.

A single-mode displaced thermal state with {\em mean field amplitude} $\alpha \in {\mathbb C}$, and {\em thermal-noise mean photon number}, ${\bar N} > 0$ has the following density operator:
\begin{equation}
\hat{\rho}_{\rm th}(\alpha, {\bar N}) = \int_{\mathbb C} \frac{1}{\pi {\bar N}}e^{-|\beta - \alpha|^2/{\bar N}}|\beta\rangle \langle \beta | d^2\beta.
\end{equation} 
For ${\bar N} = 0$, it reduces to the pure coherent state $|\alpha\rangle$. The photodetection statistics of $\hat{\rho}_{\rm th}(\alpha, {\bar N})$ is Laguerre-distributed~\cite{Hel76}. The probability that it produces zero clicks when detected with an ideal photon detector is 
\begin{equation}
\langle 0|\hat{\rho}_{\rm th}(\alpha, {\bar N}) |0\rangle = (1/({\bar N}+1))e^{-|\alpha|^2/({\bar N}+1)}. 
\end{equation}

As depicted at the top of Fig.~\ref{fig:JDR}, Bob inputs the received $M$ modes of each of the $L$ phase-modulated blocks of the Hadamard code, along with the corresponding $M$ idler modes (pre-shared with the transmitted block of $M$ signal modes), into a feed-forward (FF) SFG module. An FF-SFG module stacks $K$ SFG stages, each unitary corresponding to applying ${\hat H}_{\rm SFG}$ for a duration of $\frac{\pi}{2\sqrt{M}g}$, with $K$ beamsplitters and combiners of transmissivities $\kappa = 1/K$ and $1-\kappa$ respectively, as shown. The $K$-stage SFG ensures that the signal input of each SFG has much less than one photon per mode, so that we can use the ``qubit-approximation" analysis of the SFG from~\cite{Zhu17}. ${\hat b}_k^{(l)}$ denotes the sum-frequency mode of the $k$-th SFG, $1 \le k \le K$, of the $l$-th FF-SFG module, $1 \le l \le L$. 

In the $\kappa \ll 1/{\bar N}_{\rm B}$ limit, the sum-frequency mode ${\hat b}_k^{(l)}$ is in a displaced thermal state $\hat{\rho}_{\rm th}(\pm \alpha_k, {\bar N}_{\rm T})$~\cite{Zhu17}, where the $\pm$ sign depends on whether the mode block $i$ is modulated with phase $0$ or $\pi$. The mean $\alpha_k = \sqrt{M\kappa \eta {\bar N}_{\rm S}(1+{\bar N}_{\rm S})\mu^{k-1}}$, with $\mu = \left(1-\kappa(1+{\bar N}_{\rm S}^\prime)\right)^2$, and ${\bar N}_{\rm T} = \kappa {\bar N}_{\rm S}{\bar N}_{\rm S}^\prime$~\cite{Zhu17}. Let us also define $\bar{N}_k = |\alpha_k|^2$. 

\begin{figure*}[!ht]
\begin{minipage}[b][0.52\paperheight
][s]{2\columnwidth}
    \includegraphics[width=\textwidth]{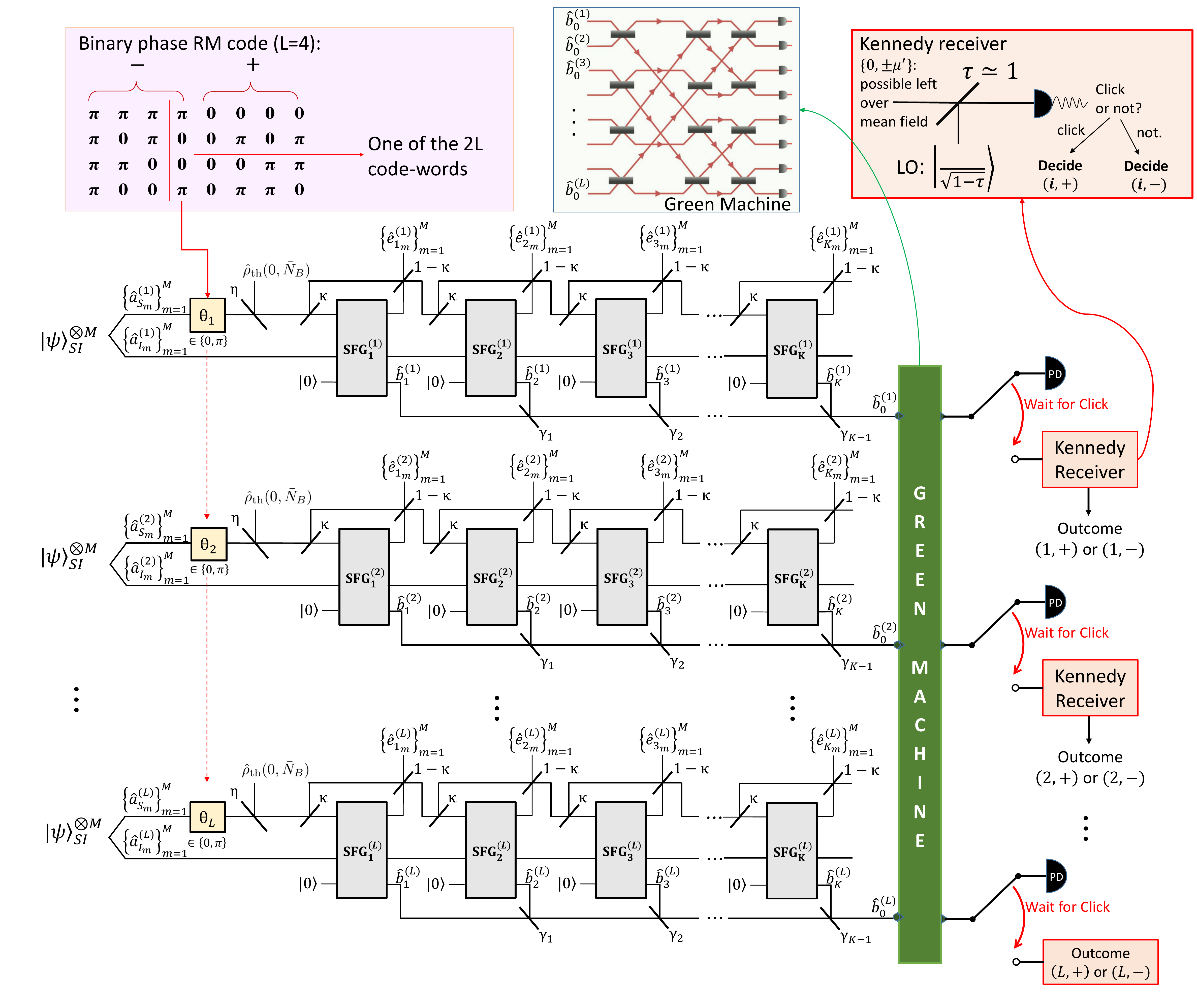}
\end{minipage}
    \caption{Our second (upgraded) transceiver design JDR2 that uses a more information-dense Reed-Muller (RM) code for the phase encoding and combines the $\{\hat{b}_k^{(l)}\}_k$ modes for each $l$ via an array of beam splitters into a single mode $\hat{b}_0^{(i)}$ before applying a single Green Machine stage and a Kennedy Receiver-augmented photon detection stage needed to decode the extra bit of information of the Reed-Muller code book (further details are contained in the text).}
    \label{fig:JDR2}
\end{figure*}

For a fixed $k$, the $L$ modes ${\hat b}_k^{(l)}$, $1 \le l \le L$, produced by the $k$-th FF-SFG gates are in a product of displaced thermal states with the same mean photon number ${\bar N}_{\rm T}$, but with mean field amplitudes $\alpha_k$ or $-\alpha_k$ corresponding to which Hadamard code word was transmitted. On the other hand, for a fixed $l$, the $K$ modes $\hat{b}_k^{(l)}$, $1\leq k\leq K$ can be approximated as having maximally correlated noise (see appendix \ref{app:sfganalysis}). Therefore, it is possible, for each $l$, to interfere the $K$ sum-frequency modes ${\hat b}_k^{(l)}$, $1 \le k \le K$ on an appropriately-tuned beam-splitter array to produce a displaced thermal state with mean thermal photon number $\approx K\bar{N}_T/7.61$, where $\bar{N}_T$ is the mean thermal photon number of each sum-frequency mode ${\hat b}_k^{(l)}$, $\forall k, l$, as described above (see Fig.~\ref{fig:JDR2} for a schematic). 

We will refer to the JDR design in Fig. \ref{fig:JDR}, proposed in~\cite{Guha2020EAJDR}, as JDR1. For each $k\in\{1,...,K\}$, the $L$ modes ${\hat b}_k^{(l)}$, $1 \le l \le L$ are input to an $L$-mode Green Machine (GM), a linear-optical circuit comprising $L\log_2(L)/2$ $50$-$50$ beasmplitters, denoted GM$_k$. The GM transforms the $L$-mode BPSK-modulated coherent-state Hadamard code word, e.g., $|\alpha_k, -\alpha_k, \ldots, \alpha_k\rangle$ into one of the $L$ code words of order-$L$ coherent-state pulse-position modulation (PPM), e.g., $|0, \ldots, \sqrt{L}\alpha_k, \ldots, 0\rangle$~\cite{Guh11b}. The bottom of Fig.~\ref{fig:JDR} shows an example binary-phase $L=8$ Hadamard code, and the circuit of an $8$-mode GM. At the output of GM$_k$, {\em one} of the $L$ output modes (based on the input Hadamard code word) is in a displaced thermal state $\hat{\rho}_{\rm th}(\sqrt{L}\,\alpha_k, {\bar N}_{\rm T})$. We call this the ``pulse-containing output" (mode). The remaining $L-1$ output modes of GM$_k$ are in the zero-mean thermal state $\hat{\rho}_{\rm th}(0, {\bar N}_{\rm T})$. 

All GM outputs are detected by single photon detectors~\cite{Guh11b}. The electrical outputs of the $l$-th detectors, $1 \le l \le L$, of each GM are classically combined---by integrating all photon clicks---into one single output that is monitored for zero or more clicks during each ($M$-temporal-mode) SPDC pulse interval. Since the $K$ modes ${\hat b}_k^{(l)}$, $1 \le k \le K$ in the $l$-th FF-SFG module are in a temporal sequence, we only need one $L$-mode GM and $L$ detectors to realize this receiver.

In Fig.~\ref{fig:JDR2}, we sketch a JDR variant, which we call JDR2. 
Here, the sum-frequency modes are combined using an array of beamsplitters before entering a single Green Machine. Namely, for each $l\in\{1,...,L\}$ the $K$ modes ${\hat b}_k^{(l)}$, $1 \le k \le K$ are combined on a set of beam-splitters with transitivities $\gamma_k$ chosen such that the modes $\hat{b}^{(l)}_0$, $1\leq l\leq L$ entering the Green Machine (i.e., output of the final beam splitters of transmissivity $\gamma_{K-1}$) are in displaced thermal states $\hat{\rho}_\text{th}(\pm\alpha_0,\bar{N}_{T0})$ where $\bar{N}_{T0}$ is the thermal mean photon number of the modes $\hat{b}_0^{(l)}$, and
\begin{equation*}%combined displacement
	|\alpha_0|^2=\sum_{k=1}^K\bar{N}_k,
\end{equation*}
which, in the limit as $K\rightarrow\infty$ ($\kappa\rightarrow0$), evaluates to
$$|\alpha_0|^2=M\eta\bar{N}_S(1+\bar{N}_S)\frac{1-e^{-2(1+\bar{N}'_S)}}{2(1+\bar{N}'_S)}.$$

Although the thermal noise in the modes ${\hat b}_k^{(l)}$, $1 \le k \le K$ are maximal-phase-insensitive classically-correlated, choosing the $\gamma$-beam splitters to combine the mean fields into a single mode, as
\begin{equation}
    \gamma_k=\frac{\sum_{i=1}^{k}\alpha_i^2}{\sum_{i=1}^{k+1}\alpha_i^2}=\frac{1-(1-\kappa(1+\bar{N}_S'))^{2k}}{1-(1-\kappa(1+\bar{N}_S'))^{2(k+1)}},
    \label{eq:gammaexpression}
\end{equation} 
is not the choice that combines the total thermal noise into a single mode. Instead, it is shown (numerically) in Appendix~\ref{app:gammanoise} that when the $\gamma_k$-values are tuned to maximally combine the mean fields (Eq.~\eqref{eq:gammaexpression}), the thermal noise on the output modes $\hat{b}_0^{(i)}$ is given by:
$$\bar{N}_{T0}\approx K\bar{N}_T/7.61=\bar{N}_S\bar{N}'_S/7.61.$$

Although one could use JDR2 with a BPSK Hadamard code as discussed in the context of JDR1, we will consider instead BPSK modulation with the 1st order Reed-Muller (RM) code. As shown in the $L=4$ example in the top-left corner of Fig.~\ref{fig:JDR2}, the RM code of code word length $L$ has $2L$ code words, i.e., twice as many code words as the Hadamard code. It has all the Hadamard code words, and each of their bit-flipped versions in the code book. Therefore, with RM code words and JDR2, the pulse-containing mode at the output of the Green Machine has a phase of $\pm 1$ depending on which of the two $L$-code word halves of the RM code book the transmitted code word belonged in. To decode this extra bit of information, JDR2's final measurement stage uses a Kennedy receiver as described in Section~\ref{sec:classicalcommJDR}~\cite{Ken73} with `exact nulling', conditioned on detecting a photon.

\subsection{Performance evaluation of JDR1}

Let us consider the case of the transmission of the $L$-symbol Hadamard phase code along with the JDR1, as described in the previous subsection. The $2^L$ possible (click, no-click) patterns at the $L$ classically-combined detector outputs are classified into $L+1$ receiver {\em outcomes}: a click at a given output and no clicks elsewhere, or an {\em erasure}, which is either zero clicks at all $L$ outputs, or clicks at multiple outputs. Our scheme thus induces an $L$-input $(L+1)$-output discrete memoryless channel between the $L$ Hadamard code words and the $L+1$ outcomes, which is identical to that induced by coherent-state pulse-position modulation (PPM) and single photon detection with non-zero background (or dark) click probability. The Shannon capacity of this channel~\cite{Jar17} divided by $ML$ is the bits-per-mode entanglement-assisted capacity attained by our design. In other words,
\begin{eqnarray}
R_{\rm E}^{(M,L)} &=& \frac{1}{ML}\left( {p_{\rm e}}\log L + (L-1)p_{\rm d}\log \frac{Lp_{\rm d}}{p_{\rm e}} \right. \nonumber \\
&-& \left.\big(p_{\rm e}+(L-1)p_{\rm d}\big)\log\left[1+\frac{(L-1)p_{\rm d}}{p_{\rm e}}\right]\right),
\end{eqnarray}
where $p_{\rm d} = (1-p_{\rm c})p_{\rm b}(1-p_{\rm b})^{L-2}$, $p_{\rm e} = p_{\rm c}(1-p_{\rm b})^{L-1}$, $p_{\rm c}$ is the click probability at the pulse-containing output, and $p_{\rm b}$ is the click probability at the non-pulse-containing output. 

To simplify the analysis, we assume that the photodetection statistics of $i$-th outputs of each of the $K$ GMs are statistically independent, so $1 - p_{\rm c} = \Pi_{k=1}^K (1-p_{\rm c}^{(k)})$, where $1-p_{\rm c}^{(k)} = \frac{1}{{\bar N}_{\rm T}+1}e^{-L\bar{N}_k/({\bar N}_{\rm T}+1)}$. The capacity evaluated under this assumption will be a lower bound to the actual capacity because in reality the photodetection outcomes of the pulse-containing modes are positively correlated, reducing the variance of the sum of the outcomes.

The expressions for the click probabilities simplify to:
\begin{eqnarray}
p_{\rm c} &=& 1-\frac{1}{(1+{\bar N}_{\rm T})^K}e^{-A\left(\frac{1-\mu^K}{1-\mu}\right)}, {\text{and}}\\
p_{\rm b} &=& 1 -\frac{1}{(1+{\bar N}_{\rm T})^K},
\end{eqnarray}
with $A = ML\kappa\eta {\bar N}_{\rm S}({\bar N}_{\rm S}+1)/({\bar N}_{\rm T}+1)$.

\begin{figure}[!ht]
\centering
\includegraphics[width=0.95\columnwidth]{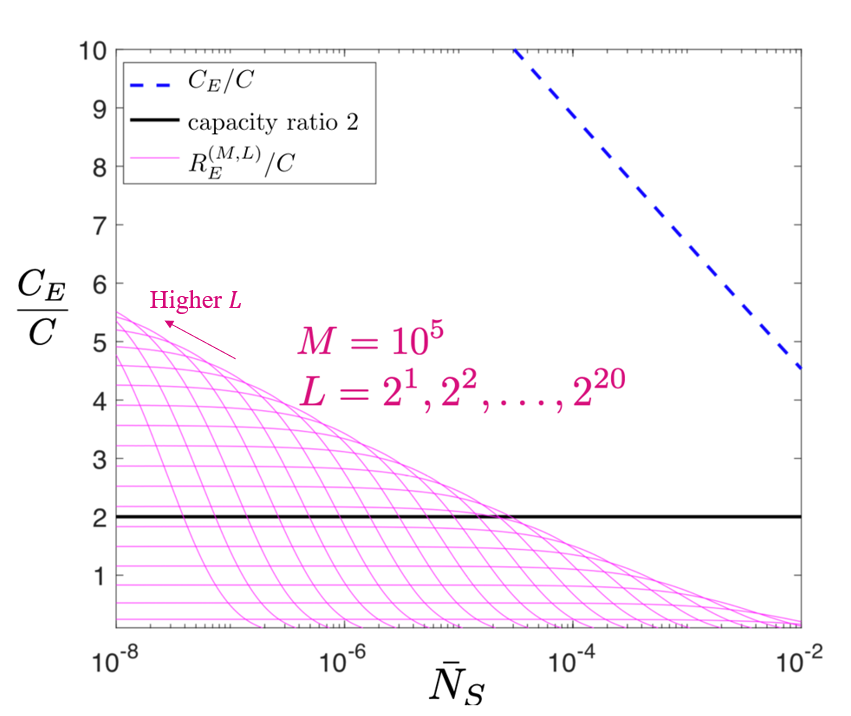}
\caption{The thin magenta lines are plots of $R_{\rm E}^{(M,L)}/C$ for $L = 2, 4, 8, \ldots, 2^{20}$ and $M = 10^5$. This shows that the capacity ratio scales as $\log(1/{\bar N}_{\rm S})$, which tends to infinity as ${\bar N}_{\rm S} \to 0$, for any $M$. However, this scheme (BPSK modulation, Hadamard code, and our proposed structured joint-detection receiver) does not achieve $C_{\rm E}$. We assume $\eta = 0.01$ and ${\bar N}_{\rm B} = 10$ photons per mode for all the plots in this figure.}
\label{fig:cap_ratio_M100000}
\end{figure}

In Fig. ~\ref{fig:cap_ratio_M100000}, we plot $C_{\rm E}/C$ as a function of ${\bar N}_{\rm S}$ in the ${\bar N}_{\rm S} \ll 1$ regime, for $\eta = 0.01$ and ${\bar N}_{\rm B} = 10$. We also plot $R_{\rm E}^{(M,L)}/C$, with $M=10^5$, $L \in \left\{2, 2^2, \ldots, 2^{20}\right\}$ and taking $\kappa=1/K$ with $K=100$, which we found to be sufficiently large such that further increasing $K$ did not affect the rate significantly. The envelope $R^{(M)}_{\rm E} = {{\rm sup}_{L}R_{\rm E}^{(M,L)}}$ shows that our transceiver achieves an entanglement-assisted capacity gain that exceeds $2$ as $\ns \to 0$, the best achievable ratio with an OPA~\cite{Guh09} or FF-SFG receiver~\cite{Zhu17,Shi19} (see Appendix~\ref{app:OPA}).

In the next subsection we apply the conditions:
\begin{equation}
\eta {\bar N}_{\rm S} \ll {\bar N}_{\rm S} \ll 1 \ll {\bar N}_{\rm B} \ll K,
\label{eq:operationalregime}
\end{equation}
and prove that our system design attains the optimal scaling of entanglement-assisted communications capacity, i.e., ${R_{\rm E}^{(M)}}/{C} \sim \ln(1/\ns)$. Despite ${R_{\rm E}^{(M)}}$ not meeting $C_{\rm E}$, it achieves the infinite-fold capacity enhancement leveraging pre-shared entanglement, but most importantly---using quantum optical states, processes and detection schemes that are readily realizable. 

\subsubsection{Capacity scaling analysis}

Let us first consider an order-$L$ pulse position modulation (PPM) alphabet over a channel with loss and noise. PPM encodes information by the position of a pulse (e.g., a coherent state of light) in one of $L$ orthogonal modes (e.g., time bins) at the input, which is direct-detected at the output (e.g., by a single photon detector). Loss attenuates the transmitted pulse amplitude, and noise results in potential detection events in one or more bins. Ignoring detection events in multiple bins (i.e., treating them as ``erasures"), and assuming an equiprobable selection over the $L$ inputs (which maximizes the throughput), the Shannon mutual information---expressed in bits per mode---of the induced $L$-input $(L+1)$-output discrete memoryless channel (DMC), is given by~\cite[Eq.~(16)]{Jar17}:
\begin{align}
I^{(L)}_{\text{PPM}} &= \frac{p_{\rm e}}{L}\log L + \frac{(L-1)}{L}p_{\rm d}\log \frac{Lp_{\rm d}}{p_{\rm e}} \nonumber \\
\label{eq:In}&\phantom{=}- \left[\frac{p_{\rm e}+(L-1)p_{\rm d}}{L}\right]\log\left[1+\frac{(L-1)p_{\rm d}}{p_{\rm e}}\right],
\end{align}
where $p_{\rm e}$ is the probability of the detection event occurring exclusively in the bin corresponding to the position of the pulse at the input, and $p_{\rm d}$ is the probability that a detection event occurs in a single bin that is different from the one containing the input pulse.  Denoting by $p_{\rm c}$ the probability of a detection event in the bin corresponding to the input pulse and by $p_{\rm b}$ the probability of a detection event in another bin \cite[Sec.~IV]{Jar17},
\begin{align}
\label{eq:pe}p_{\rm e}& = p_{\rm c}(1-p_{\rm b})^{L-1}, \,{\text{and}}\\
\label{eq:pd}p_{\rm d}& = (1-p_{\rm c})p_{\rm b}(1-p_{\rm b})^{L-2}.
\end{align}
We specialize the result by Jarzyna and Banaszek \cite{Jar17} to find the channel capacity of the DMC induced by the modulation-code-channel-receiver described in Fig. ~\ref{fig:JDR}.

Let us recall that our scheme involves BPSK-modulation of the signal modes of $M$ pre-shared two-mode-squeezed-vacuum (TMSV) states, repeating the above $L$ times, encoding an order-$L$ binary Hadamard code, and transmission of the $ML$ modulated modes over $ML$ uses of the single-mode lossy-noisy bosonic channel $\mathcal{N}_{\eta}^{{\bar N}_{\rm B}}$, followed by demodulation and detection by our joint detection receiver (JDR). This scheme results in detection events that are statistically identical to demodulating PPM in the presence of noise. Thus, we seek:
\begin{align}
\label{eq:REM}R_{\rm E}^{(M)}&=\max_{L}\frac{1}{M}I^{(L)}_{\rm PPM},
\end{align}
where we determine $p_{\rm e}$ and $p_{\rm d}$ as follows. First, let's recall the definitions. The mean number of photons per mode in the signal modes of the TMSV transmitted by Alice is ${\bar N}_{\rm S}$, and the mean photon number of the thermal noise background per transmitted mode is ${\bar N}_{\rm B}$. The modal power transmissivity of the bosonic channel is $\eta\in(0,1]$, which implies that Bob's received mean number of photons per mode is ${\bar N}_{\rm S}^\prime=\eta {\bar N}_{\rm S}+(1-\eta){\bar N}_{\rm B}$. To calculate $p_{\rm c}$ and $p_{\rm b}$, we assume the photodetection statistics of the $i$-th outputs of each of the $K$ Green Machines in the JDR are statistically independent, and $K \gg {\bar N}_{\rm B}$. Thus, $1-p_{\rm c} = 1-\prod_{k=1}^K (1-p_{\rm c}^{(k)})$, where $1-p_{\rm c}^{(k)} = \frac{1}{{\bar N}_{\rm T}+1}e^{-L{\bar N}_k/({\bar N}_{\rm T}+1)}$ with ${\bar N}_{\rm T}={\bar N}_{\rm S} {\bar N}_{\rm S}^\prime/K$, ${\bar N}_k=\frac{M\eta {\bar N}_{\rm S}(1+{\bar N}_{\rm S})\mu^{k-1}}{K}$, and $\mu=\left[1-\frac{1+{\bar N}_{\rm S}^\prime}{K}\right]^2$. Thus:
\begin{align}
p_{\rm c} &= 1-\frac{1}{(1+{\bar N}_{\rm T})^K}e^{-A\left(\frac{1-\mu^K}{1-\mu}\right)}, {\text{and}}\\
p_{\rm b} &= 1 -\frac{1}{(1+{\bar N}_{\rm T})^K},
\end{align}
with $A = \frac{ML\eta {\bar N}_{\rm S}({\bar N}_{\rm S}+1)}{K({\bar N}_{\rm T}+1)}$. Using the conditions:
\begin{align}
{\bar N}_{\rm S} \ll 1 \ll {\bar N}_{\rm B} \ll K,
\end{align}
we can make the following approximations using the limits as ${\bar N}_{\rm S}\to 0$ and $K\to\infty$:
\begin{align}
\label{eq:approxNSp}{\bar N}_{\rm S}^\prime &\approx (1-\eta){\bar N}_{\rm B},\\
(1+{\bar N}_{\rm T})^{-K}&\approx e^{-{\bar N}_{\rm S}(1-\eta){\bar N}_{\rm B}},\,{\text{and}}\\
\label{eq:approxA}\frac{A}{1-\mu}&\approx \frac{ML\eta {\bar N}_{\rm S}}{2(1+(1-\eta){\bar N}_{\rm B})}.
\end{align}
These lead to the following approximations for $p_{\rm c}$ and $p_{\rm b}$:
\begin{align}
\label{eq:pc_approx}p_{\rm c}&\approx 1-\exp\left[-{\bar N}_{\rm S}\left(\frac{ML\eta\gamma}{2(1+(1-\eta){\bar N}_{\rm B})}+(1-\eta){\bar N}_{\rm B}\right)\right]\\
\label{eq:pb_approx}p_{\rm b}&\approx 1-\exp\left[-{\bar N}_{\rm S}(1-\eta){\bar N}_{\rm B}\right],
\end{align}
where $\gamma=1-e^{-2\left(1+(1-\eta){\bar N}_{\rm B}\right)}$.
Substitution of approximations in \eqref{eq:pc_approx} and \eqref{eq:pb_approx} into \eqref{eq:pe} and \eqref{eq:pd} yields:
\begin{align}
p_{\rm e}&\approx \exp\left[-{\bar N}_{\rm S}(L-1)(1-\eta){\bar N}_{\rm B}\right]\nonumber\\
\label{eq:pe_approx}&\phantom{\approx}-\exp\left[-{\bar N}_{\rm S}L\left(\frac{M\eta\gamma}{2(1+(1-\eta){\bar N}_{\rm B})}+(1-\eta){\bar N}_{\rm B}\right)\right]\\
&\approx\exp\left[-{\bar N}_{\rm S}L(1-\eta){\bar N}_{\rm B}\right]\nonumber\\
\label{eq:pe_approx_lb}&\phantom{\approx}-\exp\left[-{\bar N}_{\rm S}L\left(\frac{M\eta\gamma}{2(1+(1-\eta){\bar N}_{\rm B})}+(1-\eta){\bar N}_{\rm B}\right)\right],\\
p_{\rm d}&\approx \exp\left[-{\bar N}_{\rm S}L\left(\frac{M\eta\gamma}{2(1+(1-\eta){\bar N}_{\rm B})}+(1-\eta){\bar N}_{\rm B}\right)\right]\nonumber\\
\label{eq:pd_approx}&\phantom{\approx}-\exp\left[-{\bar N}_{\rm S}\left(\frac{ML\eta\gamma}{2(1+(1-\eta){\bar N}_{\rm B})}+(1-\eta){\bar N}_{\rm B}(L+1)\right)\right],
\end{align}
where we assume $L\gg1$ so that $L-1\approx L$ for the approximation in \eqref{eq:pe_approx_lb}.
When ${\bar N}_{\rm S}\to0$, we can approximate $p_{\rm e}$ and $p_{\rm d}$ by the Taylor series expansions at ${\bar N}_{\rm S}=0$ of \eqref{eq:pe_approx_lb} and \eqref{eq:pd_approx}, respectively:
\begin{align}
\label{eq:pe_approx_taylor}p_{\rm e}&\approx \frac{{\bar N}_{\rm S}ML\eta\gamma}{2(1+(1-\eta){\bar N}_{\rm B})},\\
\label{eq:pd_approx_taylor}p_{\rm d}&\approx {\bar N}_{\rm S}(1-\eta){\bar N}_{\rm B}.
\end{align}
Substituting \eqref{eq:pe_approx_taylor} and \eqref{eq:pd_approx_taylor} into the last two terms of \eqref{eq:In}, and approximating $\frac{L-1}{L}\approx 1$, reveals that only the first term of \eqref{eq:In} has a significant dependence on $L$ in our regime of interest.
Thus, for the optimal order, we need:
\begin{align}
\label{eq:nopt}L^*&=\argmax_L\frac{p_{\rm e}}{L}\log L.
\end{align}
The linear approximation in \eqref{eq:pd_approx_taylor} is insufficient to find $L^*$.  We follow the methodology in \cite{Jar17} by substituting in \eqref{eq:nopt} the quadratic Taylor series expansion at ${\bar N}_{\rm S}=0$,
\begin{align}
p_{\rm e}&\approx \frac{L {\bar N}_{\rm S} M\eta\gamma}{2(1+(1-\eta){\bar N}_{\rm B})}\nonumber\\
&\phantom{\approx}-\frac{L^2 {\bar N}_{\rm S}^2 M\eta\gamma\left(M\eta\gamma + 4(1-\eta){\bar N}_{\rm B}(1+(1-\eta){\bar N}_{\rm B})\right)}{8(1+(1-\eta){\bar N}_{\rm B})^2}.\nonumber
\end{align}
Let $v\equiv\frac{{\bar N}_{\rm S}^2M\eta\gamma\left(M\eta\gamma + 4(1-\eta){\bar N}_{\rm B}(1+(1-\eta){\bar N}_{\rm B})\right)}{8(1+(1-\eta){\bar N}_{\rm B})^2\ln 2}$ and $u\equiv \frac{{\bar N}_{\rm S}M\eta\gamma}{2(1+(1-\eta){\bar N}_{\rm B})\ln 2}$. This reduces the problem in \eqref{eq:nopt} to finding the location of the extremal values of $f(L)=(u+vL)\ln L$ by solving
\begin{align}
\label{eq:dfdn}\frac{\dif f(L)}{\dif L}&=\frac{u}{vL}-1-\ln L=0
\end{align}
for $L$, which involves the principal branch of the Lambert $W$-function \cite[Sec.~4.13]{DLMF}:
\begin{align}
L^*&=\frac{u}{v}\left[W\left(\frac{u}{v}e\right)\right]^{-1},
\end{align}
where $W\left(xe^x\right)=x$ for $x\geq-1$.
Using equality $\ln W(x)=\ln(x)-W(x)$ for $x>0$ \cite[Eq.~(4.13.3)]{DLMF} and asymptotic expansion $W(x)=\ln(x)-\ln\ln (x)+o(1)$ as $\ln(x)\to\infty$ \cite[Eq.~(4.13.10)]{DLMF} in our regime of interest ${\bar N}_{\rm S}\to0$, we have:

\begin{equation}
\label{eq:lognstar}\log(L^*) \approx\log\left(\frac{w}{{\bar N}_{\rm S}}\right)-\log\left(\ln\left[\frac{we}{{\bar N}_{\rm S}}\right]\right),
\end{equation}
where $w=\frac{4(1+(1-\eta){\bar N}_{\rm B})}{M\eta\gamma+4(1-\eta){\bar N}_{\rm B}(1+(1-\eta){\bar N}_{\rm B})}$.
Substituting \eqref{eq:pe_approx_taylor} and \eqref{eq:lognstar} into \eqref{eq:REM}, we obtain:
\begin{widetext}
\begin{align}
\label{eq:REMapprox}R_{\rm E}^{(M)}&\approx \frac{\eta {\bar N}_{\rm S}\gamma}{2(1+(1-\eta){\bar N}_{\rm B})}\left[\log\left[\frac{w}{{\bar N}_{\rm S}}\right]-\log\left[\ln\left[\frac{we}{{\bar N}_{\rm S}}\right]\right]-g\left[\frac{2(1-\eta){\bar N}_{\rm B}(1+(1-\eta){\bar N}_{\rm B})}{M\eta\gamma}\right]\right],
%\label{eq:REMapprox}R_{\rm E}^{(M)}&\approx \frac{\eta {\bar N}_{\rm S}\gamma}{2(1+(1-\eta){\bar N}_{\rm B})}\left[\log\left[\frac{w}{{\bar N}_{\rm S}}\right]-\log\left[\ln\left[\frac{we}{{\bar N}_{\rm S}}\right]\right]\right.\\
%&\phantom{\approx \frac{\eta {\bar N}_{\rm S}\gamma}{2(1+(1-\eta){\bar N}_{\rm B})}\bigg[}\left.-g\left[\frac{2(1-\eta){\bar N}_{\rm B}(1+(1-\eta){\bar N}_{\rm B})}{M\eta\gamma}\right]\right],
\end{align}
\end{widetext}
where $g(x)=(x+1)\log(x+1)-x\log x$. As ${\bar N}_{\rm S}\to0$, the logarithmic term dominates \eqref{eq:REMapprox}, and we obtain the scaling:
\begin{equation}
R_{\rm E}^{(M)}=O\left({\bar N}_{\rm S}\log \left(\frac{1}{{\bar N}_{\rm S}}\right)\right).
\label{eq:optimal_scaling}
\end{equation}

In Appendix~\ref{app:PPMcrude}, we consider a cruder approximation of $R_{\rm E}^{(M)}$, providing an alternative proof of the scaling in~\eqref{eq:optimal_scaling}, but one that lets us establish a connection with a problem that was studied by Wang and Wornell in the context of coherent-state PPM modulation, where the dark click probability per mode $\lambda$ is proportional to the mean photon number per mode $\cal E$~\cite{Wan14}.

\subsubsection{Numerical rate calculations}

\begin{figure}[!h]
\centering
\includegraphics[width=.95\columnwidth]{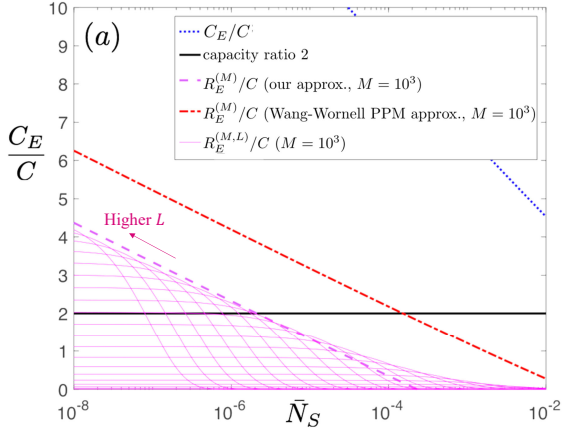}%used to be .35\columnwidth
\caption{Plot of ${R^{(M,L)}_{\rm E}}/C$ with $M=10^3$, for $L \in \left\{2, 2^2, \ldots, 2^{20}\right\}$. We assume $\eta = 0.01$ and ${\bar N}_{\rm B} = 10$ photons per mode, for all the plots.} %We compare the two approximations for the capacity-ratio envelope ${R^{(M)}_{\rm E}}/C$: the one we obtained in Eq.~\eqref{eq:REMapprox} leveraging the Jarzyna-Banaszek analysis, and the one we obtained leveraging the Wang-Wornell analysis in Eq.~\eqref{eq:REM_secondorder}. It is seen that our approximation is tighter, especially for smaller $M$.}
\label{fig:capratio_approx_compare1}
\end{figure}

\begin{figure}[!h]
\centering
\includegraphics[width=.95\columnwidth]{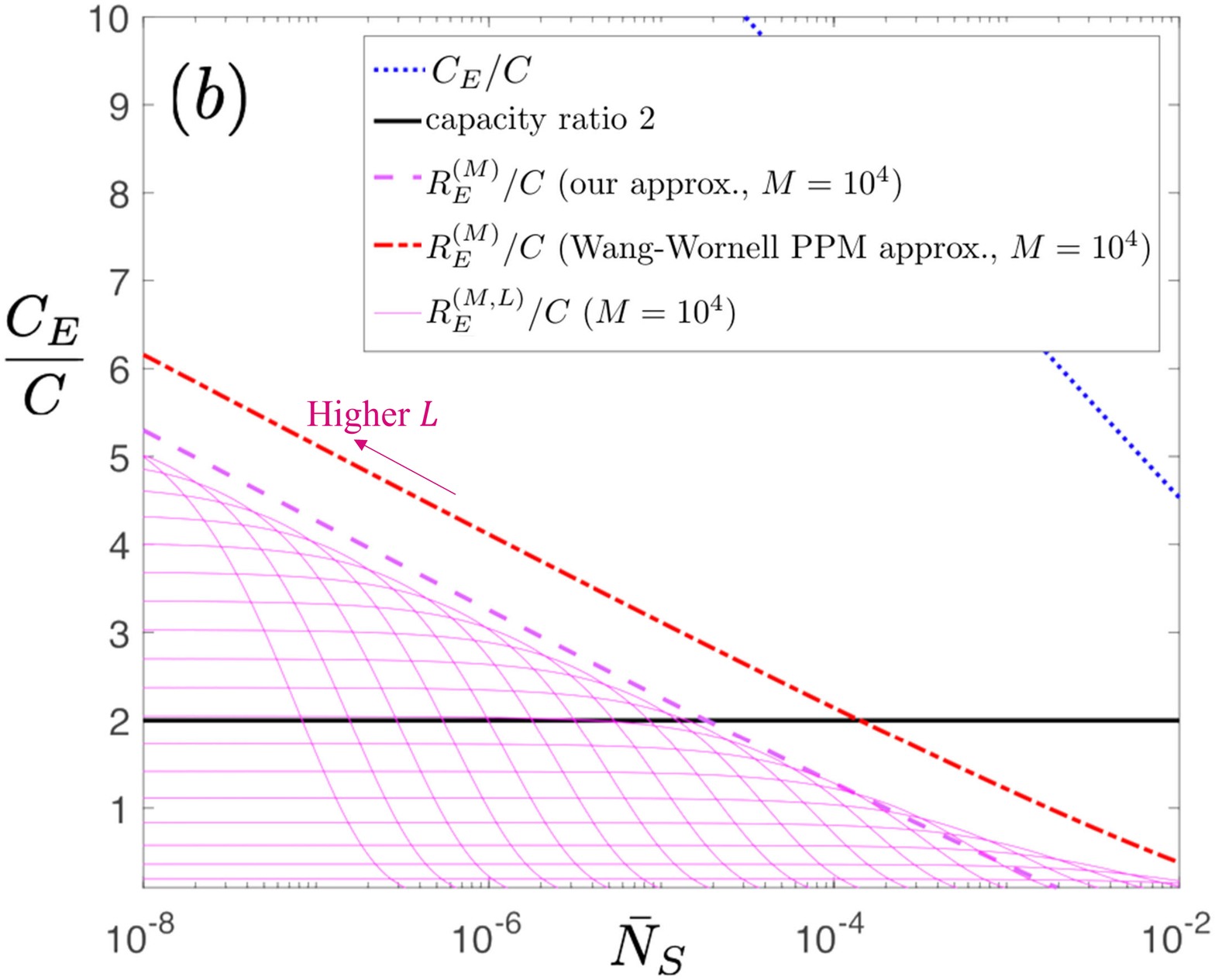}%used to be .35\columnwidth
\caption{Plot of ${R^{(M,L)}_{\rm E}}/C$ with $M=10^4$, for $L \in \left\{2, 2^2, \ldots, 2^{20}\right\}$. We assume $\eta = 0.01$ and ${\bar N}_{\rm B} = 10$ photons per mode, for all the plots.} %We compare the two approximations for the capacity-ratio envelope ${R^{(M)}_{\rm E}}/C$: the one we obtained in Eq.~\eqref{eq:REMapprox} leveraging the Jarzyna-Banaszek analysis, and the one we obtained leveraging the Wang-Wornell analysis in Eq.~\eqref{eq:REM_secondorder}. It is seen that our approximation is tighter, especially for smaller $M$.}
\label{fig:capratio_approx_compare2}
\end{figure}

\begin{figure}[!h]
\centering
\includegraphics[width=\columnwidth]{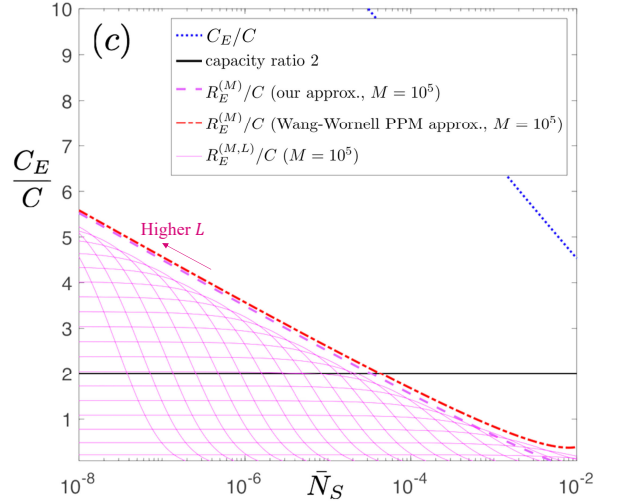}%used to be .35\columnwidth
\caption{Plot of ${R^{(M,L)}_{\rm E}}/C$ with $M=10^5$, for $L \in \left\{2, 2^2, \ldots, 2^{20}\right\}$. We assume $\eta = 0.01$ and ${\bar N}_{\rm B} = 10$ photons per mode, for all the plots.} %We compare the two approximations for the capacity-ratio envelope ${R^{(M)}_{\rm E}}/C$: the one we obtained in Eq.~\eqref{eq:REMapprox} leveraging the Jarzyna-Banaszek analysis, and the one we obtained leveraging the Wang-Wornell analysis in Eq.~\eqref{eq:REM_secondorder}. It is seen that our approximation is tighter, especially for smaller $M$.}
\label{fig:capratio_approx_compare3}
\end{figure}

In Figs. \ref{fig:capratio_approx_compare1}, \ref{fig:capratio_approx_compare2}, and \ref{fig:capratio_approx_compare3}, we compare the two approximations for ${R^{(M)}_{\rm E}}$: the one we derived by modifying the Jarzyna-Banaszek analysis of PPM applied to our problem, shown in Eq.~\eqref{eq:REMapprox} and labeled ``our approx." in Figs. ~\ref{fig:capratio_approx_compare1}, \ref{fig:capratio_approx_compare2}, \ref{fig:capratio_approx_compare3}, and the one we obtained from the Wang-Wornell PPM analysis, shown in Appendix~\ref{app:PPMcrude}, Eq.~\eqref{eq:REM_secondorder}. It is seen that the former is closer to the true envelope, especially for smaller values of $M$.

In Fig.~\ref{fig:cap_ratios} we plot (the exact) $R^{(M)}_{\rm E}$ as a function of ${\bar N}_{\rm S}$ for $M = 10, 10^2, \ldots, 10^6$. For the assumed values of $\eta = 0.01$ and ${\bar N}_{\rm B} = 10$ photons per mode used for plots in this figure, the highest capacity occurs at around $M \sim 10^5$ for JDR1 and $M\sim10^4$ for JDR2. The existence of such an optimum value of $M$ can be explained by the negative sign of the $M$-dependent second-order term in~\eqref{eq:REMapprox}.

For $\eta = 0.01$ and ${\bar N}_{\rm B} = 10$ photons per mode, our scheme achieves the maximum rate at the modulation-block length $M \approx 10^5$. For a typical SPDC entanglement source of optical bandwidth $W \sim 1$ THz, and $M \approx WT$, $M = 10^5$ modes in a signal pulse translates to a pulse duration of $T \sim 100$ ns. This means the BPSK phase-modulation bandwidth necessary would be $\sim 10$ MHz, which is readily realizable with commercial-grade electro-optical modulators (EOMs) at $1550$ nm.

\newpage
\subsection{Performance evaluation of JDR2}\label{sec:capacityJDR2}
Let us now consider the RM code ($2L$ binary-phase code words each of length $L$) and the JDR2, which together induce a 2$L$-input $2L+1$ discrete memoryless channel. The additional output is the erasure outcome in which no photons are detected. The transition matrix $\X$ is defined to be the matrix of conditional probabilities such that $\X_{ji}$ is the probability that the receiver decides on outcome $i$ given that code word $j$ was transmitted. In this case, $i$ runs from 1 to $2L+1$, indexing the possible outcomes of the receiver and $j$ runs from 1 to $2L$, indexing the code words of the Reed-Muller code. If the Reed-Muller code book is seen as a Hadamard code book appended with its sign-flipped copy with the original copy consisting of `plus-words' and the sign-flipped copy consisting of `minus-words', even indices of $\X$ correspond to minus-words and odd indices to plus-words. Because of symmetry, we observe that $\X$ has $6$ independent entries $\X_{11}$, $\X_{22}$, $\X_{12}$, $\X_{21}$, $\X_{14}$, $\X_{1,2L+1}$, corresponding respectively to the transition probabilities of: a plus-outcome on the pulse-containing mode of the GM-output given that a plus-word was transmitted, a minus-outcome on the pulse-containing mode given that a minus-word was transmitted, a minus-outcome on the pulse-containing mode given that a plus-word was transmitted, a plus-outcome on the pulse-containing mode given that a minus-word was transmitted, a minus-outcome on a non-pulse-carrier given that a plus-word was transmitted, and erasure (no clicks on any mode). Moreover, by the same symmetry, the plus-words can be assumed to have equal priors $p_{+}/L$, and similarly the minus-words can be assumed to have equal priors $p_{-}/L=(1-p_{+})/L$ where $0\leq p_{+}\leq1$. The entirety of $\mb{X}$ can then be written in terms of the above values, as:
\begin{subequations}
\begin{align}
	\mat{cc}{\mb{X}_{2i-1,2i-1}& \mb{X}_{2i-1,2i}\\ \mb{X}_{2i,2i-1}& \mb{X}_{2i,2i}}=&\mat{cc}{\mb{X}_{11}& \mb{X}_{12}\\ \mb{X}_{21}& \mb{X}_{22}}\\
	\mb{X}_{2j,2i}=\mb{X}_{2j-1,2i}\;=&\;\mb{X}_{14}\\
	\mb{X}_{k,2n+1}\;=&\;\mb{X}_{1,2L+1}\\
	\mb{X}_{2j,2i-1}=&\mb{X}_{2j-1,2i-1}\nonumber\\
	=\;\frac{1}{L-1}(1-\mb{X}_{1,2L+1}-&\mb{X}_{11}-\mb{X}_{12})-\mb{X}_{14}
	\label{tpinotj++}
\end{align}
\label{Xentries}
\end{subequations}
for $k=1,2,\cdots,2n$; $i,j=1,2,\cdots,L$; but $i\neq j$. Eq.~\eqref{tpinotj++} says that entries $\X_{2j,2i-1}$ and $\X_{2j-1,2i-1}$ are obtained for free as result of the normalization condition of the rows of $\X$.

The information rate in bits per mode is obtained by dividing the mutual information $I(\bar{N}_S,\bar{N}_B,\eta,p_+)$ associated with the transition matrix by $ML$, the total number of modes transmitted to send one code word. Therefore,
\begin{widetext}
\begin{align}
	R_E^{(M,L)}(\bar{N}_S,\bar{N}_B,\eta,p_+)=&\frac{1}{LM}I(\bar{N}_S,\bar{N}_B,\eta,p_+)\nonumber\\
	=&\frac{1}{LM}\Bigl(\;\log_2 (L) \bigl((L-1) \X_{13}+(L-1) \X_{14}+p_- (\X_{21}+\X_{22})+p_+(\X_{11}+\X_{12})\bigr)\nonumber\\
	&-((L-1) \X_{13}+p_- \X_{21}+p_+ \X_{11}) \log_2((L-1) \X_{13}+p_- \X_{21}+p_+ \X_{11})\nonumber\\
	&-((L-1) \X_{14}+p_- \X_{22}+p_+ \X_{12}) \log_2 ((L-1) \X_{14}+p_- \X_{22}+p_+\X_{12})\nonumber\\
	&+(L-1) \X_{13} \log_2 (\X_{13})+(L-1) \X_{14} \log_2 (\X_{14})\nonumber\\&+p_-\X_{21} \log_2 (\X_{21})+p_- \X_{22} \log_2 (\X_{22})+p_+ \X_{11}\log_2 (\X_{11})+p_+ \X_{12} \log_2 (\X_{12})\;\Bigr),
	\label{eq:info}
\end{align}
\end{widetext}
where the dependence of the transition probability matrix elements on $\bar{N}_S$, $L$, $M$, and $\bar{N}_B$ are given in Appendix~\ref{app:transitionentries}. The optimal $p_+$ is solved numerically to satisfy
$$\frac{\partial}{\partial p_+}I(\bar{N}_S,\bar{N}_B,\eta,p_+)=0.$$

\begin{figure}
	\centering
	\includegraphics[width=\linewidth]{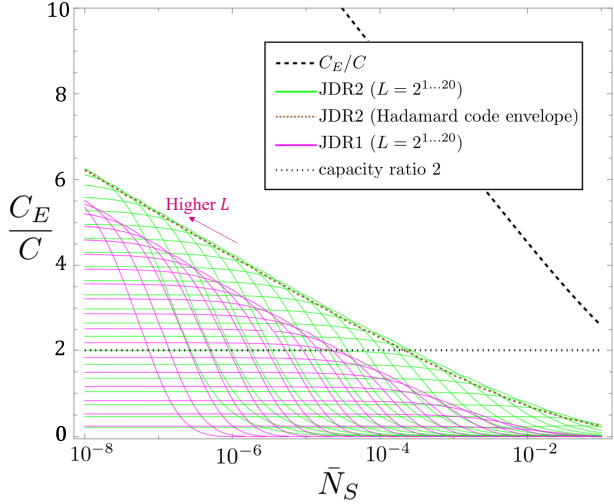}
	\caption{The capacity of JDR2 (green) and JDR1 (magenta) as functions of $\bar{N}_S$ plotted as ratios over the ultimate unassisted Holevo capacity $C$ assuming $\bar{N}_B=10$ and $\eta=0.01$. The individual green and pink lines correspond to different values of $L$, the code word length. As $\bar{N}_S$ decreases, the optimal value of $L$ increases. Also included in the plot is the envelope of the capacity of JDR2 when code words are drawn from a Hadamard code book instead of Reed-Muller code book (dotted brown)}.
	\label{fig:jdr2rate}
\end{figure}

In Fig.~\ref{fig:jdr2rate}, we plot the bits-per-mode rates of JDR2 and JDR1 as a ratio over the Holevo capacity $C$.

Note that Eq.~\eqref{eq:info} subsumes the rate associated with using a Hadamard code book, which can be obtained by setting $p_-=0$ and dividing the mutual information by $M(L-1)$ instead of $ML$, since in the Hadamard code, all code words share the same initial symbol ($\theta=0$), which can hence be appended at the receiver-end instead of consuming a channel use. The resulting capacity ratio is only slightly less than using the full Reed-Muller code, as Fig. \ref{fig:jdr2rate} shows.

\begin{figure}[!h]
\begin{minipage}[b][%.27\paperheight
][b]{\columnwidth}
    \centering
    \includegraphics[width=.95\columnwidth]{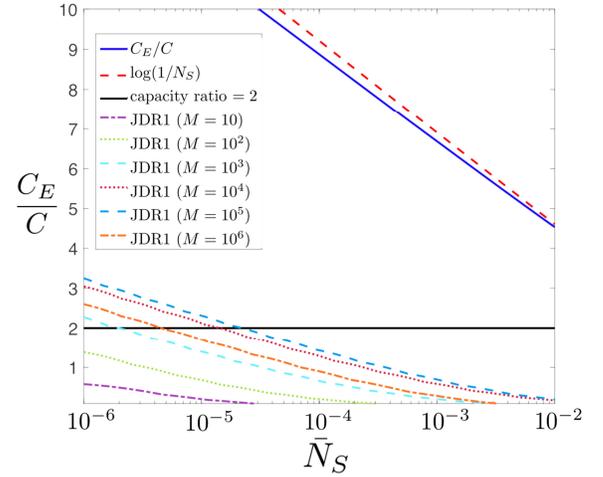}
    \subcaption{$R_E^{(M,L)}/C$ envelopes of JDR1.}
    \label{fig:cap_ratios}
\end{minipage}
\begin{minipage}[b][%.27\paperheight
][t]{\columnwidth}
    \centering
    \includegraphics[width=.95\linewidth]{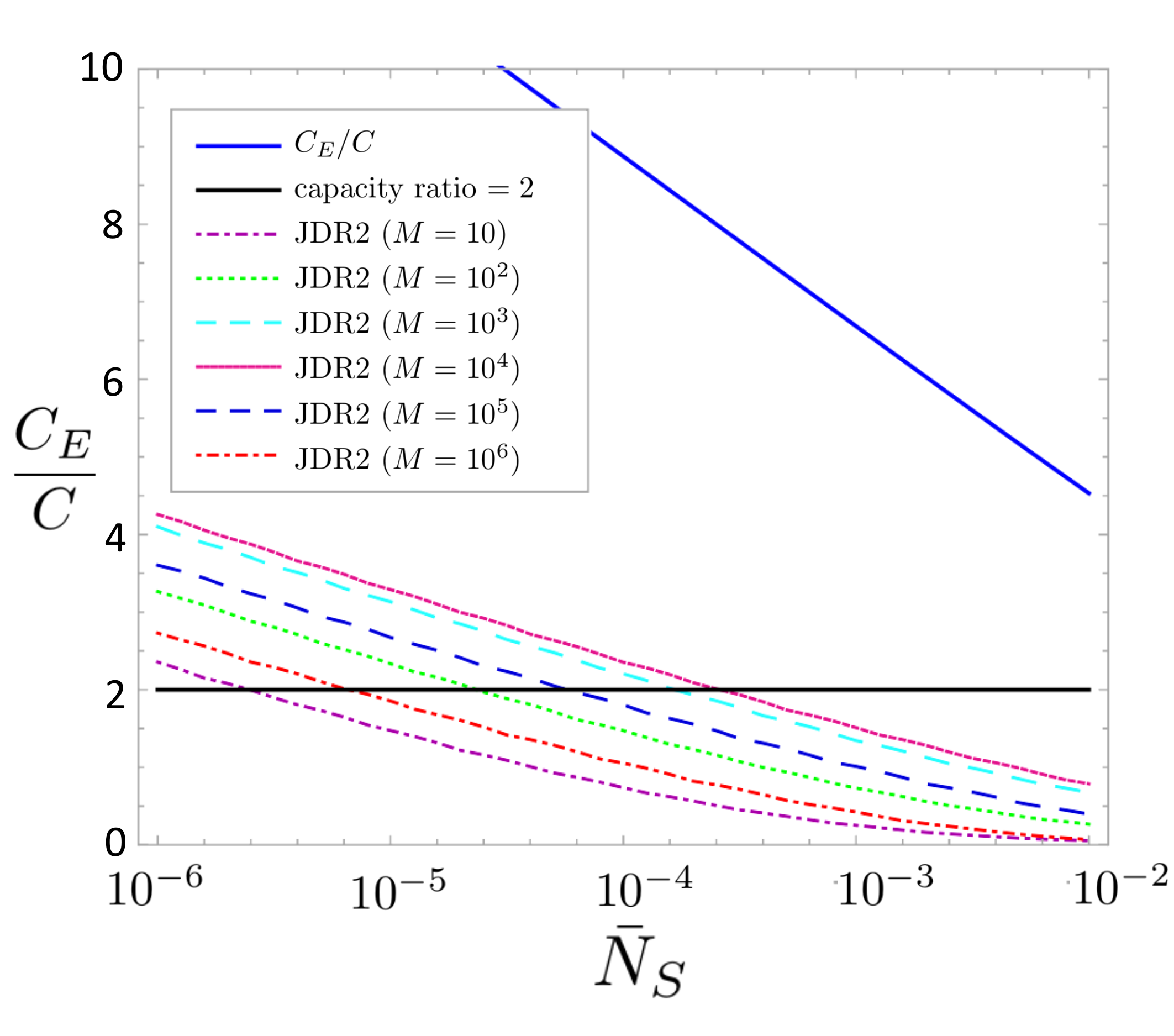}
    \subcaption{$R_E^{(M,L)}/C$ envelopes of JDR2.} 
    \label{fig:cap_ratios_manyMs_JDR2}
\end{minipage}

\caption{Here we plot the envelopes of $R_E^{(M,L)}/C$ (taken over $L=2,4,8,...,2^21$), for $M=10,100,...,10^6$. This shows that an optimum performance occurs at $M\sim10^5$ for JDR1 and $M\sim10^4$ for JDR2. We assume $\eta=0.01$ and $\bar{N}_B=10$ photons per mode, for all the plots in these figures.
}
\label{fig:JDR1andJDR2cap_ratios}
\end{figure}

Fig. \ref{fig:cap_ratios_manyMs_JDR2} shows that $M\approx10^4$ is the new optimal $M$ for JDR2 as opposed to $M\approx10^5$ for JDR1. To understand this effect, note first that repeating modes of the trasmitted code words $M$-fold effectively increases the clarity of the received code words at the expense of consuming $M$-times many channel uses. Of course, once $M$ is high enough, the code words are almost perfectly distinguishable and further increasing $M$ only hurts the communication rate by consuming unnecessary channel uses, but when the receiver's distinguishing power between two symbols is low (e.g. because of the shot-noise associated with the particular design), adding clarity to the code words is worth the extra channel uses, hence there is an optimal value of $M>1$. Depending on the signal-to-noise (SNR) ratio of the output of the receiver's front-end, this optimal value of $M$ will be higher or lower. Note that evaluating the Holevo information of the recieved ensemble $\hat{\rho}^{(RI)}$ for $\theta\in\{0,\pi\}$, corresponding to $M=1$, in section \ref{sec:EA_binarymodulation} resulted in $C_E$. Thus a lower value for the optimal $M$ is indicative of a receiver being closer to optimal. By combining the sum frequency modes, JDR2 effectively filters out a significant portion of the noise (into the unused outputs of the $\gamma$-beam splitters) and reduces the SNR, which brings down the optimal value of $M$. Since this corresponds to less channel uses, JDR2 achieves a higher rate than JDR1. 

\section{PPM and OOK modulation formats for Entanglement-Assisted Communications}

%In the main paper, we described an entanglement-assisted quantum-optical transmitter and a joint detection receiver, for communicating classical information. The scheme begins with pre-sharing SPDC entanglement, and the transmitter modulating an $M$-temporal-mode signal pulse of a pre-shared two-mode squeezed vacuum state $|\psi\rangle_{\rm SI}^{\otimes M}$, with a binary phase, i.e., $0$ or $\pi$. The transmitter uses a sequence (code word) of $L$ successive such $M$-mode pulses to modulate a length-$L$ binary-phase-shift keying (BPSK) Hadamard code word, and transmits those modulated signal pulses, consuming $nM$ uses of the single-mode lossy-noisy bosonic channel ${\cal N}_\eta^{{\bar N}_{\rm B}}$ to Bob. Bob combines the received modulated modes---corrupted by the noisy channel---and his pre-shared entangled half, the $nM$ losslessly-retained idler modes, in a feed-forward sum-frequency-generation (FF-SFG) receiver, as explained in the main paper. Each $L$ signal-idler pair ($2M$ modes) is processed by an FF-SFG module comprised of $K$ SFG stages. The output of each FF-SFG module resembles a noisy coherent-state Hadamard-coded BPSK modulation, which is fed into $K$ Green Machine (GM) circuits. 

In Section~\ref{sec:EA_binarymodulation}, we discussed the entanglement assisted capacity achievable with on-off based modulation formats, while leveraging continuous-variable SPDC-based pre-shared entanglement, and showed that the $\log(1/{\bar N}_{\rm S})$ capacity-ratio improvement over the Holevo capacity is attainable with such modulation formats. Despite the capacity not being as good as phase-only modulation formats, and the need for more pre-shared entanglement, the on-off modulation formats are easier to realize experimentally. Further, the pulse-position modulation (PPM) scheme, a modulation-code over the on-off keying (OOK) alphabet, has a close connection to quantum ranging, which also saturates the $\log(1/{\bar N}_{\rm S})$ scaling~\cite{Zhu21}. The optimal receiver design for OOK and PPM modulation formats are not known. In this section, we discuss these two on-off based modulation formats for entanglement-assisted communications, in further detail. 

\subsection{Pulse position modulation (PPM)}\label{sec:PPM}

At the $L$ output modes of the $K$ Green Machine (GM) circuits of JDR1, shown in Fig. ~\ref{fig:JDR}, the state of the $LK$ output modes resembles pulse-position modulation (PPM): One block of $K$ modes carries displaced thermal states $\hat{\rho}_{\rm th}(\sqrt{L {\bar N}_k}, {\bar N}_{\rm T})$, where ${\bar N}_k = M\kappa \eta {\bar N}_{\rm S}(1+{\bar N}_{\rm S})\mu^{k-1}$, $1 \le k \le K$, with $\mu = (1-\kappa(1+{\bar N}_{\rm S}^\prime))^2$, ${\bar N}_{\rm T} = \kappa {\bar N}_{\rm S} {\bar N}_{\rm S}^\prime$, ${\bar N}_{\rm S}^\prime = \eta {\bar N}_{\rm S} + (1-\eta){\bar N}_{\rm B}$. The remainder $L-1$ of the $K$-mode blocks are excited in zero-mean thermal states $\hat{\rho}_{\rm th}(0, {\bar N}_{\rm T})$.

%Instead of BPSK Hadamard code words followed by GM stages, one can use PPM to achieve the same performance. This requires pre-sharing brighter SPDC signal-idler mode pairs of mean photon number per mode $L\bar{N}_{\rm S}$. Alice would send an $M$-temporal-mode signal pulse (of mean photon number $L\bar{N}_{\rm S}$) in one ``pulse slot" ($M$ temporal modes) and vacuum in the other $L-1$ pulse slots, exciting only $M$ out of $ML$ transmitted modes. Demodulation would use FF-SFG stages as before, but no GM stages would be needed. The optimal PPM order is $L \approx ({\cal E}\log(1/{\cal E}))^{-1}$ with ${\cal E} = M\eta {\bar N}_{\rm S}/(2{\bar N}_{\rm B})$, which yields  $L{\bar N}_{\rm S} \approx \frac{\bar{N}_0}{\log(\bar{N}_0/{\bar N}_{\rm S})}$ with $\bar{N}_0 = 2{\bar N}_{\rm B}/(M\eta)$. For the parameters in Fig. ~\ref{fig:cap_ratio_M100000}, $\bar{N}_0 = 0.2$, implying that for ${\bar N}_{\rm S} < 0.01$, $L{\bar N}_{\rm S} \lesssim 0.07$. Thus, the ``qubit approximation" analysis of the SFG~\cite{Zhu17} is still valid. However, even though removal of GM stages reduces the receiver complexity, in addition to higher peak power usage, this scheme consumes more entanglement. 

An alternative to this scheme described in Section~\ref{sec:JDRdesign} is for Alice to directly modulate PPM code words. Alice and Bob need to pre-share (brighter) SPDC signal-idler mode pairs of mean photon number per mode $L{\bar N}_{\rm S}$, and Alice sends an $M$-temporal mode signal pulse (of mean photon number $L{\bar N}_{\rm S}$) and nothing (vacuum) in $L-1$ pulse slots. Thus, only $M$ modes are occupied by signal pulses out of each $ML$ transmitted modes. FF-SFG stages are used to demodulate, as before, but no GM stages are needed. The state of the $LK$ output modes of the $L$ $K$-stage FF-SFG modules are identical to the above: one block of $K$ modes carries displaced thermal states $\hat{\rho}_{\rm th}(\sqrt{L {\bar N}_k}, {\bar N}_{\rm T})$, and the remainder $L-1$ of the $K$-mode blocks will be excited in zero-mean thermal states $\hat{\rho}_{\rm th}(0, {\bar N}_{\rm T})$. 

The mean transmit photon number of both schemes are identical. The discrete-memoryless channel (DMC) induced by the modulation-code-receiver combination for both schemes are identical. Hence, the capacities achieved by the two schemes are identical. The optimal PPM order for the second scheme is the optimal Hadamard-code length for the first scheme. That optimal PPM-order (or Hadamard code length) is given by: $L \approx ({\cal E}\log(1/{\cal E}))^{-1}$ with ${\cal E} = M\eta {\bar N}_{\rm S}/(2{\bar N}_{\rm B})$, which translates to $L{\bar N}_{\rm S} \approx \frac{{\bar N}_0}{\log({\bar N}_0/{\bar N}_{\rm S})}$ with ${\bar N}_0 = 2{\bar N}_{\rm B}/(M\eta)$. For the parameters in Fig. ~\ref{fig:cap_ratios}, i.e., $\eta=0.01$, ${\bar N}_{\rm B}=10$, $M=10^4$, we get ${\bar N}_0 = 0.2$, and optimal $L \approx 7$. This implies that that, for ${\bar N}_{\rm S} < 0.01$, $L{\bar N}_{\rm S} \lesssim 0.07$, and that the idler pulses are still in the regime that the implicit ``qubit approximation" analysis of the SFG~\cite{Zhu17} is valid. 

There are key operational differences, however, between the two schemes, which are described below:
\begin{enumerate}
\item {\textbf{Peak power usage}}---Even though the mean photon number that is transmitted over the channel is identical for both schemes, the peak power is not. The PPM scheme uses $L$ times more peak power than the BPSK scheme. For the aforementioned parameters, the optimal PPM order $L \approx 7$, which implies the peak power is $7$ times that of BPSK. However, the BPSK scheme is slightly more restrictive since Hadamard codes exist only for $L$ that is an integer power of $2$. However, it is possible to redesign the BPSK scheme with complex-valued Hadamard codes that would work for all integer $L$.
\item {\textbf{Entanglement consumption}}---More important than the peak power advantage the BPSK scheme enjoys is that its entanglement consumption is lower. Despite the fact that the mean photon number per transmitted mode is ${\bar N}_{\rm S}$ for both schemes, in the PPM scheme, every $M$-mode SPDC pulse that needs to be pre-shared must have $L{\bar N}_{\rm S}$ photons per mode. This is true, even though $(L-1)/L$ fraction of the signal pulses of the pre-shared entangled states will never be transmitted in the PPM scheme. This is a major drawback for this scheme.
\item {\textbf{Receiver complexity}}---The BPSK scheme needs the $K$ Green Machine circuits in JDR1 and one in JDR2 in addition to the FF-SFG modules. That is an added receiver complexity for the BPSK scheme over the PPM scheme.
\item {\textbf{Using the noise modes of FF-SFG stages}}---In the BPSK scheme described in Section~\ref{sec:JDRdesign}, we ignore the $LK$ ``noise modes" labeled ${\hat e}_k^{(i)}$ in Fig. ~\ref{fig:JDR}. In our operational regime, for both the BPSK and PPM schemes, the state of ${\hat e}_k^{(i)}$ is close to zero-mean thermal state of the same mean photon number as that of the corresponding sum-frequency mode, ${\hat b}_k^{(i)}$. The capacity analyses (for both BPSK and PPM) above ignores  ${\hat e}_k^{(i)}$ modes. There is information about the transmitted code word in them, which can only increase the achievable capacity. For the PPM scheme, one can simply do photon counting on all the ${\hat e}_k^{(i)}$ modes. For the pulse-containing block of $K$ noise modes ${\hat e}_k^{(i)}$, $1 \le k \le K$, on-off direct detection of those modes effectively doubles the energy of the ``on" PPM pulse, causing the capacity-ratio plots to shift right by $\log_{10}2$. This is a small improvement, but one that only needs additional single-photon detectors. A similar capacity improvement for the BPSK scheme leveraging the ${\hat e}_k^{(i)}$ modes requires a feedback-based scheme like in~\cite{Zhu17}, where, based on photon-detection events at the noise modes, one adaptively applies two-mode squeezing before and after each of the SFG stages within the FF-SFG modules. 
%Analyzing this scheme requires a second-order characterization of SFG, since the photo-detection statistics across the ${\hat e}_k^{(i)}$ modes are correlated.
\end{enumerate}

\subsection{On-off keying (OOK)}
Finally, PPM can be thought of as a modulation code over an on-off keying (OOK) alphabet, and hence its capacity is strictly inferior to that of OOK, although it is very close to OOK when ${\cal E} \ll 1$. This means that an OOK version of our modulation format also attains the $\log(1/{\bar N}_{\rm S})$ capacity ratio. The ``on" symbol (transmission of the $M$-mode signal pulse) is associated with a prior probability $p$ and the ``off" symbol (no signal transmission) with a prior probability $1-p$, with $p \sim {\cal E}\log(1/{\cal E})$ assuming the role of the inverse-order $1/L$ of PPM, except that there is now no restriction that there must be exactly one ``on" pulse in every $L$-pulse block. 

As described in Section~\ref{sec:EA_binarymodulation}, the entanglement assisted capacity attainable with PPM and OOK based modulation (with TMSV pre-shared entanglement) is strictly inferior to that attainable with BPSK modulation on TMSV pre-shared entanglement. However, BPSK modulation on TMSV, paired with the GM-based JDR1 described in Section~\ref{sec:JDRdesign}, attains the same entanglement assisted capacity achievable with PPM with TMSV but without the GM. Despite this, the BPSK-based JDRs from Section~\ref{sec:JDRdesign} may be more practical in the near term compared to PPM and OOK formats, which require Alice and Bob to pre-share more entanglement (i.e., the pre-shared signal-idler mode pairs need a higher mean photon number per mode). High-rate fault-tolerant entanglement distribution to pre-share the resource necessary for supporting entanglement-assisted communications is likely to be the most expensive process in a future implementation. 

\section{Receiver designs to attain $C_E$}\label{future}

As discussed in section \ref{sec:EAcap_review}, BPSK modulation on TMSV pre-shared entanglement suffices to closely attain $C_E$ in the ${\bar N}_S \ll 1$, ${\bar N}_B \gg 1$ regime (the regime where the most entanglement assisted gain $C_E/C \sim \ln(1/{\bar N}_S)$) holds. In Section~\ref{sec:JDRdesign}, we developed and analyzed two receivers for BPSK modulation on TMSV, the JDR1 and JDR2, both of which attain entanglement-assisted communication rates that exceed the $2C$ limit ($C$ being the unassisted Holevo capacity) associated with symbol-by-symbol measurements. However, these JDRs are not optimal, as there is still a large gap to $C_E$. We refer to the JDR2 design depicted in Fig.~\ref{fig:JDR2} for notation, in this section.

There are two distinct sources of information inefficiencies that must be addressed to close this capacity gap:

\begin{figure}[ht]
\centering
\includegraphics[width=\linewidth]{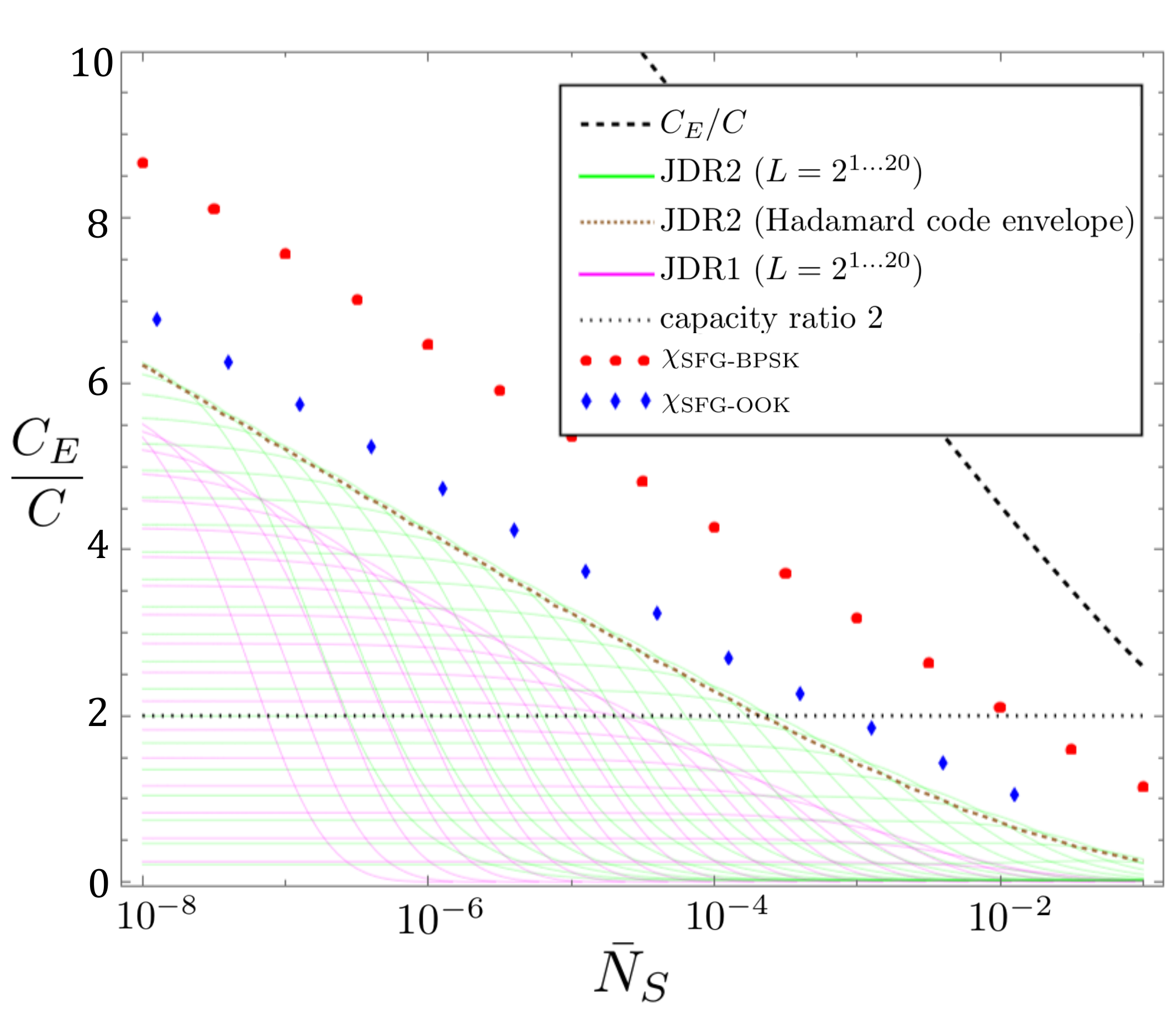}%CEandSFGHolevoPlot.pdf}           
\caption{Plot (red dots) of the Holevo capacity of the effective output ensemble $\{\hat{\rho}_\text{th}(-\alpha_0,\bar{N}_{T0}),\hat{\rho}_\text{th}(\alpha_0,\bar{N}_{T0})\}$ of a single SFG-module of JDR2 (at the output of the beam splitter $\gamma_{K-1}$ of Fig. \ref{fig:JDR2}), as a ratio to the ultimate unassisted capacity $C$. For comparison, the plot also includes (blue diamonds) the Holevo capacity of the ensemble $\{\hat{\rho}_\text{th}(0,\bar{N}_{T0}),\hat{\rho}_\text{th}(\alpha_0,\bar{N}_{T0})\}$ (corresponding to OOK modulation rather than BPSK), as a ratio to $C$. Both plots are shown overlaid on Fig. \ref{fig:jdr2rate} and drawn assuming $\bar{N}_B=10$ and $\eta=0.01$. The thick dashed black line is $C_E/C$.}
\label{fig:sfgholevo}
\end{figure}

\subsection{Receiver back-end that achieves the Holevo-capacity of unassisted communications}\label{sec:optimalbackend}
Even if the FF-SFG receiver front-end was designed to optimally convert the modulated phase in a symbol to the displacement of a single BPSK coherent state without losing any Holevo information content, receivers such as JDR1 and JDR2 that are based on passive linear optics, coherent-state local oscillators and shot-noise-limited photon number detection, are known to be insufficient to achieve the Holevo capacity of the induced phase-modulated coherent-state constellation~\cite{Chu17}, and hence fall short of achieving $C_E$. The various receiver designs we discussed in Section~\ref{sec:receiverforHolevo} ---in the context of achieving the quantum (Holevo) limit of classical communications---can be used as a back-end for an entanglement-assisted receiver design to close the gap between the green envelope and the red dots in Fig. \ref{fig:sfgholevo}.

\begin{figure*}[ht]
\centering
\begin{minipage}[b][0.52\paperheight
][s]{2\columnwidth}
    \includegraphics[width=\textwidth]{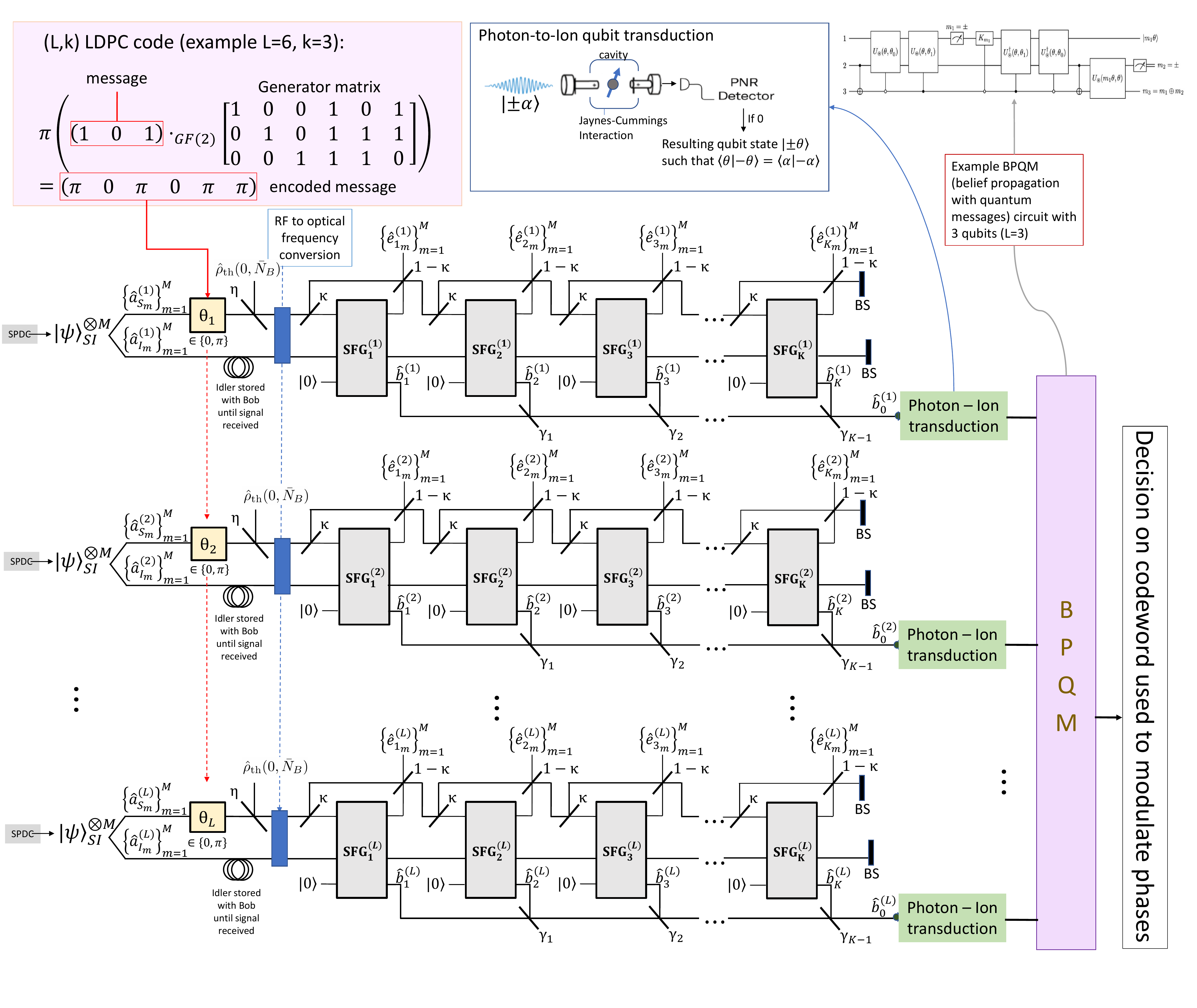}
\end{minipage}
\caption{An alternative to JDR2 to attain the FF-SFG BPSK Holevo capacity (red dots in Fig. \ref{fig:sfgholevo}). See \cite{Delaney21} for definitions of the example circuit components.}
\label{fig:bpqmreceiver}
\end{figure*}

Note that the envelope of the JDR2 capacity in Fig. \ref{fig:sfgholevo} falls short of the Holevo capacity of the FF-SFG output ensemble (red dots) in the same way that the Reed-muller-GM-Dolinar receiver of \cite{Guh11a} falls short of the BPSK Holevo capacity for coherent-state modulated communication (see Fig. \ref{fig:classcaps}). To close the gap, one could implement a quantum belief-propagation receiver, recently proposed in \cite{Rengas20} as the back-end of JDR2. The Reed-muller code would get replaced by an LDPC code, and the demodulation phase would require a photon-to-qubit conversion step before running a quantum message-passing circuit to decode the code words, exemplified by a photon-to-ion transduction in Fig.~\ref{fig:bpqmreceiver}.

Just as JDR2, a BPQM-based receiver for EA communications has the added complication that the detection front-end has nonzero thermal noise compared to noiseless classical communication. The analysis is left as future work.

%This is good reason to explore the limits of Gaussian receiver designs for entanglement-assisted communication in the meantime. So far no Gaussian receiver has been able to exceed $R/C$ above 2 in the low signal brightness limit. One Gaussian receiver idea comes from implementing a multi-mode interferometer-based receiver was implemented where interfering multiple entangled mode pairs before applying phases is part of Alice's modulation scheme? Then Bob would reverse Alice's initial interference transformation and perform photo detection and decide on Alice's code word based on the click pattern. This is essentially a multi-mode generalization of interferometer-based phase inference \cite{Yurke86}.

\subsection{Receiver front-end that extracts information from the noise modes}
The red dots in Fig. \ref{fig:sfgholevo} reveal that the Holevo capacity of the ensemble $\{\hat{\rho}_\text{th}(\pm\alpha_0,\bar{N}_{T0})\}$ at the output of the beam splitter $\gamma_{K-1}$ combining the outputs of the sum-frequency modes does not attain $C_E$, either in value or in scaling. Yet the input ensemble does attain $C_E$, as shown in section \ref{sec:EAcap_review} (Fig. \ref{fig:eacaps}). Given that the $\kappa$-beam splitters and SFG operations are unitary, and hence do not decrease the information content of the input, the information missing in the combined sum-frequency modes warrants tracking down. 

\begin{figure}[h]
\centering
\includegraphics[width=\linewidth]{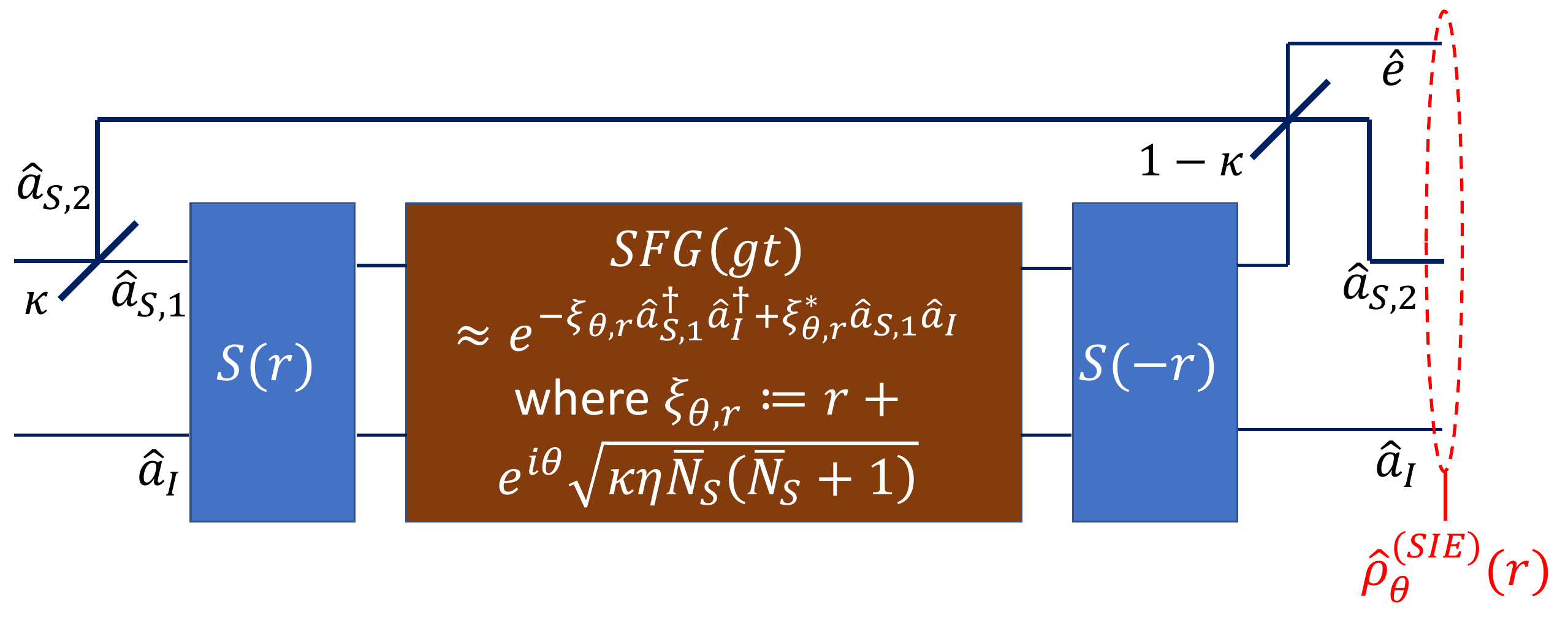}
\caption{Diagram of the effective action of an SFG gate on the signal and idler modes (ignoring the sum-frequency mode) in the qubit approximation, sandwiched by TMS operations with squeezing amount $r$.}
\label{fig:lostinfo}
\end{figure}

One candidate for this missing information are the noise modes ${\hat e}$ in Fig.~\ref{fig:JDR2}. But, they are in a zero-mean thermal state that has no information about the modulated binary phase on the symbol. We claim that sandwiching each SFG gate with two-mode-squeezing gates $S(r)$ and $S(-r)$, whose action on the joint state of the signal and idler modes is described by the unitary
$$
S(r):=e^{\text{sinh}^{-1}(r)(\hat{a}_S^\dagger\hat{a}_I^\dagger-\hat{a}_S\hat{a}_I)},
%\label{eq:tmsdef}
$$
with the associated Heisenberg mode transformation given by $\hat{a}_S\longrightarrow \sqrt{1+r^2}\,\hat{a}_S+r\,\hat{a}_I^\dagger$ (swapping $S$ for $I$ for the idler mode transformation), enables the extraction of the information missing in the sum frequency modes from the photon detection statistics of the noise mode $\hat{e}$. To see how, we consider the effect of a single sandwiched SFG gate on the received signal and idler in the qubit approximation as depicted in Fig. \ref{fig:lostinfo}, ignoring the sum-frequency mode. 
Let $\hat{a}_{S,1}, \hat{a}_{S,2}$ and $\hat{e}$ be associated with modes as depicted in Fig. \ref{fig:lostinfo}. It was shown in \cite{Zhu17} that if the input signal and idler are dim enough, the evolution time $t$ can be chosen appropriately, in particular $gt=\pi/2$, such that the SFG gate effectively acts as a two-mode squeezer whose squeezing coefficient $-\xi_{\theta,r}$ depends on $\theta$ and $r$, being equal to the negative of the phase-dependent cross correlation $\langle\hat{a}_{S,1}\hat{a}_I\rangle$ at the input of the SFG gate. Up to first order in $\bar{N}_S$ and $r$,
\begin{equation}
\xi_{\theta,r}=e^{i\theta}\sqrt{\kappa\eta\bar{N}_S(\bar{N}_S+1)}+r.
\label{eq:SFGsqueeze}
\end{equation}

Note that even though all of the operations in Fig. \ref{fig:lostinfo} including the beam splitters are unitary, the information content of the ensemble $$\{\hat{\rho}_\theta^{\text{(SIE)}}\,\mid\,\theta\in\{0,\pi\}\}$$ of the joint state of the signal (S), idler (I) and noise (E) modes (a Gaussian state in the qubit approximation for the SFG gate) after the gates and beam splitters is less than that of the input ensemble. The reason is that in order to ignore the sum-frequency mode, the SFG gate has been reduced to a two-mode unitary whose action is a function of $\theta$, the  very parameter in which the information is encoded. If the full SFG unitary is considered as a $\theta$-independent operation then one cannot neglect the sum-frequency mode when evaluating the information content at the output, at which point one finds that some of the information in the received state has been funneled into the sum-frequency mode, as evaluated in detail in section \ref{sec:JDRdesign}.
\begin{figure}[h]
\centering
\includegraphics[width=\linewidth]{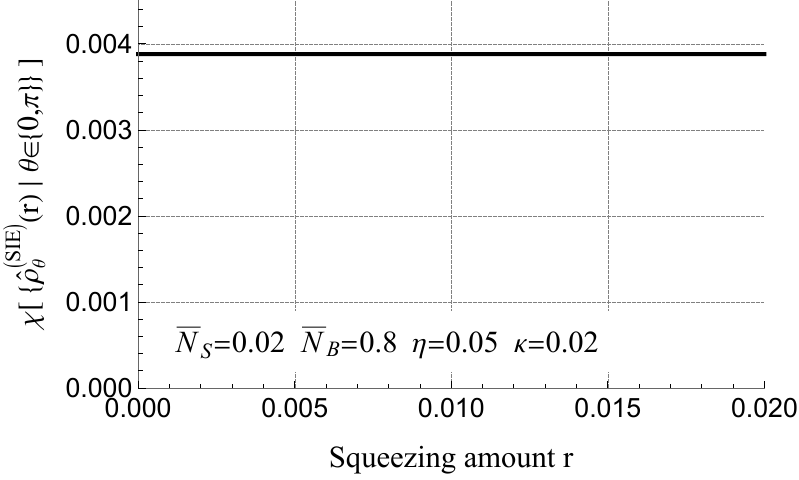}
\caption{Holevo information of the BPSK ensemble $\hat{\rho}_\theta^{\text{(SIE)}}$ for $\theta\in\{0,\pi\}$ (in bits/channel use), plotted as a function of $r$, where $\hat{\rho}_\theta^\text{(SIE)}(r)$ is the state of the signal, idler and noise modes after the operations depicted in Fig.\ref{fig:lostinfo}. $\chi$ was evaluated numerically by diagonalizing $\frac{1}{2}(\hat{\rho}_0^{(SIE)}+\hat{\rho}_\pi^{(SIE)})$ in the Fock basis with cutoff $n=18$.}
\label{fig:lostinfoplot}
\end{figure}

Moreover, it is not obvious that the information content of $\hat{\rho}_\theta^{(\text{SIE})}$ is independent of $r$. Since the effective SFG action is also a function of $r$, the amount of information removed via the $\theta$-dependence of the effective SFG could in principle be parameterized by $r$. However, when the Holevo capacity of $\hat{\rho}_{\theta}^{(\text{SIE})}$ is evaluated as a function of $r$ (Fig. \ref{fig:lostinfoplot}), it is seen to be independent of $r$.

Consider the case when $r=0$ in Fig.\ref{fig:lostinfo}. If increasing $r$ to a nonzero value does in fact activate the noise mode $\hat{e}$ so that the value of $\theta$ is reflected in the $\hat{e}$ mode's photon detection statistics, which we have yet to show, Fig. \ref{fig:lostinfoplot} raises the question of how the $\theta$-information is embedded in the joint state of the modes $\hat{a}_{S,1}, \hat{a}_I$ and $\hat{e}$ before the post-SFG TMS gate. The answer is that the $\theta$ dependence is in the phase-sensitive cross-correlation $\langle\hat{a}_{S,2}\hat{a}_I\rangle$, whose expression for any value of $r$ is given by
\begin{equation}
\begin{split}
    \langle\hat{a}_{S,2}\hat{a}_I\rangle=\pm(1-\kappa(\frac{1}{2}+\bar{N}_B))\sqrt{\eta\bar{N}_S} \\+ \mathcal{O}(\bar{N}_S^{3/2})+\mathcal{O}(\kappa r^2)+\mathcal{O}(\kappa^{3/2})
    \label{eq:idlernoisecorr}
\end{split}
\end{equation}
where $\pm$ corresponds to $\theta=0$ and $\theta=\pi$.

Note that neither the $\hat{a}_{S,1}$-mode nor the $\hat{a}_{S,2}$-mode's mean photon number depends on $\theta$, so when the modes $\hat{a}_{S,2}$ and $\hat{a}_{S,1}$ recombine on the $1-\kappa$ beam splitter to produce the $\hat{e}$-mode, a $\theta$-dependent mean photon number on the $\hat{e}$-mode can only result if the $\hat{a}_{S,1}$ and $\hat{a}_{S,2}$-modes are classically correlated by an amount that depends on $\theta$. At the input of the post-SFG TMS gate in Fig. \ref{fig:lostinfo}, the actual expression for this classical correlation is
\begin{align}
    \langle\hat{a}_{S,2}^\dagger\hat{a}_{S,1}\rangle=&\sqrt{\kappa}\bar{N}_B\nonumber+\mathcal{O}(\kappa\bar{N}_S^{3/2})\\
    &+\mathcal{O}(\kappa^{3/2})+\mathcal{O}(\sqrt{\kappa}(\bar{N}_S^2+r^2)),
    \label{eq:signal12corr}
\end{align}
which is clearly independent of $\theta$.

\begin{figure}[!ht]
    \includegraphics[width=\linewidth]{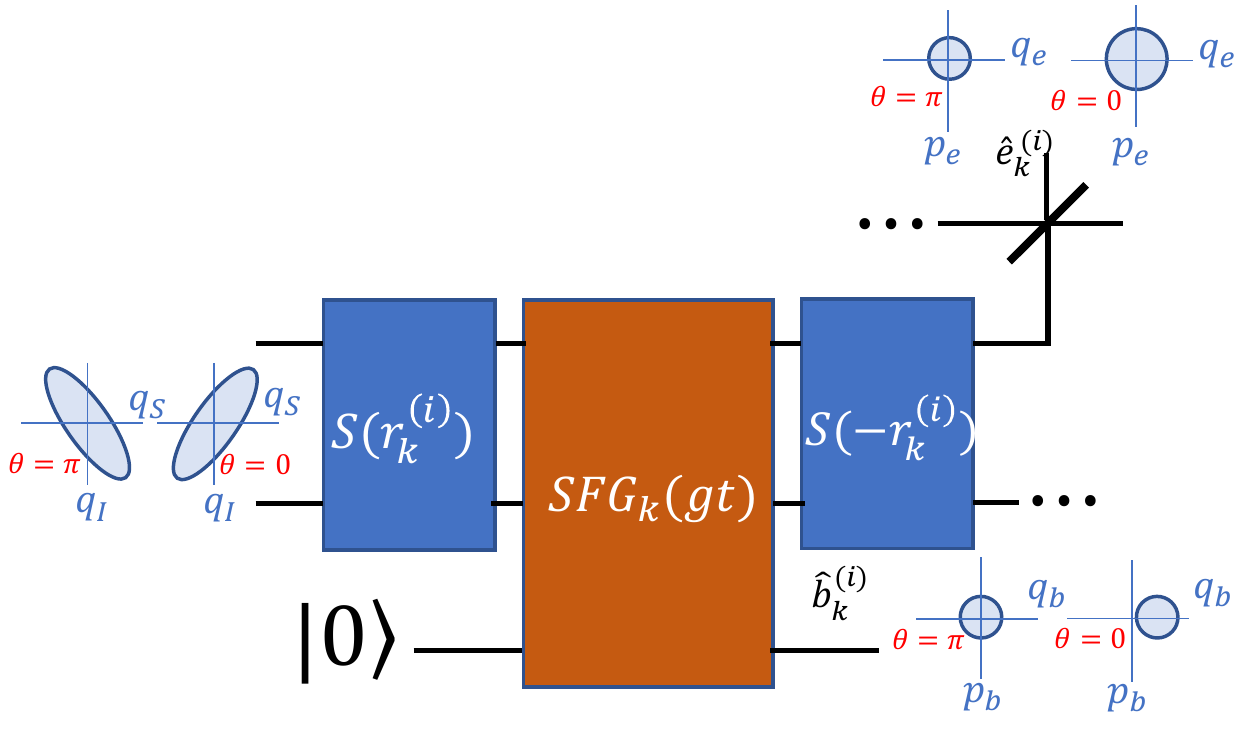}
    \caption{Sandwiching the SFG gate with squeezers allows for extraction of information from the environment modes as well as the sum-frequency modes. The effect of the phase of the cross-correlation between the signal and idler modes on the SFG output modes $\hat{e}$ and $\hat{b}$ when the squeezing parameters $r_k$ are adjusted to null one of the BPSK symbols is depicted by cartoons of the Wigner function representation of the states.}
    \label{fig:sanwich}
\end{figure}

Eq.s~\eqref{eq:signal12corr} and \eqref{eq:idlernoisecorr} reveal the crucial role of the post-SFG two-mode squeezing gate depicted in Fig. \ref{fig:lostinfo}. By linearly mixing the quadratures $\hat{a}_{S,1}^\dagger$ and $\hat{a}_I$, the post-SFG TMS gate transfers $\theta$-dependence from the correlation $\langle\hat{a}_ {S,2}\hat{a}_I\rangle$ onto the correlation $\langle\hat{a}_{S,2}^\dagger\hat{a}_{S,1}\rangle$ so that upon recombining on the $1-\kappa$ beam splitter, the mean photon number of the $\hat{e}$ mode is given by
\begin{equation}\begin{split}
    \langle\hat{e}^\dagger\hat{e}\rangle=\pm2r\sqrt{\eta\kappa\bar{N}_S}\\+\kappa\eta\bar{N}_S+r^2\\+\mathcal{O}((\bar{N}_S-\kappa)r^2),
    \label{eq:nnoiseatend}
    \end{split}
\end{equation}
where $\pm$ corresponds to $\theta=0$ and $\theta=\pi$. This matches the coherent contribution to the mean photon number of the sum-frequency mode whose coherent displacement, after the pre-SFG squeezing, is approximately $\langle\hat{b}\rangle\approx r\pm\sqrt{\kappa\eta\bar{N}_S}$. The distillation of the phase of the signal-idler cross-correlation into the sum-frequency and noise modes by the squeezer-sandwiched SFG operation is sketched out in phase-space in Fig. \ref{fig:sanwich}

%As a reminder, if the $\hat{e}$-modes are excluded, as in case of JDR1 and JDR2, the effective output state of the FF-SFG module is $\hat{\rho}_\text{th}(\pm\alpha_0,\bar{N}_{T0})$, without the factor of $\sqrt{2}$ on the displacement.
Without the squeezers sandwiching the SFG gates, the information bridging the gap between $C_E$ and the Holevo capacity of the ensemble of the combined sum-frequency modes (red dots in Fig. \ref{fig:sfgholevo}) would be trapped in the cross-correlations $\langle\hat{e}^{(i)}_k\hat{a}_I\rangle$ between the noise modes and the idler mode, undetectable by photon counters on the $\hat{e}^{(i)}_k$ modes for any $i$ and $k$.

\begin{figure}[!ht]
    \centering
    \includegraphics[width=\linewidth]{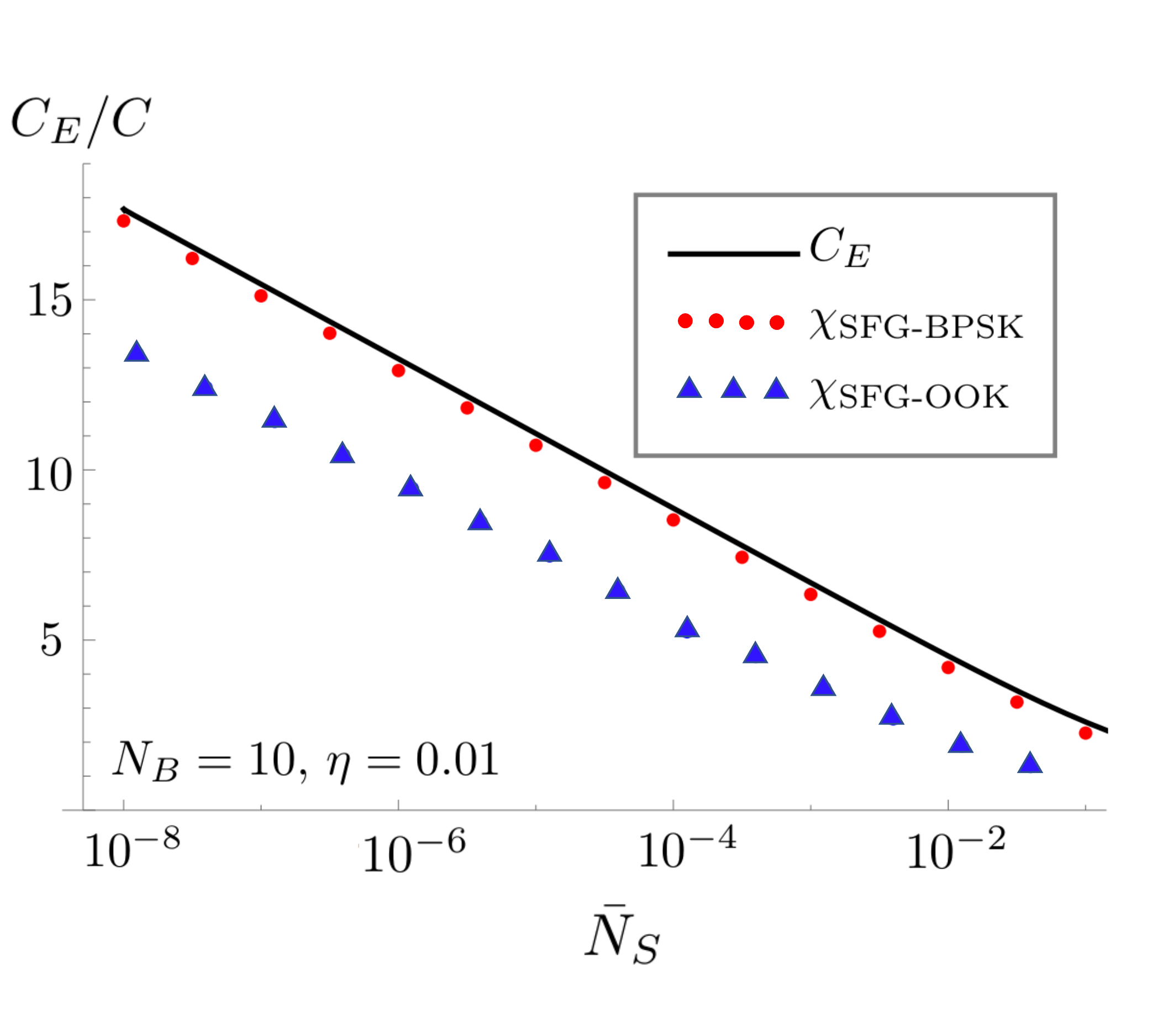}
    \caption{Plot (red dots) of the Holevo capacity of the effective output ensemble $\{\hat{\rho}_\text{th}(-\sqrt{2}\alpha_0,\bar{N}_{T0}),\hat{\rho}_\text{th}(\sqrt{2}\alpha_0,\bar{N}_{T0})\}$ of a single SFG-module of JDR2 (capturing both the output of $\gamma_{K-1}$ and the noise modes $\{\hat{e}^{(i)}_k\}_{k=1}^K\}$), as a ratio over the ultimate classical capacity $C$. The plot also includes the Holevo capacity (blue triangles) of the ensemble $\{\hat{\rho}_\text{th}(0,\bar{N}_{T0}),\hat{\rho}_\text{th}(\sqrt{2}\alpha_0,\bar{N}_{T0})\}$ (corresponding to OOK modulation).}
    \label{fig:fullsfgholevo}
\end{figure}

With the squeezing parameters $r_k^{(i)}=\pm\alpha_k^{(i)}$ chosen to null one of the hypotheses, the effective output state of an FF-SFG module (a multimode state consisting of $\hat{b}_0^{(i)}$ and i.i.d. $\hat{e}_k^{(i)}$-modes for $k$ ranging from $1$ to $K$) is $\hat{\rho}_\text{th}(\pm\sqrt{2}\alpha_0,\bar{N}_{T0})$, as shown in Appendix~\ref{app:sqrt2}. The Holevo capacity of this ensemble achieves $C_E$, as Fig. \ref{fig:fullsfgholevo} shows. A receiver back-end design that actually achieves the full $C_E$ latent in the front-end output however, must be a JDR that is not only Holevo-capacity-attaining for BPSK modulated coherent states (with a small amount of additive thermal noise), but also one that is able to see the effective output of the front-end as $\hat{\rho}_\text{th}(\pm\sqrt{2}\alpha_0,\bar{N}_{T0})$ when the noise modes are included.
\iffalse
On the other hand, the full Holevo capacity of the FF-SFG output (encompassing the noise modes and corresponding to the BPSK ensemble $\hat{\rho}_\text{th}(\pm\sqrt{2}\alpha_0,\bar{N}_{T0})$ where $\bar{N}_{T0}$ is evaluated in appendix \ref{app:gammanoise}) turns out to match $C_E$ almost exactly, as Fig. \ref{fig:fullsfgholevo} shows. This shows that the body of JDR2 has the potential to attain $C_E$.\fi

\begin{figure}[!ht]
\centering
\hspace*{-0.5cm}
\includegraphics[width=1.05\linewidth]{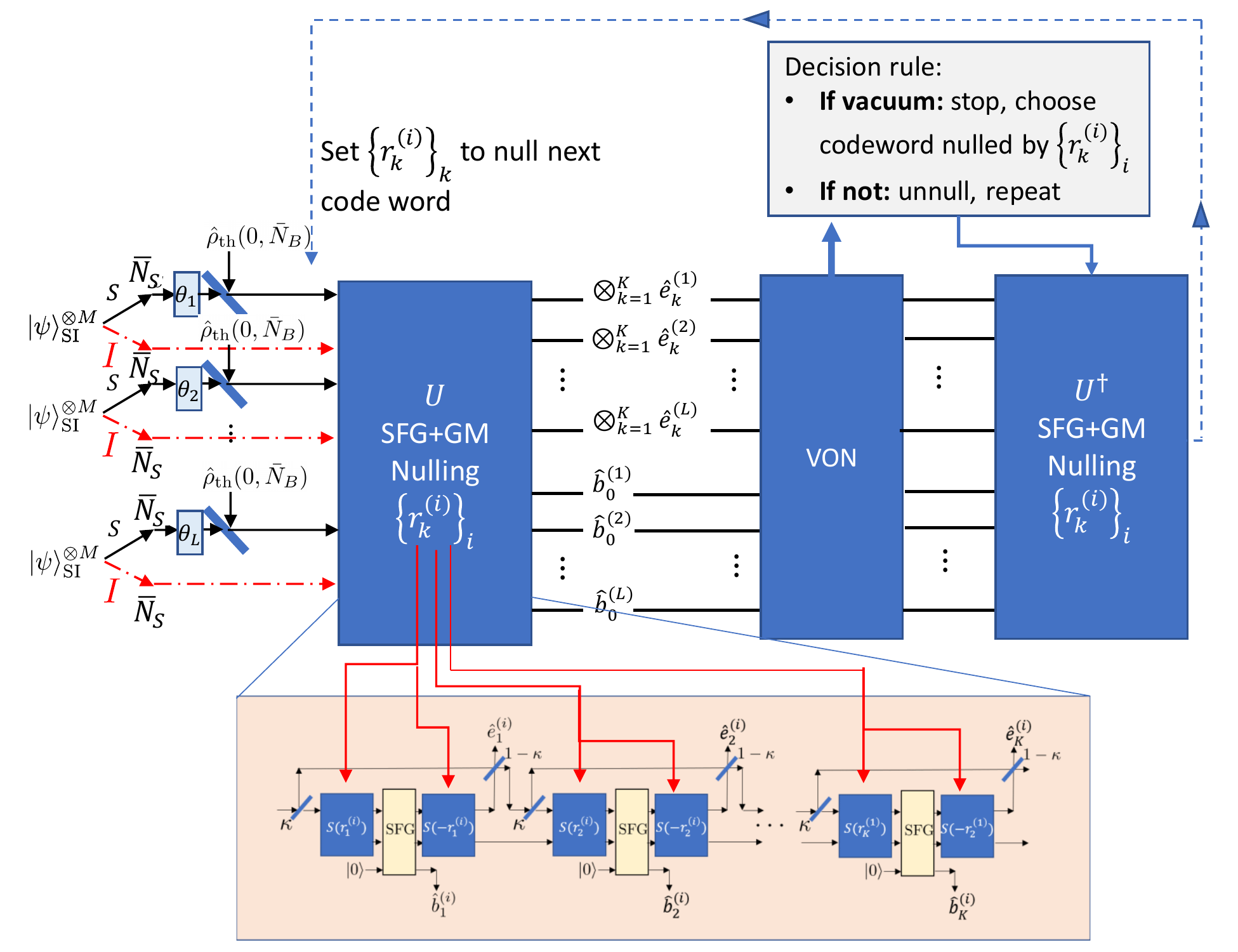}
\caption{Vacuum-or-not joint receiver design idea for entanglement-assisted communications to attain the entanglement-assisted \iffalse Holevo-Werner \fi capacity $C_E$.}
\label{fig:vonreceiverdiag}
\end{figure}

One possible complete structured receiver design that could potentially attain $C_E$ is inspired by the Holevo-capacity attaining vacuum-or-not JDR for coherent-state modulation proposed in Ref.~\cite{Wil12}. The receiver would successively null the patterns of cross-correlations on the encoded mode-pair-blocks using the appropriate squeezing amounts on the sandwiching squeezers, then perform a multi-mode vacuum-or-not (VON) measurement on the joint output state of the SFG-Green-Machine module, as pictured in Fig. \ref{fig:vonreceiverdiag}.

The analysis of such a receiver is complicated by the fact that correct nulling of the code words does not result in perfect vacuum, as for the case of noiseless coherent state encoded communication~\cite{Wil12}. Moreover, while the idea of combining an SFG-based JDR with a VON measurement possibly attains $C_E$ in principle, it will be difficult to construct in practice. Two major experimental challenges associated with building such a receiver are the successful implementation of a multi-mode vacuum-or-not measurement, e.g., leveraging the V-STIRAP interaction in a light-atom interaction~\cite{Oi2013}, and the ability to push the entire multi-mode output state back through the SFG in reverse losslessly after each round of the vacuum-or-not measurement. 

In Appendix~\ref{app:sqrt2}, we provide strong evidence to this vacuum-or-not (VON) receiver back-end achieving $C_E$. We calculate the block vacuum-probability with correct and incorrect codeword nulling, showing that these probabilities are exactly those of the VON receiver of~\cite{Wil12} applied on the noisy BPSK coherent state codewords with symbols $\hat{\rho}_\text{th}(\pm\sqrt{2}\alpha_0,\bar{N}_{T0})$.

Notably, the sandwiching squeezers were an essential component in~\cite{Zhu17} for closing the 6dB gap to the quantum Chernoff exponent for the illuminated target-detection problem so the TMSV-sandwiched SFG gates are in a sense, natural to consider as a component for the task of communication. Even so, the optimal receiver designs for the tasks of sensing and communication are different, the former implementing a feed-forward structure paralleling the Dolinar receiver for optimal symbol-by-symbol coherent-state discrimination, and the latter implementing a joint multi-mode non-destructive vacuum-or-not module described above.

\section{Conclusion}

Bridging the gap to $C_{\rm E}$ will require better codes and more complex quantum joint detection receivers. Ideas for follow on works include (1) developing theoretical work on receiver performance classification based on constituent optical elements as in Ref.~\cite{Chu17}, (2) using information in the extra modes ${\hat e}_k^{(l)}$'s in Fig. ~\ref{fig:JDR} as in~\cite{Zhu17}, and (3) FF-SFG modules consisting of SFG gates sandwiched by two-mode-squeezing stages as in~\cite{Zhu17} to successively null the mean field amplitudes of the thermal states of the sum-frequency modes and the mean thermal photon numbers of the environment modes unused by JDR1 and JDR2. In addition, we would like to note that as ${\bar N}_{\rm B}\to\infty$, $C_{\rm E}/(C \ln {\bar N}_{\rm S}) \to 1$, while $R_{\rm E}^{(M)}/(C \ln {\bar N}_{\rm S}) \to 1/2$~\cite{Gagatsos2020}, which indicates a possible check, to see if the entanglement-assisted capacity attained by an improved receiver design improves this ratio from $1/2$ to $1$. Finally, as an alternative to the originally-proposed random-coding method to achieve $C_{\rm E}$~\cite{BSST}, ``position based encoding" was proposed to achieve $C_{\rm E}$ over a general quantum channel~\cite{Qi18,Din17,Ans19,Osk19,Zhu21}. However, a structured optical receiver of the associated receiver's joint-detection measurement is unknown, and is an interesting topic for future work.

Pre-shared entanglement improves capacity when the transmitted power is low and thermal-noise mean photon number is high, despite the entanglement not surviving these conditions. While typically uncommon, this regime corresponds to \emph{covert communications}, a security modality where the mere attempt to transmit data must be hidden. Covert transmission using $n$ total modes is constrained to per-mode photon number ${\bar N}_{\rm S}= c/\sqrt{n}$ for a constant $c$ that depends on ${\bar N}_{\rm B}$, $\eta$, and the desired stringency of covertness~\cite{Bas15,Bul19}. Thus, one can transmit covertly and reliably only ${O}(\sqrt{n})$ bits using $n$ modes without entanglement assistance, however, pre-shared entanglement breaks this \emph{square root law}, allowing transmitting $\mathcal{O}(\sqrt{n}\log n)$ bits covertly~\cite{Gagatsos2020}. %In fact, while the transceiver structure described in this paper is suboptimal, in the limit as $n\to\infty$ it achieves half the covert capacity.

\section*{Acknowledgments}

SG, CNG, and AC acknowledge General Dynamics Mission Systems (GDMS) for supporting this research. AC and BB was partially funded by the National Science Foundation (NSF) grant CCF-2006679. QZ and BB further acknowledge the Army Research Office (ARO) Grant Numbers W911NF-19-1-0418 and W911NF-19-1-0412, respectively. The authors acknowledge William Clark of GDMS, as well as Michael Bullock and Zheshen Zhang of University of Arizona for many valuable discussions.

S.G. has outside interests in Guha LLC, Xanadu Quantum Technologies, Quantum Network Technologies, and SensorQ Technologies. These have been disclosed to the University of Arizona and reviewed in accordance with its conflict of interest policies. Any resulting conflicts of interest from these interests will be managed by The University of Arizona in accordance with its policies.

\appendix

\section{OPA receiver analysis}\label{app:OPA}

In the low photon number regime (${\bar N}_{\rm S}\ll 1$) the communication capacities are well-approximated by the Taylor series expansion around ${\bar N}_{\rm S} = 0$. For example, the Holevo capacity $C(\eta, {\bar N}_{\rm S}, {\bar N}_{\rm B})$ is:
\begin{equation}
C(\eta, {\bar N}_{\rm S}, {\bar N}_{\rm B}) = \eta {\bar N}_{\rm S} \log\left(1+\frac{1}{(1-\eta){\bar N}_{\rm B}}\right)+o({\bar N}_{\rm S}).
\end{equation}

Here we derive the Taylor series expansion of the entanglement-assisted communication capacity with an SPDC source, BPSK modulation, and the OPA receiver~\cite{Guh09} of gain $G$. We use it to evaluate the entanglement-assisted capacity gain achieved by an OPA receiver over the Holevo capacity. This channel's capacity is the classical mutual information between the random binary phase input $\theta\in\{0,\pi\}$, $P(\theta=0)=q$, modulating the block of $M$ transmitted symbols (i.e., $M$-fold tensor product of TMSV states) and the photon-count output $L$ of Bob's detector, optimized over the probability distribution of the input defined by $q$:
  \begin{align}
  C_{\text{EA-OPA}}(\eta,{\bar N}_{\rm S},{\bar N}_{\rm B})=\max_q I(\theta;{\bar N}_{\rm S}).
  \end{align}
The probability that the photon counter records $k$ photons over $M$ modes is: 
\begin{align}
\label{eq:Pk}
P(k|\theta;M) = \frac{1}{(1+{\bar N}_{\theta})^M}{k+M-1 \choose k}\left(\frac{{\bar N}_{\theta}}{1+{\bar N}_{\theta}}\right)^k.
\end{align}
When phase $\theta$ is transmitted, the mean received photon number per mode is:
\begin{align}
{\bar N}_{\theta} = G {\bar N}_{\rm S} + (G-1){\bar N}_{\rm S}'+2C_p\sqrt{G(G-1)}\cos(\theta),
\end{align}
where ${\bar N}_{\rm S}$ is the mean photon number in each signal and idler mode, ${\bar N}_{\rm B}$ is the mean thermal noise injected by the environment, $\eta$ is the channel transmissivity, ${\bar N}_{\rm S}'\equiv \eta {\bar N}_{\rm S}+(1-\eta){\bar N}_{\rm B}+1$, $G$ is the gain of the OPA, and $C_p\equiv \sqrt{\eta {\bar N}_{\rm S}({\bar N}_{\rm S} + 1)}$.
	
%	Since the receiver and transmitter are fixed, the analysis of this channel is classical. Thus, the one-shot bounds \cite{wang09coding}, \cite{polyanskiy09coding} apply in the same way they applied in \cite{wang15covert}. Therefore, $c_{\mathrm{rel},EA}$ is the coefficient corresponding to $\ns$ in the Taylor series expansion of the mutual information.
The Taylor series of mutual information $I(\theta;{\bar N}_{\rm S})$ at ${\bar N}_{\rm S}=0$ is:
\begin{align}
\nonumber I(\theta;{\bar N}_{\rm S})=-{\bar N}_{\rm S}\sum_{k=0}^{\infty}\sum_{\theta\in\{0,\pi\}}\left.Q_\theta(k,{\bar N}_{\rm S})\right|_{{\bar N}_{\rm S}=0}+o({\bar N}_{\rm S}),
\end{align}
%\ns\sum_{n=0}^{\infty}&-q\frac{\d\atermM}{\d{\ns}}\log\left(q+\frac{(1-q)\btermM}{\atermM}\right)\\+ \nonumber&(1-q)\frac{\d\btermM}{\d{\ns}}\log\left((1-q)+\frac{q\atermM}{\btermM}\right) \\ & + \mathrm{o}(\ns^2),
where 
\begin{widetext}
\begin{align}
Q_\theta(k,{\bar N}_{\rm S}) =  
	\begin{cases} 
      q\frac{\d\atermM}{\d{{\bar N}_{\rm S}}}\log\left(q+(1-q)\frac{\btermM}{\atermM}\right), & \theta = 0 \\
      (1-q)\frac{\d\btermM}{\d{{\bar N}_{\rm S}}}\log\left((1-q)+q\frac{\atermM}{\btermM}\right), & \theta = \pi 
   \end{cases}.
\end{align}
\end{widetext}
\vspace{30pt}
Substitution of \eqref{eq:Pk} and evaluation of 
  $Q_\theta(k, {\bar N}_{\rm S})\big\rvert_{{\bar N}_{\rm S} = 0}$ by taking the limit 
  $\lim_{{\bar N}_{\rm S} \to 0} Q_\theta(k,{\bar N}_{\rm S})$ yields:\newpage
\begin{widetext}
\begin{align}
&I(\theta;{\bar N}_{\rm S})={\bar N}_{\rm S}8q(1-q)\eta\nonumber\\
&\times\sum_{k=0}^{\infty} G(G-1)^{k-1}({\bar N}_{\rm B}')^{k-2}(G+(1-\eta)(G-1){\bar N}_{\rm B})^{k-M-2}(k+(G-1)M({\bar N}_{\rm B}')^2{k+M-1 \choose k}\nonumber \\ & 	+ o({\bar N}_{\rm S}),\label{eq:sumI}	
\end{align}
\end{widetext}
where ${\bar N}_{\rm B}'\equiv1+(1-\eta){\bar N}_{\rm B}$.
%Evaluating the sum yields 
%\begin{align}
%I(\theta;N)=\frac{8q(1-q){\eta}GM}{(1+\nt-\eta{\nt})(G+(1-\eta)(G-1)\nt)}\ns+ \mathrm{o}(\ns).
%\end{align}
Well-known results for the moments of binomial distribution are used to evaluate the sum in \eqref{eq:sumI}. Maximizing over $q$ yields:
\begin{align}
C_{\text{EA-OPA}}(\eta,{\bar N}_{\rm S},{\bar N}_{\rm B})=&\frac{2{\eta}GM{\bar N}_{\rm S}}{{\bar N}_{\rm B}'(G+(1-\eta)(G-1){\bar N}_{\rm B})}\nonumber \\&+ o({\bar N}_{\rm S}).
\end{align}
%Thus, $c_{\mathrm{rel},EA}$ is the coefficient of the $\ns$ term in the Taylor series. 
%\begin{align}
%\label{eq:crelea}c_{\mathrm{rel},EA} = \frac{2{\eta}GK}{(1+\nt-\eta{\nt})(G+(1-\eta)(G-1)\nt)}
%\end{align}

% This coefficient when $K=1$ is given as follows:
%\begin{align}
%\label{eq:crelea}c_{\mathrm{rel},EA}=\frac{2\eta G}{\left(1+(1-\eta)\nt\right)\left(G+(G-1)(1-\eta)\nt\right)}
%\end{align}
The maximum gain from using the SPDC source, BPSK modulation and the OPA receiver over the Holevo capacity when ${\bar N}_{\rm S}\ll 1$ and ${\bar N}_{\rm B}\gg 0$ is thus:
\begin{align}
\lim_{G \downarrow 1}\lim_{{\bar N}_{\rm B} \to \infty} \frac{C_{\text{EA-OPA}}(\eta,{\bar N}_{\rm S},{\bar N}_{\rm B})}{M\times C(\eta,{\bar N}_{\rm S},{\bar N}_{\rm B})} = 2,
\end{align}
where $\lim{G\downarrow1}$ indicates a one-sided limit taken from above, and we normalize the denominator by $M$ to account for employing block encoding of $M$ symbols.
We note that, with such normalization, the gain does not depend on $M$.  There is also no dependence on the transmissivity~$\eta$.

\section{Connection with PPM with dark-click rate proportional to mean energy per slot}\label{app:PPMcrude}

In this Appendix, we consider a cruder approximation of $R_{\rm E}^{(M)}$, providing an alternative proof of the scaling in~\eqref{eq:optimal_scaling}, but one that lets us establish a connection with a problem that was studied by Wang and Wornell in the context of coherent-state PPM modulation, where the dark click probability per mode $\lambda$ is proportional to the mean photon number per mode $\cal E$~\cite{Wan14}.

Recall that $R^{(M)}_{\rm E} = {{\rm sup}_{L}R_{\rm E}^{(M,L)}}$ is the envelope of capacities attained by our scheme over all $L$, for a given $M$. 
%Applying the conditions pertinent to our problem setting, $\kappa {\bar N}_{\rm S} \ll {\bar N}_{\rm S} \ll 1 \ll {\bar N}_{\rm B} \ll 1/\kappa$, we get ${\bar N}_{\rm S}^\prime \to {\bar N}_{\rm B}$, $1/(1+{\bar N}_{\rm T})^K \to e^{-{\bar N}_{\rm S}{\bar N}_{\rm B}}$ and $A/(1-\mu) \to nM\eta {\bar N}_{\rm S}/2(1+(1-\eta){\bar N}_{\rm B})$, which 
Approximations in \eqref{eq:approxNSp}-\eqref{eq:approxA}
lead to the following simplified asymptotic expressions: $1 - p_{\rm c} \approx e^{-(L{\cal E}+\lambda)}$, and $1-p_{\rm b} \approx e^{-\lambda}$, $\lambda = c{\cal E}$, with
\begin{align}
{\cal E}& = \frac{M\eta {\bar N}_{\rm S}}{2(1+(1-\eta){\bar N}_{\rm B})}\\
c &= \frac{2(1-\eta){\bar N}_{\rm B}(1+(1-\eta){\bar N}_{\rm B})}{M\eta}.
\end{align} 
This is exactly the setting of $L$-mode coherent-state PPM modulation and direct detection, where the dark click probability per mode $\lambda$ is proportional to the mean photon number per mode ${\cal E}$~\cite{Wan14}. The leading-order terms of the optimal capacity for this setting, in the regime of ${\cal E} \ll 1$, is given by:
\begin{equation}
C_{\rm PPM}({\cal E}) \approx {\cal E}\log\frac{1}{\cal E} - {\cal E}\log \ln\frac{1}{\cal E} - {\cal E}\ln(1+c),
\end{equation}
with the optimal PPM order, $L = \lfloor \left({{\cal E}\log(1/{\cal E})}\right)^{-1} \rfloor$~\cite{Wan14}. 
Applying this result to our problem, we get
\begin{eqnarray}
\label{eq:REM_secondorder}{R^{(M)}_{\rm E}}&=&\frac{C_{\rm PPM}({\cal E})}{M},
\end{eqnarray}
%\begin{eqnarray}
%{R^{(M)}_E} &\approx& \frac{\eta {\bar N}_{\rm S}}{2(1+(1-\eta){\bar N}_{\rm B})}\left[\log\left(\frac{2(1+(1-\eta){\bar N}_{\rm B})}{M\eta {\bar N}_{\rm S}}\right) - \log\left(\ln\left(\frac{2(1+(1-\eta){\bar N}_{\rm B})}{M\eta {\bar N}_{\rm S}}\right)\right)-\ln(1+c)\right].
%\label{eq:REM_secondorder}
%\end{eqnarray}
with ${R^{(M)}_{\rm E}}\approx (\eta {\bar N}_{\rm S}/(2(1+(1-\eta){\bar N}_{\rm B})))\log(2(1+(1-\eta){\bar N}_{\rm B})/(M\eta {\bar N}_{\rm S}))$ in the leading order. In the same regime as above, $\kappa {\bar N}_{\rm S} \ll {\bar N}_{\rm S} \ll 1 \ll {\bar N}_{\rm B}$, the leading order term for the Holevo capacity (attained using coherent states and Gaussian amplitude-and-phase modulation), $C \approx \eta {\bar N}_{\rm S}/{\bar N}_{\rm B}$, and that of the entangled-assisted capacity (achieved via an SPDC transmitter and phase-only modulation), $C_{\rm E} \approx (\eta {\bar N}_{\rm S}/{\bar N}_{\rm B})\log(1/{\bar N}_{\rm S})$~\cite{Shi19}. It therefore follows that,
\begin{equation}
\frac{R^{(M)}_{\rm E}}{C} \sim \log\left(\frac{1}{{\bar N}_{\rm S}}\right), \, \forall M,
\end{equation}
proving that our transmitter-receiver structure attains the optimal capacity scaling.

%The figure below belongs to the following appendix
\begin{figure}[!ht]
\centering
\begin{subfigure}{.8\columnwidth}
    \includegraphics[width=\linewidth]{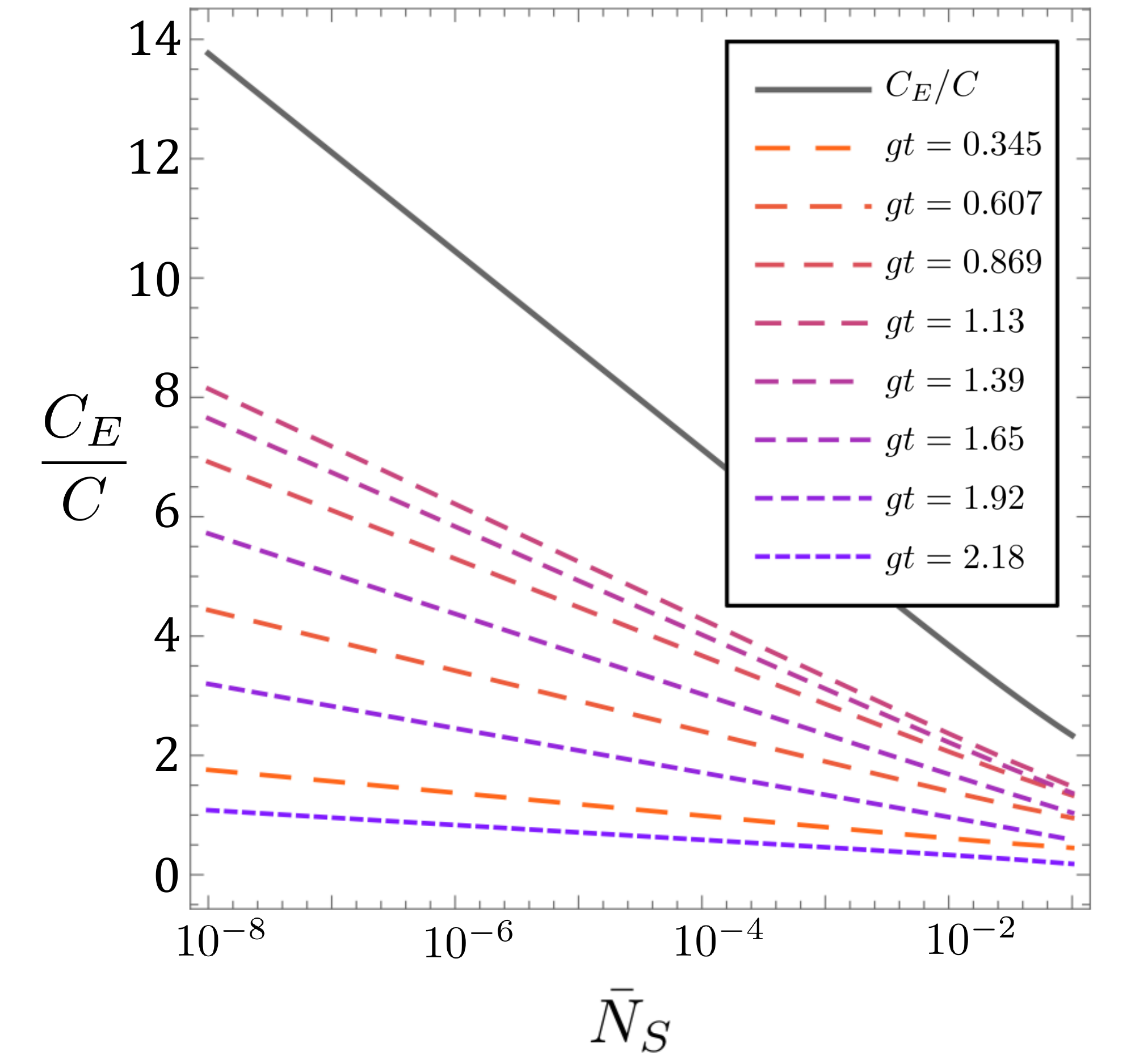}
    \caption{$\kappa=1$}
    \label{fig:nokapsfghol}
\end{subfigure}
\begin{subfigure}{.85\columnwidth}
    \includegraphics[width=\linewidth]{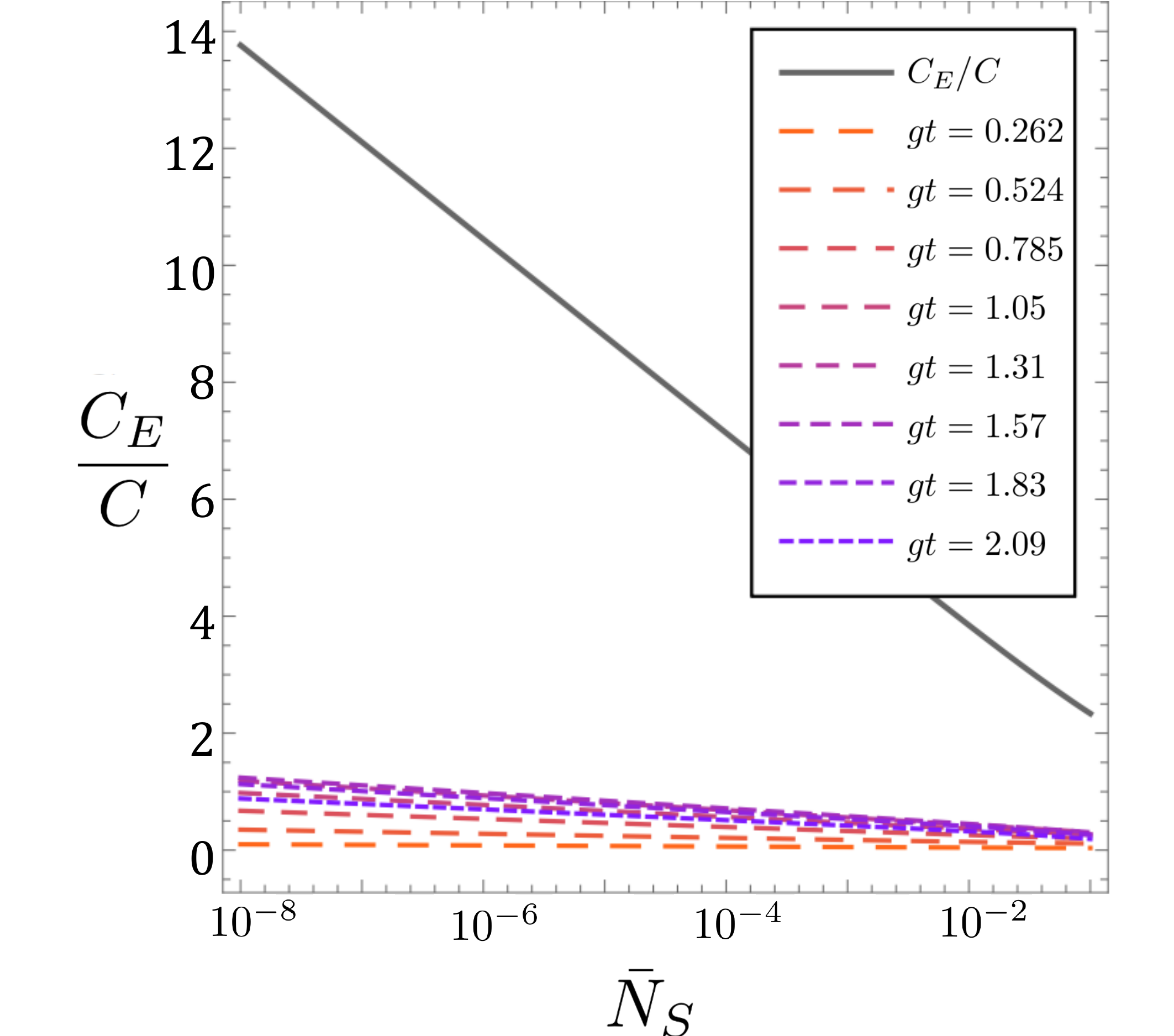}
    \caption{$\kappa=.05$}
    \label{fig:smallkapsfghol}
\end{subfigure}
    \caption{The Holevo capacity (ratio to ulimate classical Holevo capacity) at the output of the first sum-frequency mode. Notice that when $\kappa=1$, i.e. when a single SFG-gate is fed all of the signal mode without splitting off a part of it using a beam splitter, the optimal SFG evolution time is not $gt=\pi/2$ but less. Moreover even for the optimal value of $gt\approx1.13$, the capacity does not attain $C_E$, as the nonlinearity of the SFG gate prevents complete conversion of the signal-idler cross-correlation into the sum-frequency mode mean field. Only in the regime in which the qubit approximation is valid does the optimal value of $gt$ approach $\pi/2$. Plots are drawn for $\bar{N}_B=0.8$ using Fock-basis cutoff $n=3$.}
\label{fig:nokapholevo}
\end{figure}
\section{Analysis of the sum-frequency modes produced by the SFG-operations}\label{app:sfganalysis}
\begin{figure}[h]
\centering
\includegraphics[width=\linewidth]{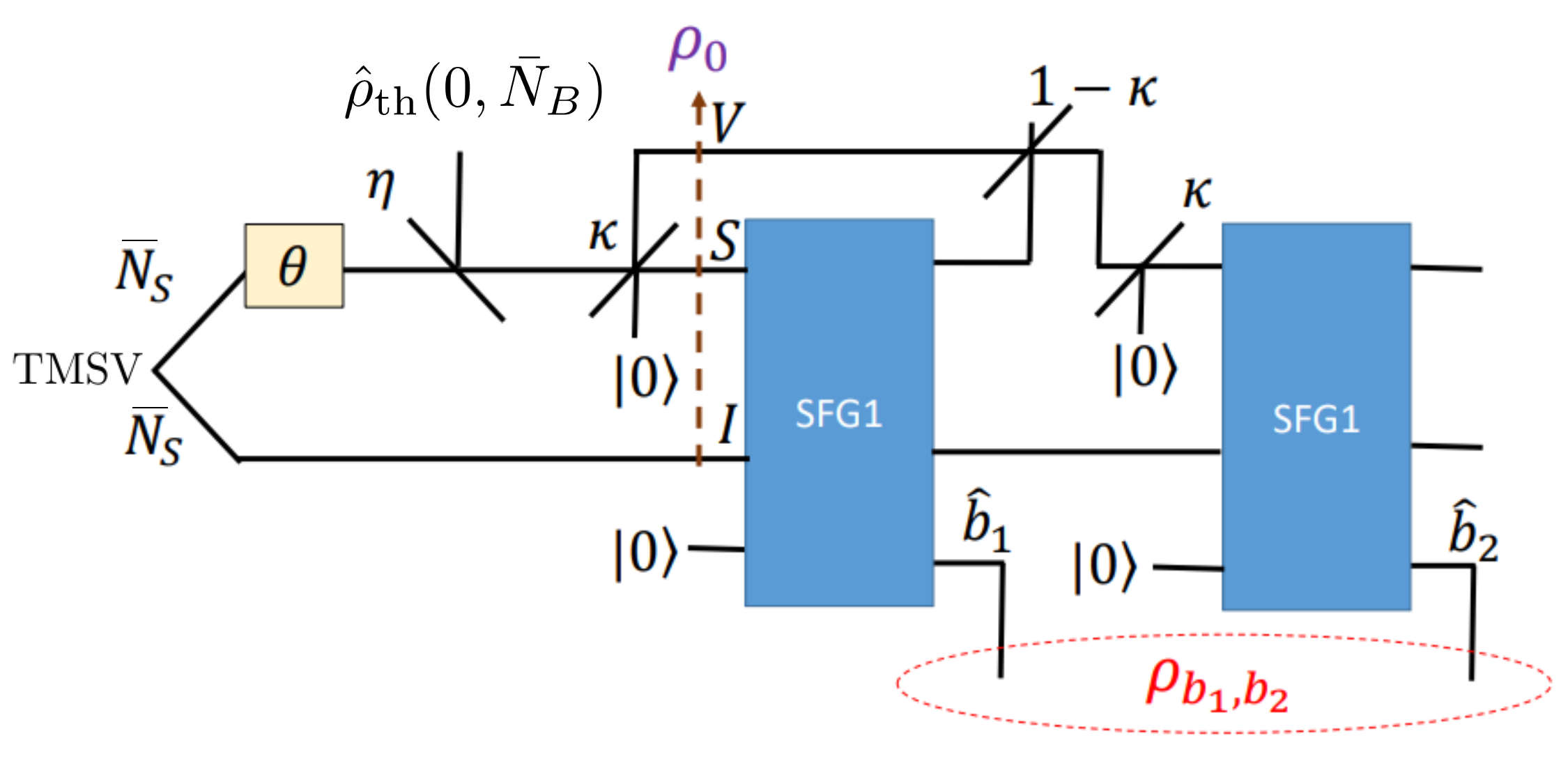}
\caption{Setup for the evaluation of the joint sum-frequency state}
\label{fig:twosfgsetup}
\end{figure}
The analysis of JDR2 relies on the key assumptions that 1) each sum-frequency output mode is individually a displaced thermal state and 2) two successive sum-frequency output states are in a maximally classically correlated joint Gaussian state with covariance matrix given by
\begin{align}
\sigma_{th2}&=\frac{1}{2}\langle\{\overset{\longrightarrow}{\Delta\hat{a}},\overset{\longrightarrow}{\text{\raisebox{0 pt}[6.2pt]{$\Delta\hat{a}^\dagger$}}}\}\rangle\nonumber\\
&=
\begin{spmatrix}
     \kappa \bar{N}_S \bar{N}_S'+\frac{1}{2} &\kappa \bar{N}_S \bar{N}_S' & 0 & 0 \\
     \kappa \bar{N}_S \bar{N}_S' &\kappa \bar{N}_S \bar{N}_S'+\frac{1}{2} & 0 & 0 \\
     0 & 0 & \kappa \bar{N}_S \bar{N}_S'+\frac{1}{2} & \kappa \bar{N}_S \bar{N}_S' \\
     0 & 0 & \kappa \bar{N}_S \bar{N}_S' & \kappa \bar{N}_S \bar{N}_S'+\frac{1}{2},
\end{spmatrix}
\label{eq:covth2}
\end{align}
and mean field given by 
\begin{equation}
\mu_{th2}=\langle(\hat{a}_1,\hat{a}_2,\hat{a}_1^\dagger,\hat{a}_2^\dagger)\rangle=(\alpha_1,\alpha_2,\alpha_1*,\alpha_2*)
\label{eq:muth2}
\end{equation}
where $\alpha_1=\sqrt{\kappa\eta \bar{N}_S(1+\bar{N}_S)}$ and $\alpha_2=\sqrt{\kappa\eta \bar{N}_S(1+\bar{N}_S)(1-\kappa(1+\bar{N}_S'))^2}$.

\begin{figure}[!ht]
\centering
\begin{subfigure}{.5\linewidth}
\includegraphics[width=\columnwidth]{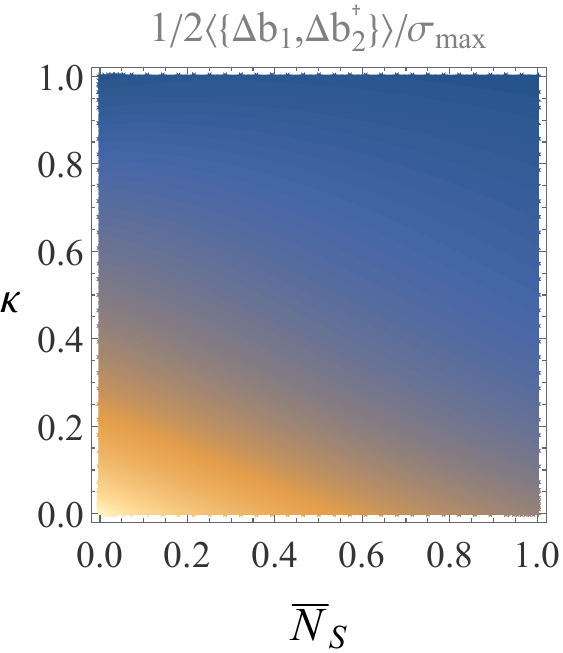}
\caption{Large scale plot}
\end{subfigure}
\begin{subfigure}{.13\linewidth}
\includegraphics[width=\columnwidth]{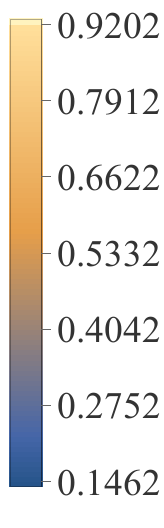}
\end{subfigure}
\begin{subfigure}{.6\columnwidth}
\includegraphics[width=\linewidth]{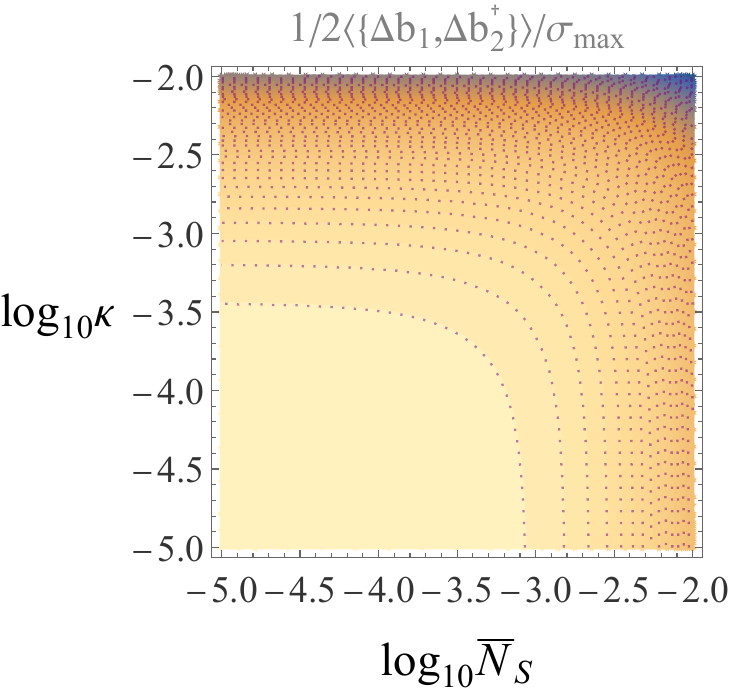}
\caption{Small scale plot zooming in on the region where $\bar{N}_S\ll 1$ and $\kappa\ll 1$}
\end{subfigure}
\begin{subfigure}{.25\columnwidth}
\includegraphics[width=.5\linewidth]{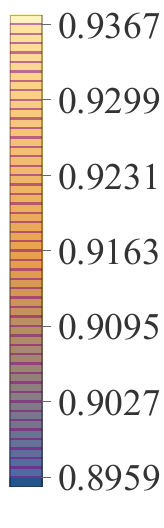}
\end{subfigure}
\caption{Ratio between phase-insensitive cross-correlation of $\hat{\rho}_{b1,b2}$ and maximally allowed phase-insensitive cross-correlation $\sigma_\text{max}=\bar{N}_B \bar{N}_S (1 - \eta) \kappa + \bar{N}_S^2 \eta \kappa$. Both plots use values $\bar{N}_B=1$, $\theta=\pi$ and Fock-basis cutoff $n=6$.}
\label{fig:corratios}
\end{figure}

In order to justify these assumptions independently from \cite{Zhu17}, we compute the fidelity of the actual sum-frequency output gate with the state whose covariance matrix is given in Eq.~\eqref{eq:covth2} and compute the ratio $\frac{1}{2}\langle\{\hat{b}_1\hat{b}_2^\dagger\rangle\}/\sigma_{\text{max}}$ where $\sigma_{\text{max}}=(\sigma_{th2})_{12}=\kappa \bar{N}_S \bar{N}_S'$ between the actual cross-correlation and the maximal classical cross-correlation. The ratio is plotted in Fig. \ref{fig:corratios}

The evolution of the signal, idler and sum-frequency modes through the sum-frequency generation is done in the Schrödinger picture according to 
\begin{equation}
    \tilde{\hat{\rho}}_t=e^{-it\hat{H}_{SFG}}\tilde{\hat{\rho}}_0e^{it\hat{H}_{SFG}}.
    \label{eq:sfgevolution}
\end{equation}
where $\tilde{\hat{\rho}}$ is the state consisting of the signal, idler, and sum-frequency modes represented in the Fock-basis with cutoff at $n=6$. The SFG Hamiltonian $\hat{H}_{SFG}$ (see section \ref{sec:JDRdesign} for definition) in Eq.~\eqref{eq:sfgevolution} is also represented in the Fock basis as a matrix.

Excluding the initial vacuum sum-frequency mode, the state $\hat{\rho}_0$ at the input of the SFG receiver indicated by the dashed line in Fig. \ref{fig:twosfgsetup} is a 3-mode zero-mean Gaussian state with covariance matrix $\sigma:=\frac{1}{2}\langle\{\Delta\vec{\hat{r}},\Delta\vec{\hat{r}}\}\rangle$ given by
\begin{widetext}
\begin{equation}
\text{\fontsize{9}{12}\selectfont$
\left(
\begin{array}{cccccc}
A&0&B&0&C\cos(\theta)&-\sqrt{\kappa} C\sin(\theta)\\
0&A&0&B&-\sqrt{\kappa} C\sin(\theta)&-\sqrt{\kappa}C\cos(\theta)\\
B&0&D&0&\sqrt{1-\kappa}C\cos(\theta)&-\sqrt{1-\kappa}C\sin(\theta)\\
0&B&0&D&-\sqrt{1-\kappa}C\sin{\theta}&-\sqrt{1-\kappa}C\cos(\theta)\\
\sqrt{\kappa} C\cos(\theta)&-\sqrt{\kappa} C\sin(\theta)&\sqrt{1-\kappa}C\cos(\theta)&-\sqrt{1-\kappa}C\sin(\theta)&\bar{N}_S+\frac{1}{2}&0\\
-\sqrt{\kappa}C\sin(\theta)&-\sqrt{\kappa}C \cos(\theta)&-\sqrt{1-\kappa}C\sin(\theta)&-\sqrt{1-\kappa}C\cos(\theta)&0&\bar{N}_S+\frac{1}{2}
\end{array}
\right)$}
\end{equation}
\end{widetext}
where 
\begin{align*}
\Delta\vec{\hat{r}}:=(&\hat{q}_S-\langle\hat{q}_S\rangle,\hat{p}_S-\langle\hat{p}_S\rangle,\\
&\hat{q}_V-\langle\hat{q}_V\rangle,\hat{p}_V-\langle\hat{p}_V\rangle,\\
&\hat{q}_I-\langle\hat{q}_I\rangle,\hat{p}_I-\langle\hat{q}_I\rangle).
\end{align*}
and 
\begin{align*} 
A:=& \bar{N}_B (\kappa -\eta  \kappa )+\eta  \kappa  \bar{N}_S+\frac{1}{2}\\
B:=& \sqrt{\kappa(1-\kappa)} ((1-\eta)\bar{N}_B+\eta  \bar{N}_S)\\
C:=& \sqrt{\eta}\sqrt{\bar{N}_S(\bar{N}_S+1)}\\
D:=&  (1-\eta) (1-\kappa) \bar{N}_B+\bar{N}_S(\eta -\eta  \kappa )+\frac{1}{2}.
\end{align*}
To evolve $\hat{\rho}_0$ through two iterations of the SFG cycle, as diagrammed in Fig. \ref{fig:twosfgsetup}, we first represent $\hat{\rho}_0$ in the Fock-basis with Fock cutoff $n=6$ using the formula from \cite{Queseda19}:
\begin{equation}
    \langle\mb{m}|\hat{\rho}_0|\mb{n}\rangle=T\times \text{lhaf}(\bar{A})
    \label{fockrepformula}
\end{equation}
where $|\mb{n}\rangle$ is a photon number product state represented by photon number vector $\mb{n}$. $T$ is a scalar factor computed from the displacement and the covariance matrix of $\hat{\rho}_0$ (see \cite{Queseda19} for expression). $\bar{A}$ is derived from the covariance matrix by repeating rows and columns of $\sigma$ and entries of mean $\mb{\mu}$ according to $\mb{m}$ and $\mb{n}$.

To compute the loop-Hafnian lhaf in Eq.~\eqref{fockrepformula} we use a function from Python library \textit{the-walrus}, based on multi-dimensional Hermite polynomials \cite{Hermite}.

\begin{figure}[!ht]
\centering
    \begin{subfigure}{0.35\linewidth}
        \flushleft
        \includegraphics[width=.9\columnwidth]{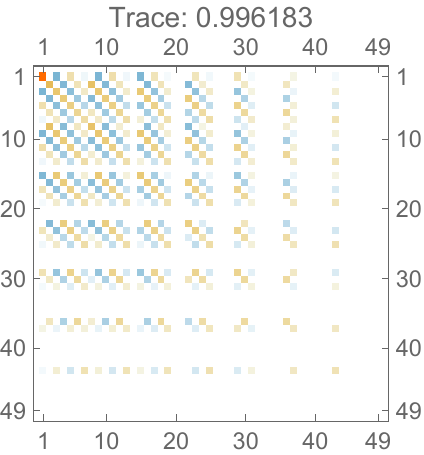}
        \caption{Real part of the joint sum-frequency state $\hat{\rho}_{b1,b2}$}
        \label{rejoint}
    \end{subfigure}
    \begin{subfigure}{.1\linewidth}
        \includegraphics[width=\columnwidth]{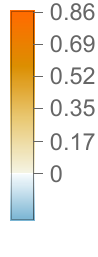}
    \end{subfigure}
    \begin{subfigure}{0.35\linewidth}
        \flushright
        \includegraphics[width=.9\columnwidth]{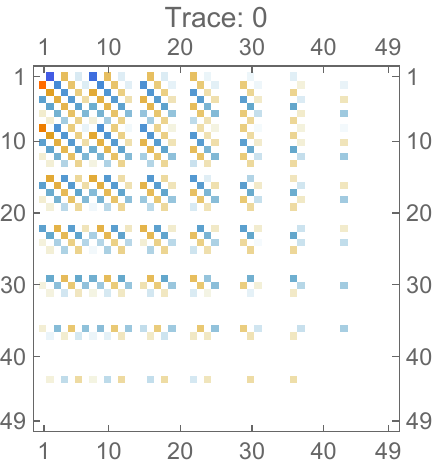}
        \caption{Imaginary part of the joint sum-frequency state $\hat{\rho}_{b1,b2}$}
        \label{imjoint}
    \end{subfigure}
    \begin{subfigure}{.12\linewidth}
        \includegraphics[width=\columnwidth]{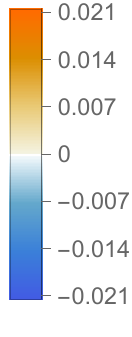}
    \end{subfigure}
    \begin{subfigure}{.35\linewidth}
        \flushleft
        \includegraphics[width=.9\columnwidth]{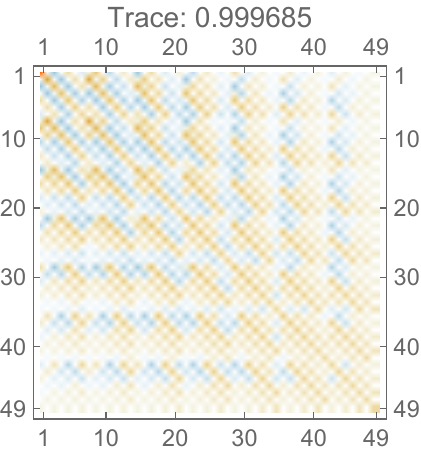}
        \caption{Real part of the maximally classically correlated Gaussian state}
        \label{recorr}
    \end{subfigure}
    \begin{subfigure}{.1\linewidth}
        \includegraphics[width=\columnwidth]{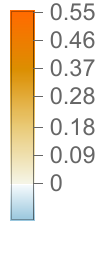}
    \end{subfigure}
    \begin{subfigure}{.35\linewidth}
        \flushright
        \includegraphics[width=.9\columnwidth]{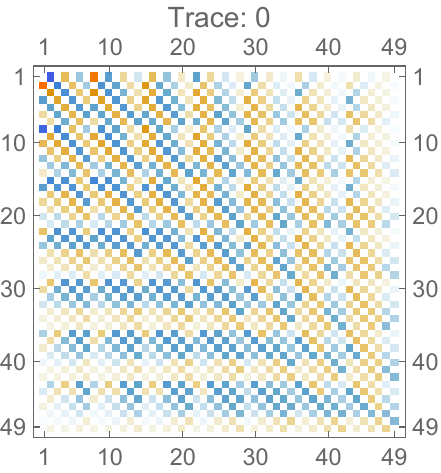}
        \caption{Imaginary part of the maximally classically correlated Gaussian state}
        \label{imcorr}
    \end{subfigure}
    \begin{subfigure}{.16\linewidth}
        \includegraphics[width=\columnwidth]{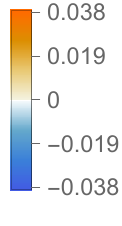}
    \end{subfigure}
\caption{Plots of the actual state (top row) at the output of the first two sum-frequency modes and the maximally classically correlated two-mode displaced thermal state (bottom row) approximating the actual joint-sum-frequency state with means $\alpha_1=\sqrt{\kappa\eta \bar{N}_S(1+\bar{N}_S)}$ and $\alpha_2=\sqrt{\kappa\eta \bar{N}_S(1+\bar{N}_S)(1-\kappa(1+\bar{N}_S'))^2}$ and mean thermal noise $\kappa \bar{N}_S \bar{N}_S'$ on each mode, where $\bar{N}_S'=\eta \bar{N}_S+(1-\eta \bar{N}_B)$. The fixed channel parameters are  $\bar{N}_S=.5$, $\bar{N}_B=1$, $\theta=\pi$, $\eta=.01$, $\kappa=.8$, and $gt=\pi/2$. If the plots are seen as $7\times7$ block-matrices, then the row and column indices of the blocks correspond to the Fock-numbers of the mode $\hat{b}_2$, and the row and column indices within each block correspond to the Fock-numbers of the mode $\hat{b}_1$.}
\label{stateplots}
\end{figure}

$\hat{\rho}_0$ is upgraded to a 4-mode state via Kronecker product with a vacuum initial sum-frequency mode. The state is then successively evolved via the SFG unitary with evolution coefficient $gt=\pi/2$, a beam-splitter unitary of transmissivity $(1-\kappa)$, a beam-splitter unitary of transmissivity $\kappa$, and a final SFG unitary of evolution coefficient $gt=\pi/2$. The state is partially traced out and Kronecker-multiplied with vacuum at each step appropriately to maintain a 4-mode form throughout the evolution. Two of the four modes are traced out at the end to produce the final joint-sum-frequency state, which are plotted as matrices in the Fock basis representation in Fig. \ref{stateplots} for $\bar{N}_S=.5$, $\bar{N}_B=1$, $\theta=\pi$, $\eta=.01$, $\kappa=.8$, and $gt=\pi/2$.

$\pi/2$ is the value of $gt$ which corresponds to all cross-correlation between signal and idler modes having been converted to a displacement of the sum-frequency mode in the qubit approximation. This is verified numerically by Fig. \ref{fig:smallkapsfghol}, which plots the Holevo capacity of the BPSK ensemble at the first sum-frequency mode.

\begin{figure}[!ht]
\centering
    \begin{subfigure}{.35\linewidth}
        \flushleft
        \includegraphics[width=.9\columnwidth]{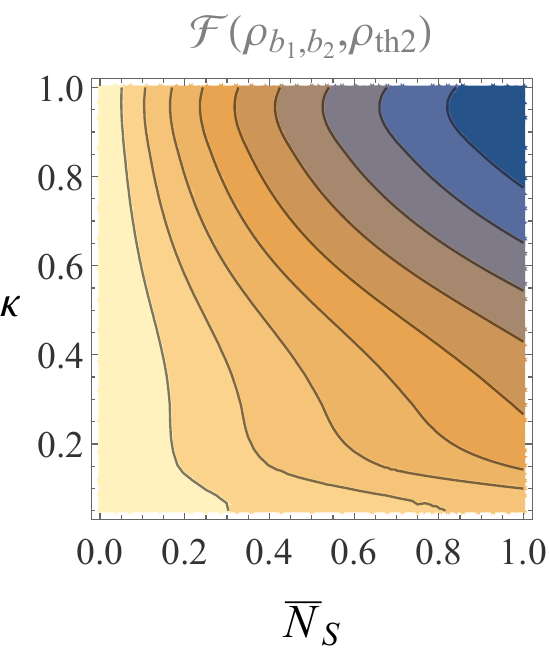}
        \caption{Fidelity plot between actual joint sum-frequency state and the Gaussian approximation assumed by \cite{Zhu17}}
        \label{jointfidel}
    \end{subfigure}
    \begin{subfigure}{.1\linewidth}
        \includegraphics[width=\columnwidth]{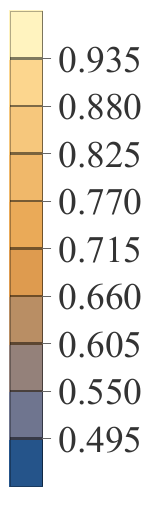}
    \end{subfigure}
    \begin{subfigure}{.35\linewidth}
        \flushright
        \includegraphics[width=.9\columnwidth]{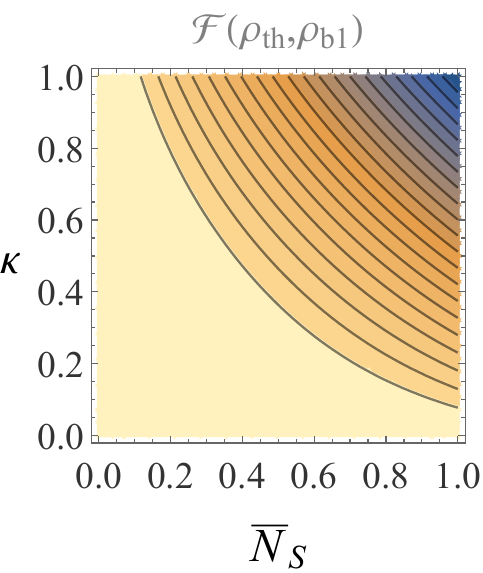}
        \caption{Fidelity plot between single sum-frequency state and the displaced thermal state assumed to approximate it in \cite{Zhu17}}
        \label{singlefidel}
    \end{subfigure}
    \begin{subfigure}{.1\linewidth}
        \includegraphics[width=\columnwidth]{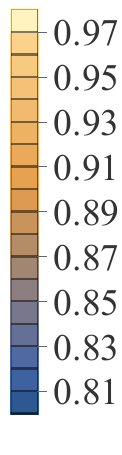}
    \end{subfigure}
\caption{Plots of the fidelity between output modes and the Gaussian approximation assumed by \cite{Zhu17}}
\label{fidelities}
\end{figure}

The similarity between the actual joint state $\hat{\rho}_{b1,b2}$ and maximally classically correlated two-mode displaced thermal state $\hat{\rho}_\text{th2}$ with covariance matrix given by Eq.~\eqref{eq:covth2} and plotted in Fig. \ref{stateplots} can be measured by the Uhlmann fidelity defined as $\mathcal{F}(\hat{\rho}_1,\hat{\rho}_2):=\text{Tr}(\sqrt{\sqrt{\hat{\rho}_1}\hat{\rho}_2\sqrt{\hat{\rho}_1}})^2$. The fidelities between  $\hat{\rho}_{b1,b2}$ and  $\hat{\rho}_\text{th2}$ and between single sfg-output $\hat{\rho}_{b_1}$ and $\hat{\rho}_\text{th}$ is plotted in Fig. \ref{fidelities} as a function of $\kappa$ and $\Bar{N}_S$.

\begin{figure}[!ht]
\centering
\begin{subfigure}{.5\columnwidth}
    \includegraphics[width=\linewidth]{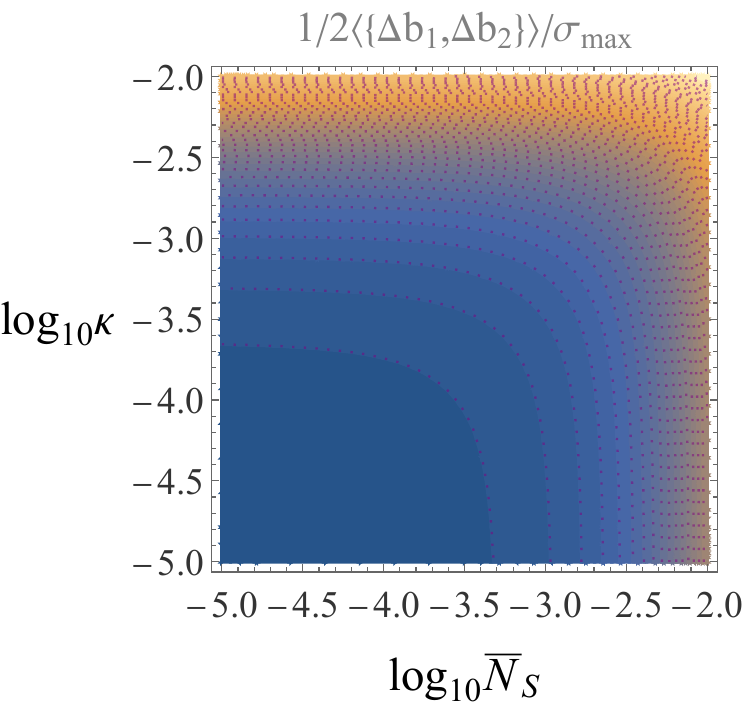}
\end{subfigure}
\begin{subfigure}{.15\columnwidth}
    \includegraphics[width=\linewidth]{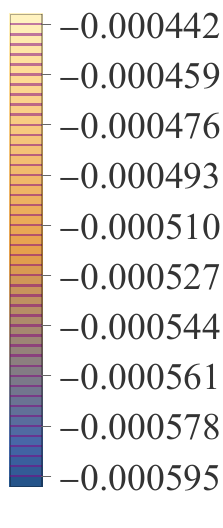}
\end{subfigure}
\caption{Ratio between the phase-sensitive cross-correlation of $\hat{\rho}_{b1,b2}$ and the maximally allowed phase-insensitive cross-correlation $\sigma_\text{max}=\kappa \bar{N}_S\bar{N}_S'$. Since the ratio is negligible in the small $\bar{N}_S$, $\kappa$ regime, $\hat{\rho}_{b1,b2}$ is not entangled, which is consistent with the qubit approximation \cite{Zhu17a}}
\label{fig:sensitivecorratios}
\end{figure}

Since both the approximation and the actual joint state are close to vacuum for small $\kappa$ and $\bar{N}_S$, the fidelity might not capture the full difference in the nature of the states. To check that the sum-frequency modes are indeed maximally classically correlated, the ratio between the phase-insensitive cross-correlation $\langle\{\hat{b}_1,\hat{b}_2^\dagger\}\rangle/2$ and the maximum allowed phase-insensitive cross correlation $\sigma_{\text{max}}=\kappa \bar{N}_S\bar{N}_S'$ is plotted in Fig. \ref{fig:corratios}. The phase-sensitive correlation of two successive sum-frequency modes is also plotted in Fig. \ref{fig:sensitivecorratios}.
The phase-insensitive cross-correlation of the actual state matches the maximal allowed phase-insensitive (classical) correlation $\sigma_\text{max}$ of the Gaussian approximation for small $\bar{N}_S$ and $\kappa$. Moreover, the phase-sensitive correlation becomes negligible in comparison to $\sigma_\text{max}$.

Plots \ref{fig:sensitivecorratios}, \ref{fig:corratios} and \ref{fidelities} together are sufficient to justify the claims made in the qubit approximation in \cite{Zhu17a}, as well as the assumption that the thermal photons of the $\hat{b}$-modes are maximally correlated. 

\section{Calculation of the combined thermal noise of the states entering the Green-Machine of JDR2} \label{app:gammanoise}
In order to combine the mean fields $\alpha_k$ of the sum-frequency modes, the $K$ $\gamma_k$ beam splitters of JDR2 pictured in Fig. \ref{fig:JDR} must be adjusted to have transmissivities
\begin{equation}
	\gamma_k=\frac{\sum_{i=1}^{k}\alpha_i^2}{\sum_{i=1}^{k+1}\alpha_i^2}=\frac{1-(1-\kappa(1+\bar{N}_S'))^{2k}}{1-(1-\kappa(1+\bar{N}_S'))^{2(k+1)}}.
	\label{gammaexpression}
\end{equation}

\begin{figure}[h]
\centering
	\begin{subfigure}{\linewidth}
		\flushright
		\includegraphics[width=0.9\linewidth]{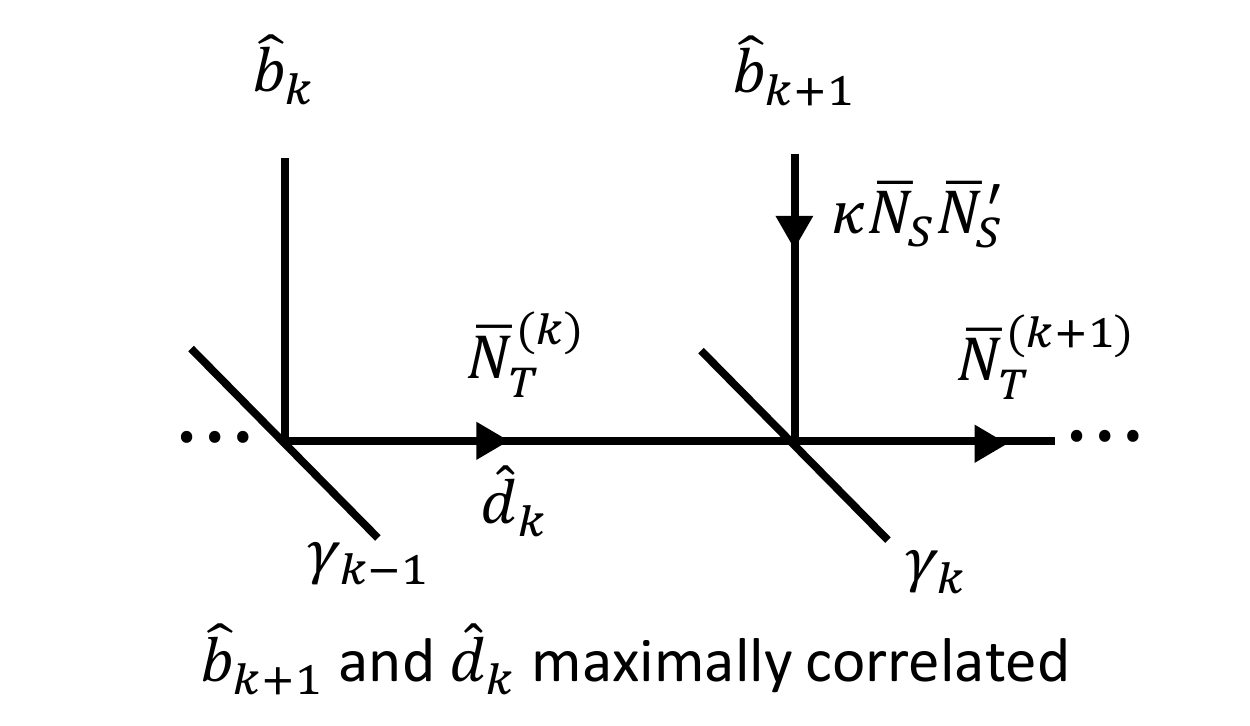}
		\caption{The thermal part of the SFG-modes interacting under the $k^{th}$ $\gamma$ beam splitter}
		\label{fig:gammadiag}
	\end{subfigure}
	\vspace{1em}
	\begin{subfigure}{\linewidth}
		\flushleft
		\includegraphics[width=0.9\linewidth]{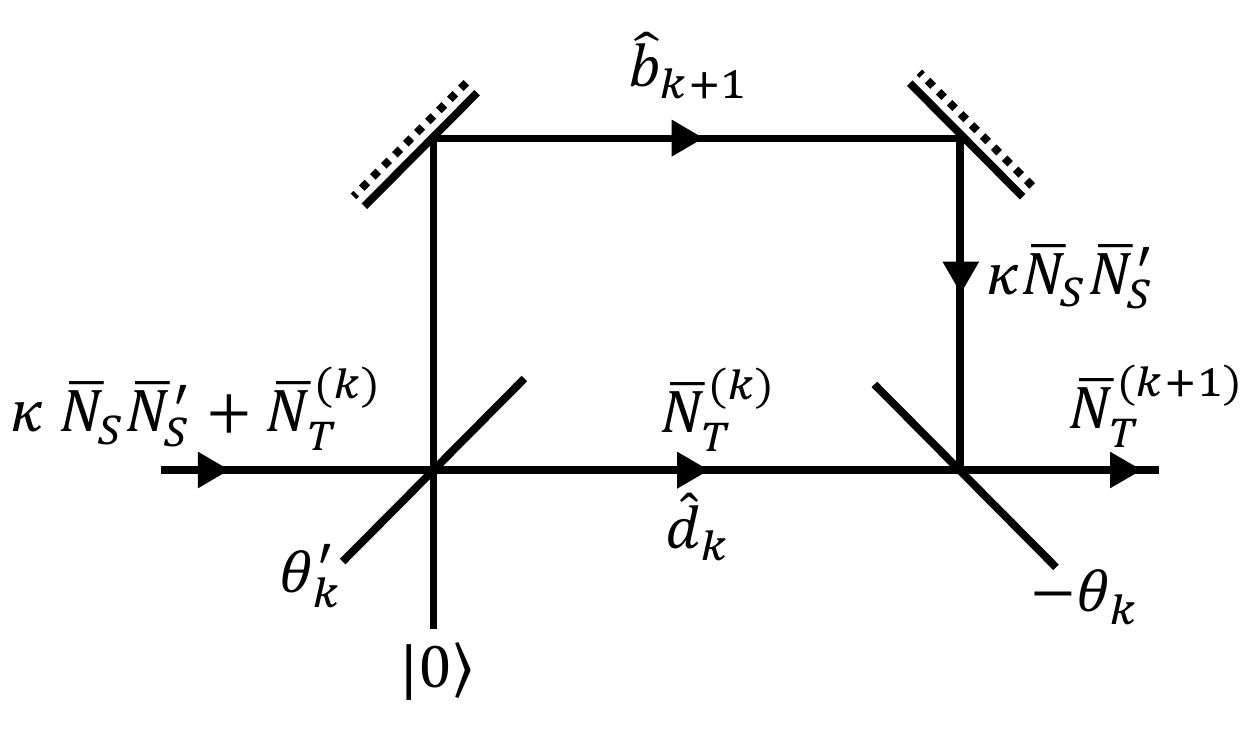}
		\caption{Representation of the imaginary splitting of a parent beam with the combined mean thermal photon number of the modes entering $\gamma_k$. The beam splitters are now represented by their rotation parameters because the direction of the rotation is important when determining the net effect of $\gamma_k$ and $\gamma_k'$. $\theta_k$ and $\theta_k'$ are taken to be positive.}
		\label{fig:gammadiag2}
	\end{subfigure}
\caption{Diagram of interference of maximally correlated thermal noise on the $\gamma$ beam-splitters.}
\end{figure}

To analyze how the thermal mean photons of the $\hat{b}_k$ modes combine under the assumption that they have maximally classically correlated covariance matrices, we associate to every beam splitter $\gamma_k$ an imaginary beam-splitter with transmissivity $\gamma_k'$ responsible for splitting an imaginary parent beam (also a displaced thermal state) with the combined thermal noise of modes $\hat{d}_k$ and $\hat{b}_{k+1}$ shown in the diagram of the $k^\text{th}$ beam splitter (Fig. \ref{fig:gammadiag}).

Denote the mean thermal noise of mode $\hat{d}_k$ by $\bar{N}_T^{(k)}$, as in Fig. \ref{fig:gammadiag}. Then since $\gamma_{k-1}'$ splits the parent beam into modes $\hat{b}_k$ with thermal mean photon number $\kappa \bar{N}_S\bar{N}_S'$ and $\hat{d}_k$ with thermal mean photon number $\bar{N}_T^{(k)}$,
\begin{equation}
\gamma_{k}'=\frac{\bar{N}_T^{(k)}}{\kappa \bar{N}_S\bar{N}_S'+\bar{N}_T^{(k)}}.
\label{gammap}
\end{equation}

\begin{figure}
	\centering
	\includegraphics[width=.9\linewidth]{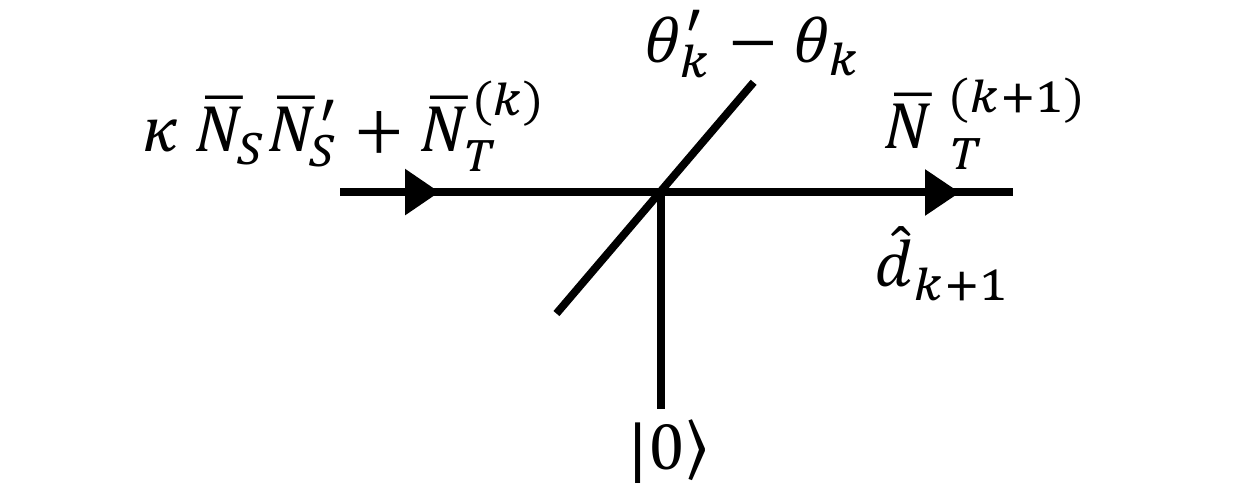}
	\caption{The effective splitting operation corresponding to Fig. \ref{fig:gammadiag2}}
	\label{fig:gammadiag3}
\end{figure}

With the equivalence of diagrams \ref{fig:gammadiag} and \ref{fig:gammadiag2}, the question of determining $\bar{N}_T^{(k)}$ assuming maximally correlated inputs $\hat{b}_k$ and $\hat{d}_{k-1}$ is reduced to interfering a thermal state with vacuum on a beam splitter of transmissivity $\gamma_k'\circ\gamma_k$, as pictured in Fig. \ref{fig:gammadiag3}.

The value of the effective transmissivity $\cos(\theta_k'-\theta_k)^2$ is {\small$$\cos(\text{arcos}(\sqrt{\gamma_k'})-\text{arcos}(\sqrt{\gamma_k}))^2=\Bigl(\sqrt{\gamma_k'\gamma_k}+\sqrt{(1-\gamma_k')(1-\gamma_k)}\Bigr)^2$$}, where $\gamma_k'$ and $\gamma_k$ are themselves functions of $\bar{N}_T^{(k)}$ and $\bar{N}_S$ as given by Eq.s~\eqref{gammaexpression} and ~\eqref{gammap}. This leads to a nonlinear recurrence relation for $\bar{N}_T^{(k)}$ given by
\begin{align}
	&\bar{N}_T^{(k+1)}\nonumber\\
    &=\Bigl(2\bigl(\gamma_k'\gamma_k+\sqrt{\gamma_k(1-\gamma_k)\gamma_k'(1-\gamma_k')}\,\bigr)+1-\gamma_k-\gamma_k'\Bigr)\nonumber\\
    &\times(\bar{N}_T^{(k)}+\kappa\bar{N}_S\bar{N}_S'),
	\label{NTrecurrence}
\end{align}
which is difficult to solve analytically, so we resort to a numerical solution. $\bar{N}_{T0}$, the thermal mean photon number of the displaced thermal states entering the Green Machine is $\bar{N}_T^{(K)}$ in the notation of Fig. \ref{fig:gammadiag}. When iterating the recurrence relation given by Eq.~\eqref{NTrecurrence}, we choose an arbitrary large $K$ value $(5000)$ to emulate taking the limit as $K\rightarrow\infty$.
\begin{figure}[h]
	\centering
	\includegraphics[width=\linewidth]{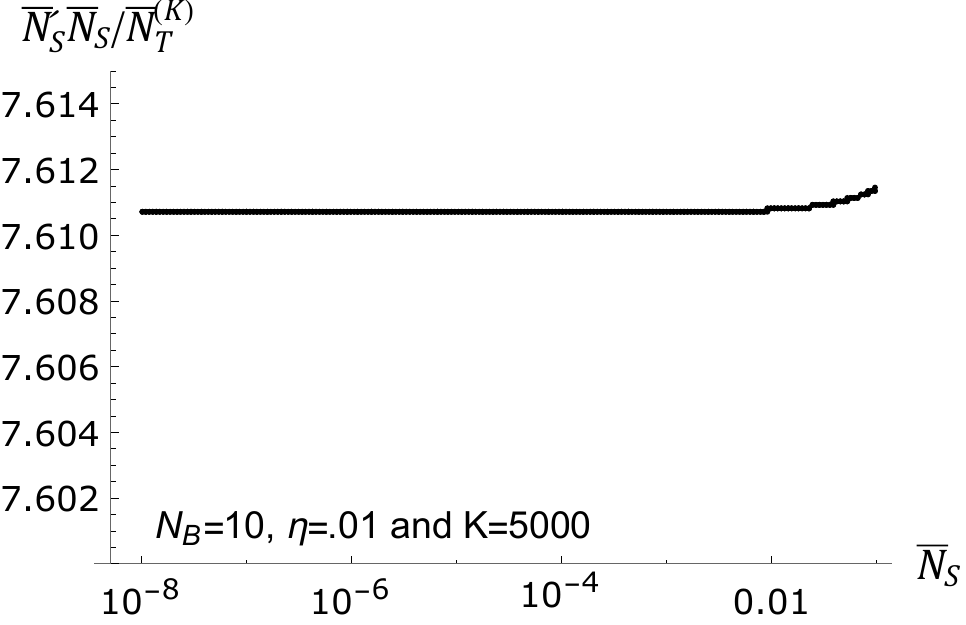}
	\caption{Plot of $\bar{N}_S\bar{N}_S'/\bar{N}_T^{(K)}$ as a function of $\bar{N}_S$}
	\label{fig:NTtotalbyNT}
\end{figure} 
As Fig. \ref{fig:NTtotalbyNT} shows, the resulting value of $\bar{N}_T^{(K)}$ is only fraction of the total thermal noise of the sum-frequency modes, approximately fixed at $\bar{N}_S\bar{N}_S'/7.611$ in the low signal brightness regime where entanglement-assisted communication has the most advantage over classical communication.

\newcommand{\mbf}{\mathbf}
\newcommand{\R}{\mbb{R}}
\newcommand{\C}{\mbb{C}}
\newcommand{\Z}{\mbb{\Z}}
\section{Argument for the factor of $\sqrt{2}$ in the collective treatement of the SFG output noise modes and sum-frequency modes}\label{app:sqrt2}
Let $r_k^{(i)}$, the squeezing parameter of the $k^{\text{th}}$ sandwiching squeezer (depicted by dark blue boxes in Fig. \ref{fig:sanwich}) of the $i^{\text{th}}$ FF-SFG module (depicted without squeezers by light blue boxes in Fig. \ref{fig:JDR2}), be proportional to $\alpha_k$, the magnitude of the coherent amplitude of the $k^\text{th}$ SFG gate in the SFG module, as
\begin{equation}
    r_k^{(i)}=\lambda_i\alpha_k
\end{equation}
where $\lambda_i\in\R$ is the proportionality constant. The squeezing parameter is defined as in \cite{Zhu17}, such that $S(r_k^{(i)})$ applied on the quadratures of the signal and idler modes $\hat{a}_S$ and $\hat{a}_I$ results in
\begin{gather}
    \hat{a}_S^{'}=\sqrt{1+(r_k^{(i)})^2}\;\hat{a}_S+r_k^{(i)}\hat{a}_I^\dagger\\
    \hat{a}_I^{'}=\sqrt{1+(r_k^{(i)})^2}\;\hat{a}_I+r_k^{(i)}\hat{a}_S^\dagger.
\end{gather}
It can be shown based on the single-cycle analysis in the supplementary material of \cite{Zhu17} that under BPSK phase modulation, the true output of the $i^\text{th}$ FF-SFG module of JDR2 is approximately
\begin{equation}
\hat{\rho}^{\pm}_{\text{true}}:=\left[\bigotimes_{k=1}^K\hat{\rho}_{\text{th}}(0,((\lambda_i\pm 1)\alpha_k)^2)\right]\otimes\hat{\rho}_{\text{th}}((\lambda_i\pm 1)\alpha_0,\bar{N}_{T0}).
\label{trueout}
\end{equation}

What follows is an argument that when evaluating the Holevo capacity of the FF-SFG receiver's output, $\hat{\rho}^\pm_{\text{true}}$ can effectively be treated as a single mode displaced thermal state
\begin{equation}
    \hat{\rho}^{\pm}_{\text{eff}}:=\hat{\rho}_{\text{th}}(\pm\sqrt{2}\alpha_0,\bar{N}_{T0}).
\end{equation}
This argument is based on two lemmas, the first of which we leave as a conjecture:
\begin{enumerate}
    \item The Holevo capacity for BPSK amplitude modulation of a displaced thermal state is approachable with a joint-detection receiver that uses sequential code word nulling followed by a nondestructive multi-mode vacuum-or-not measurement (known to attain the Holevo capacity for coherent state bpsk modulation \cite{Chen12}) such as the receiver depicted in Fig. \ref{fig:vonreceiverdiag}.
    \item The transition probability matrix of the channel induced by the vacuum-or-not (VON) receiver where received code words are encoded in the states $\hat{\rho}^{\pm}_{\text{eff}}$ is identical with that of the channel whose received code words are encoded in the states $\hat{\rho}^{\pm}_{\text{true}}$.
\end{enumerate}
\begin{proof}[Proof of Lemma 2]
First note that overall displacement of an ensemble will not affect its Holevo capacity since displacement can be thought of as the limit of a unitary transformation, so $\hat{\rho}^\pm_\text{eff}$ can be redefined without loss of generality to be $\hat{\rho}_\text{th}((\lambda_i\pm1)\sqrt{2}\alpha_0,\bar{N}_{T0})$.

Let $\mbf{c}\in\{\pm 1\}^L$ be a random code word. Then set $\vec{\lambda}$ to null $\hat{\rho}^{c_l}_\text{true}$ and $\hat{\rho}^{c_l}_\text{eff}$, i.e. $\lambda_l=-c_l\;\forall l\in\{1...L\}$. Let $\mbf{b}\in\{\pm1\}^L$ be the transmitted code word. Then the true and effective received states are
\begin{align}
    &\hat{\rho}_{\mbf{b},\text{true}}:=\nonumber\\
    &\bigotimes_{l=1}^L\left(\left[\bigotimes_{k=1}^K\hat{\rho}_\text{th}(0,(2\alpha_k\delta_{b_l,c_l})^2)\right]\otimes\hat{\rho}_\text{th}(2\alpha_0(b_l-c_l),\bar{N}_{T0})\right)\\
    &\hat{\rho}_{\mbf{b},\text{eff}}:=\bigotimes_{l=1}^L\hat{\rho}_\text{th}(2\sqrt{2}\alpha_0(b_l-c_l),\bar{N}_{T0}),
\end{align}
where $\delta$ is the Kronecker-Delta symbol.

Let $|\mbf{0}_n\rangle$ denote the n-mode vacuum state and $|\alpha\rangle$ a coherent state with mean field $\alpha$. Calling the channel in which the received states are $\hat{\rho}_{\mbf{b}, \text{true}}$ the true channel and the channel in which the received states are $\hat{\rho}_{\mbf{b}, \text{eff}}$ the effective channel, the transition probability matrix of the true channel induced by the VON receiver is determined, up to asymptotically insignificant residual terms associated with the small perturbation of the ``not vacuum'' outcome, by the probabilities
\begin{align}
    \langle\mbf{0}_{L(K+1)}|&\hat{\rho}_{\mbf{b},\text{true}}|\mbf{0}_{L(K+1)}\rangle\nonumber\\
    &=\prod_{l=1}^L\Bigl(\left[\prod_{k=1}^K\langle0|\hat{\rho}_{\text{th}}(0,(2\alpha_k\delta_{b_l,c_l})^2)|0\rangle\right]\nonumber\\
    &\hspace{.5in}\times\langle0|\hat{\rho}_\text{th}(2\alpha_0\delta_{b_l,c_l},\bar{N}_{T0})|0\rangle\Bigr) \label{prodtrue}
\end{align}
and that of the effective channel is determined entirely by the probabilities
\begin{equation}
    \langle\mbf{0}_{L}|\hat{\rho}_{\mbf{b},\text{eff}}|\mbf{0}_{L}\rangle=\prod_{l=1}^L\langle0|\hat{\rho}_\text{th}(2\sqrt{2}\alpha_0\delta_{b_l,c_l},\bar{N}_{T0})|0\rangle \label{prodeff},
\end{equation}
which are enumerated by varying the transmitted code word $\mbf{b}$.

So the proof of assumption 2 reduces to showing that the factors in the outermost products of Eq.s~\eqref{prodtrue} and \eqref{prodeff} equal each other. 

It is shown in section \ref{subapp:thermalvac} that the product over $k$ in Eq.~\eqref{prodtrue} approaches $\langle0|2\alpha_0\delta_{b_l,c_l}\rangle\langle2\alpha_0\delta_{b_l,c_l}|0\rangle$ in the large $K$ limit. 
\iffalse
Starting with eq. \ref{prodtrue}, note that in the large $K$ limit, $\kappa\ll1$ so $\alpha_k=\sqrt{\kappa\eta\bar{N}_S^{'}(1+\bar{N}_S^{'})\mu^{k-1}}\ll 1$, and when the mean photon number is small, the photon count statistics of a thermal state matches that of a coherent state having the same mean photon number, so $\langle0|\hat{\rho}_{\text{th}}(0,(2\alpha_k\delta_{b_l,c_l})^2)|0\rangle$ is approximately equal to $\langle0|2\alpha_k\delta_{b_l,c_l}\rangle\langle2\alpha_k\delta_{b_l,c_l}|0\rangle$. \fi Then eq. \ref{prodtrue} can be reduced to write
\begin{widetext}
\begin{align}
    \langle\mbf{0}_{L(K+1)}|\hat{\rho}_{\mbf{b},\text{true}}|\mbf{0}_{L(K+1)}\rangle=&\prod_{l=1}^L\langle0|2\delta_{b_l,c_l}\alpha_0\rangle\langle2\delta_{b_l,c_l}\alpha_0|0\rangle\times\langle0|\hat{\rho}_\text{th}(2\alpha_0\delta_{b_l,c_l},\bar{N}_{T0})|0\rangle\nonumber\\
    =&\prod_{l=1}^Le^{-4\alpha_0^2\delta_{b_l,c_l}}\langle0|\hat{\rho}_\text{th}(2\alpha_0\delta_{b_l,c_l},\bar{N}_{T0})|0\rangle\nonumber\\
    =&\prod_{l=1}^L\frac{e^{-4\alpha_0^2\delta_{b_l,c_l}}}{1+\bar{N}_{T0}}e^{-\frac{(2\alpha_0)^2\delta_{b_lc_l}}{1+\bar{N}_{T0}}}\nonumber\\
    =&\prod_{l=1}^L\frac{1}{1+\bar{N}_{T0}}e^{-\frac{(8+4\bar{N}_{T0})\alpha_0^2\delta_{b_lc_l}}{1+\bar{N}_{T0}}}\nonumber\\
    \approx&\prod_{l=1}^L\frac{1}{1+\bar{N}_{T0}}e^{-\frac{8\alpha_0^2\delta_{b_lc_l}}{1+\bar{N}_{T0}}},
    \label{prodtruereduced}
\end{align}
\end{widetext}
where the approximation in the last step is justified since $\bar{N}_S\ll1$, $\bar{N}_B\sim10$, and $\bar{N}_S^{'}\sim\bar{N}_B$ so that $\bar{N}_{T0}=\bar{N}_S\bar{N}_S^{'}/7.61\ll1$.

On the other hand,
\begin{align}
\langle\mbf{0}_{L}|\hat{\rho}_{\mbf{b},\text{eff}}|\mbf{0}_L\rangle
&=\prod_{l=1}^L\frac{1}{1+\bar{N}_{T0}}e^{-\frac{(2\sqrt{2}\alpha_0)^2\delta_{b_l,c_l}}{1+\bar{N}_{T0}}}\label{prodeffreduced}
\end{align}
and it is apparent that the final expressions \ref{prodtruereduced} and \ref{prodeffreduced} are equal.
\end{proof}

\subsection{Vacuum probability of $K$ thermal states}\label{subapp:thermalvac}

Let $y=(1+\bar{N}_S^{'})$ and $x=(1-\frac{1}{K}y)$. Then $\alpha_k^2=\alpha_1^2x^{2(k-1)}$ and $\alpha_0^2=\alpha_1^2\frac{1-e^{-2y}}{2y}$. 
\begin{figure}[!h]
\centering
\includegraphics[width=\linewidth]{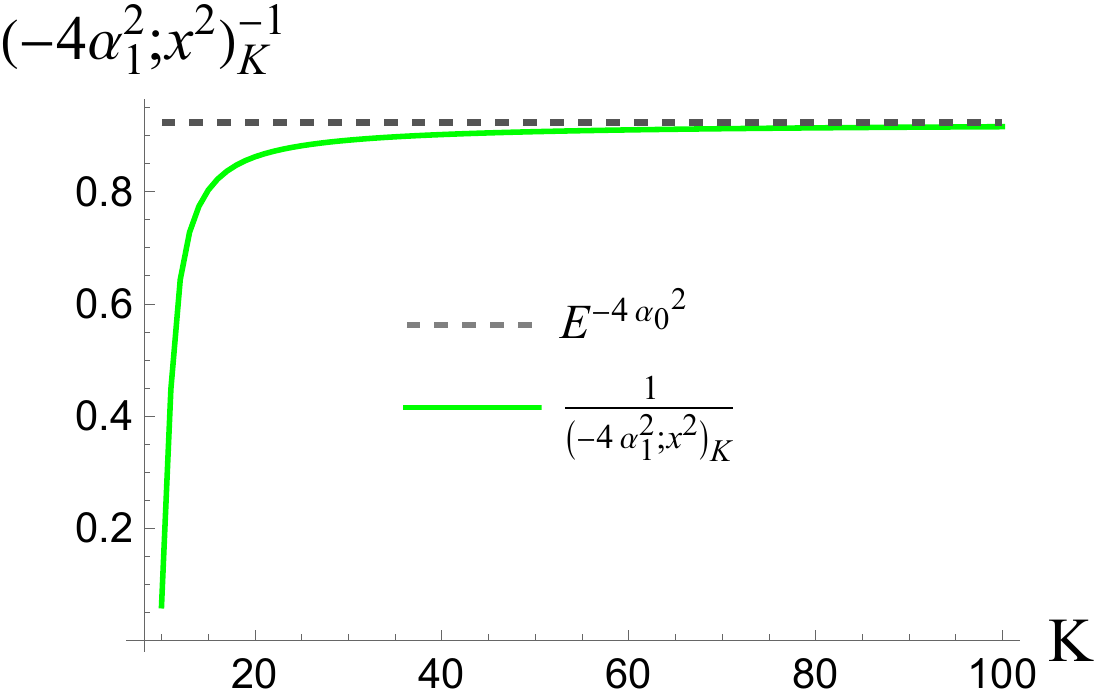}
\caption{Numerical plot of expression \ref{eq:qpoch} plotted as a function of $K$. The values for $y$ and $\eta\bar{N}_S(1+\bar{N}_S)$ are chosen to be $20$ and $0.8$ respectively. The convergence occurs faster for smaller values of $\eta\bar{N}_S(1+\bar{N}_S)$.}
\label{fig:qpochandalpha0}
\end{figure}
Then, with the $q$-Pochhammer symbol $(a;q)_n$ defined as
$$
(a;q)_n:=\prod_{k=0}^{n-1}\left(1-a q^k\right),$$
 the product over $k$ in Eq.~\eqref{prodtrue} is
\begin{align}
&\prod_{k=1}^K\langle0|\hat{\rho}_\text{th}(0,2\delta_{b_l,c_l}\alpha_k)|0\rangle\nonumber\\
=&\prod_{k=0}^{K-1}\frac{1}{1+4\alpha_k^2\delta_{b_l,c_l}}\nonumber\\
=&\left(\prod_{k=0}^{K-1}1+4\alpha_1^2x^{2k}\delta_{b_l,c_l}\right)^{-1}\nonumber\\
=&\begin{cases}
(-4\alpha_1^2;x^2)_K^{-1}&,\text{if $b_l=c_l$}\\
1&,\text{otherwise}.
\end{cases}\label{eq:qpoch}
\end{align}

Notice that $(-4\alpha_1^2;x^2)_K^{-1}$ depends on $K$ not only via its subscript but also via $x$ and $\alpha_1=\sqrt{\frac{1}{K}\eta\bar{N}_S(1+\bar{N}_S)}$.

On the other hand,
\begin{align}
\langle0|2\delta_{b_l,c_l}\alpha_0\rangle\langle2\delta_{b_l,c_l}\alpha_0|0\rangle=&e^{-4\alpha_0^2\delta_{b_l,c_l}}\nonumber\\
=&
\begin{cases}
e^{-4\alpha_0^2}&,\text{if $b_l=c_l$}\\
1&,\text{otherwise.}
\end{cases}\nonumber
\end{align}
 
Figure \ref{fig:qpochandalpha0} shows that $\prod_{k=1}^K\langle0|\hat{\rho}_\text{th}(0,2\alpha_k\delta_{b_l,c_l})|0\rangle$ approaches $\langle0|2\delta_{b_l,c_l}\alpha_0\rangle\langle2\delta_{b_l,c_l}\alpha_0|0\rangle$ for large $K$.

\newcommand{\ac}{\alpha_0}
\newcommand{\gmo}{\sqrt{L}\ac}
\section{Evaluation of the entries of the transition probability matrix for JDR2}\label{app:transitionentries}
The transition probability matrix $\mb{X}$ of the channel consists of entries
\begin{widetext}
\begin{subequations}
\begin{align}
	\mb{X}_{2j-1,2i-1}&=P((i,+)|RM_{(j,+)}):=\int_{0}^{T} P(i;t|RM_{(j,+)}) P_{\text{Ken}_{\scriptstyle i}}(+|RM_{(j,+)};t)\label{tp++}\\
	\mb{X}_{2j-1,2i}&=P((i,-)|RM_{(j,+)}):=\int_{0}^{T} P(i;t|RM_{(j,+)}) P_{\text{Ken}_{\scriptstyle i}}(-|RM_{(j,+)};t)\label{tp+-}\\
	\mb{X}_{2j,2i-1}&=P((i,+)|RM_{(j,-)}):=\int_{0}^{T} P(i;t|RM_{(j,-)}) P_{\text{Ken}_{\scriptstyle i}}(+|RM_{(j,-)};t) \label{tp-+}\\
	\mb{X}_{2j,2i}&=P((i,-)|RM_{(j,-)}):=\int_{0}^{T} P(i;t|RM_{(j,-)}) P_{\text{Ken}_{\scriptstyle i}}(-|RM_{(j,-)};t) \label{tp--}
\end{align}
\label{tpentries}
\end{subequations}
for $i,j=1,2,\cdots, L$, and
\begin{equation}
	\mb{X}_{k,2L+1}=P(\text{no clicks}|RM_{(\pm,k)})=P_{\text{erasure}}
	\label{erasureprob}
\end{equation}
\end{widetext}
for $k=1,2,\cdots,2L$, where $P(i;t|RM_{(j,\pm)})$ is the probability that the first click is detected by the ideal photon detector of the $i^{\text{th}}$ output mode at time $t$ given that the RM code word used for modulation was the one inducing a $\pm$-signed pulse in the $j^{\text{th}}$ mode, $P_{\text{Ken}_{\scriptstyle i}}(\text{sign}_{\text{out}}|RM_{(j,\text{sign}_{\text{in}})};t)$ is the probability that the $i^{\text{th}}$ Kennedy receiver decides sign$_{\text{out}}$ following a first click in mode $i$ at time $t$ given that the RM code word used for modulation was the one that induces a sign$_{\text{in}}$-signed displacement in the $j^{\text{th}}$ mode (with $\text{sign}_\text{in},\;\text{sign}_\text{out} \in\{+,-\}$), and $T$ is the pulse duration.

Note that $P(i;t|RM_{(j,\pm)})$ and $P_{\text{Ken}_{\scriptstyle i}}(\text{sign}_{\text{out}}|RM_{(j,\text{sign}_{\text{in}})};t)$ are independent of $i$ and $j$ for $i\neq j$. The latter is also independent of sign$_{\text{in}}$ when $i\neq j$. Similarly $P(i;t|RM_{(i,\pm)})$ and $P_{\text{Ken}_{\scriptstyle i}}(\text{sign}_{\text{out}}|RM_{(i,\text{sign}_{\text{in}})};t)$ are independent of $i$. So really there are 6 independent entries of $\mb{X}$ that need to be worked out. Restricting $j$ to $\{1,2\}$ and $i$ to $\{1,2,2n+1\}$ is sufficient to cover all 6:
\begin{align*}
	\mb{X}_{11}, \mb{X}_{12}, \mb{X}_{21}, \mb{X}_{22},\quad\leftarrow\,&(i=j=1)\\
	\mb{X}_{14},\quad\quad\quad\leftarrow\,&(i=2; j=1)\\
	\text{and }\mb{X}_{1,2L+1}=P_{\text{erasure}}
.&
\end{align*}
\begin{figure}[b]
    \centering
    \includegraphics[width=\linewidth]{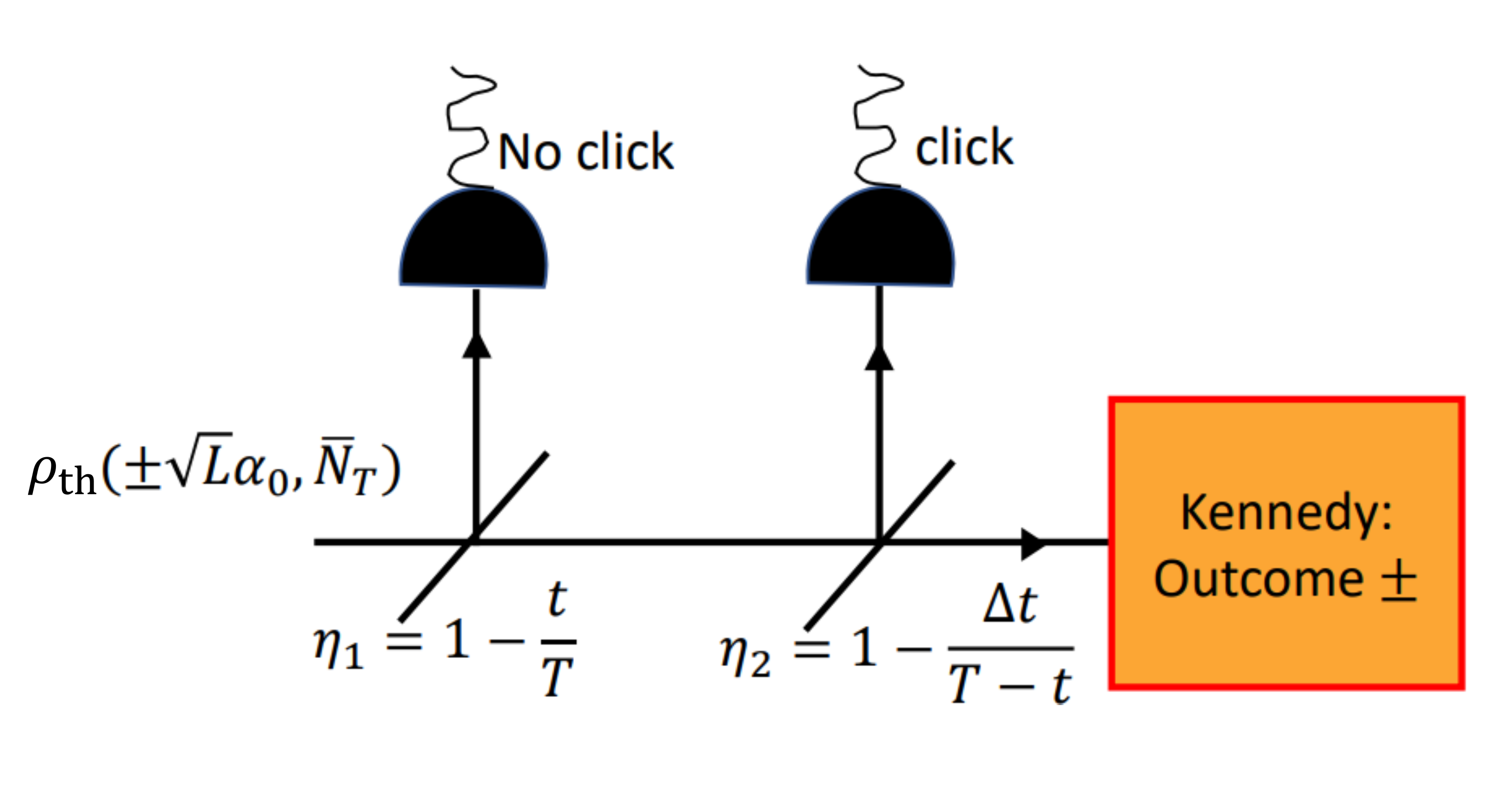}
    \caption{Time slicing diagram corresponding to the event that the photo-detector of the pulse-containing mode detects a click at time $t$, the probability of which contributes to $P(i;t|RM_{(i,\pm)})$. The two photo detectors and beam splitters is a (equivalent) representation of an individual photo detector measuring the pulse during different windows of time.}
    \label{timeslicediag}
\end{figure}

The pulse carrying mode at the output of the Green Machine is in a displaced thermal state $\hat{\rho}_\text{th}(\gmo,\bar{N}_{T0})$
To evaluate the integrands in Eq.s~\eqref{tpentries}, the Green Machine output modes must be sliced at time t after its onset. In general, this is equivalent to using a beam splitter of transitivity $t/T$ to split the pulse into two weaker ones. For a coherent state, this produces a product of two weaker coherent states, leading to the relatively easy evaluation of the conditional probabilities, but for a displaced thermal state, the time slicing produces a two-mode mixed state.

A beam splitter of transmissivity $t/T$ input with vacuum in one port and $\hat{\rho}_\text{th}(\sqrt{L}\ac,\bar{N}_{T0})=\frac{1}{\pi \bar{N}_{T0}}\int_{\mbb{C}}d^{2}\gamma\;e^{-\frac{|\gamma-\sqrt{L}\ac|^{2}}{\bar{N}_{T0}}}\,|\gamma\rangle\langle\gamma|$ in the other results in the state
\begin{align*}
	\frac{1}{\pi \bar{N}_{T0}}\int_{\mbb{C}}d^{2}\gamma\;&e^{-\scriptstyle\frac{|\gamma-\sqrt{L}\ac|^{2}}{\bar{N}_{T0}}}\\
	\times&|\ssd{\sqrt{\frac{t}{T}}}\,\gamma\rangle\langle\ssd{\sqrt{\frac{t}{T}}}\,\gamma|\otimes
	|-\ssd{\sqrt{1-\frac{t}{T}}}\,\gamma\rangle\langle-\ssd{\sqrt{1-\frac{t}{T}}}\,\gamma|.
\end{align*}

Evaluating the conditional probabilities appearing in the integrals in Eq.s~\eqref{tpentries} requires splitting the output modes of the GM into three temporal slices (a succession of two beam splitters), finding the probability that no clicks occur within the first slice among all output modes of the GM and some clicks occur in the middle slice of only one output, and taking the limit as the middle slice size goes to zero. Thus, the conditional probabilities constituting the integrands of Eq.s~\eqref{tpentries} are
\newcommand{\id}{\mathbf{\hat{I}}}
\newcommand{\M}[1]{\Hat{\mathcal{M}}_{|0\rangle}^{#1}}
\newcommand{\Md}[1]{\Hat{\mathcal{M}}_{|0\rangle}^{#1\,\dagger}}
\newcommand{\D}[2]{\Hat{\mathcal{D}}_{#2}^{#1}}
\newcommand{\Dd}[2]{\Hat{\mathcal{D}}_{#2}^{#1\,\dagger}}
\newcommand{\rosliced}[1]{\hat{\rho}_{(t;\Delta t)}^{(#1)}}%time-sliced pulse with displacement sign given by the argument
\newcommand{\ronpsliced}{\hat{\rho}_{(t;\Delta t)}^{(2)}}
%\begin{equation}
\newcommand{\pPD}[2]{P(#1;t|RM_{#2})}
\newcommand{\pKen}[3]{P_{\text{Ken}_{\scriptstyle #1}}\left(#2|RM_{#3};t\right)}
\begin{widetext}
\begin{subequations}
\begin{align}
	\pPD{1}{(1,\pm)}&=\nonumber\\
	\us{\Delta t\rightarrow0}{\lim}\text{Tr} \Bigl(\Md{2}&\M{2}\rosliced{0}\Bigr)^{L-1}\text{Tr}\Bigl((\id^{\otimes 3}-\Md{2}\M{2})\rosliced{\pm1}\Bigr)\label{pfirstclickpgen}\\
	\pPD{2}{(1,\pm)}&=\nonumber\\
	\us{\Delta t\rightarrow0}{\lim}\text{Tr}\Bigl(\Md{2}&\M{2}\rosliced{0}\Bigr)^{L-2}\text{Tr}\Bigl(\Md{2}\M{2}\rosliced{\pm1}\Bigr)
	%\nonumber\\&\hspace{0.26\textwidth}
	\text{Tr}\left((\id^{\otimes 3}-\Md{2}\M{2})\rosliced{0}\right)\label{pfirstclicknpgen}\\
	\nonumber\\
	\pKen{1}{-}{(1,\pm)}&=\us{\Delta t\rightarrow0}{\lim}\text{Tr}\left(\Md{3}\M{3}\D{3}{\beta(t)}\frac{(\id^{\otimes3}-\M{2})\rosliced{\pm1}(\id^{\otimes3}-\Md{2})}{\text{Tr}\left((\id^{\otimes3}-\Md{2}\M{2})\rosliced{1}\right)}\Dd{3}{\beta(t)}\right)\label{pkennoisy}\\
	\pKen{1}{+}{(1,\pm)}&=1-\pKen{1}{-}{(1,\pm)}\\\nonumber\\
	%\\&\pKen{1}{+}{(1,-)}=1-\pKen{1}{+}{(1,-)},
	P_{\text{erasure}}&=\text{Tr}\bigl(|0\rangle\langle0|\hat{\rho}_\text{th}(\gmo,\bar{N}_{T0})\bigr)\text{Tr}\bigl(|0\rangle\langle0|\hat{\rho}_\text{th}(0,\bar{N}_{T0})\bigr)^{L-1}
\end{align}
and additional non-zero probabilities for the non-pulse-containing outputs
\begin{align}
	&\pKen{2}{-}{(1,\pm)}=\us{\Delta t\rightarrow0}{\lim}\text{Tr}\left(\Md{3}\M{3}\D{3}{\beta(t)}\frac{(\id^{\otimes3}-\M{2})\rosliced{0}(\id^{\otimes3}-\Md{2})}{\text{Tr}\left((\id^{\otimes3}-\Md{2}\M{2})\rosliced{0}\right)}\Dd{3}{\beta(t)}\right)\label{p+nonpulse}\\
	&\pKen{2}{+}{(1,\pm)}=1-\pKen{2}{-}{(1,\pm)}\label{p-nonpulse}
\end{align}
\label{Xintegrands}
\end{subequations}%\end{equation}
where

\begin{equation*}
\begin{split}
	\rosliced{\sigma}=e^{\frac{\scriptstyle\eta L\ac^{2}}{\scriptstyle\eta \bar{N}_{T0}+1}}\frac{\eta \bar{N}_{T0}+1}{\pi \bar{N}_{T0}}\int_{\mbb{C}}d^{2}\gamma\;e^{-\frac{|\gamma-\sigma\cdot\sqrt{L}\ac|^{2}}{\bar{N}_{T0}}}e^{-\eta|\gamma|^2}\,|0\rangle\langle0| &\otimes
	|-\ssd{\sqrt{\frac{\Delta t}{T}}}\,\gamma\rangle\langle-\ssd{\sqrt{\frac{\Delta t}{T}}}\,\gamma|\\&\otimes 
	|\ssd{\sqrt{1-\frac{t+\Delta t}{T}}}\,\gamma\rangle\langle\ssd{\sqrt{1-\frac{t+\Delta t}{T}}}\,\gamma|
\end{split}
\end{equation*}
\end{widetext}
is a green machine output mode (with $\sigma\in\{-1,0,1\}$) sliced at times $t$ and $t+\Delta t$ (with a succession of splitting operations of transmissivity $t/T$ and $\frac{\Delta t}{T-t}$ respectively as shown in Fig. \ref{timeslicediag} whose first slice has been measured to have no clicks, $\D{3}{\beta(t)}$ is the displacement operator acting on the third slice with Kennedy nulling $\beta(t)$, and $\M{i}$ with $i\in\{1,2,3\}$ is the measurement operator of the vacuum (no clicks) outcome for the $i^\text{th}$ slice of the mode. For example $\M{1}=|0\rangle\langle0|\otimes\id\otimes\id$\footnote{One can check that $\{\M{i},\id^{\otimes 3}-\M{i}\}$ is a valid measurement, as it satisfies the POVM condition. In fact, it is a projective measurement, as $(\Md{i}\M{i})^{2}=\Md{i}\M{i}$ and $((\id^{\otimes3}-\M{i})^{\dagger}(\id^{\otimes3}-\M{i}))^{2}=(\id^{\otimes3}-\M{i})^{\dagger}(\id^{\otimes3}-\M{i})$.}.

Equations \ref{Xintegrands} evaluate to\footnote{Traces were taken in the coherent state basis}
\newcommand{\oneminuseta}{\ssd{\frac{T-t}{T}}}
\begin{widetext}
\begin{subequations}
\begin{align}
	\pPD{1}{(1,\pm)}&=\frac{e^{-\frac{\ssd{t}}{t\bar{N}_{T0}+T}L\ac^2}}{T}\left(\frac{T}{t\bar{N}_{T0}+T}\right)^{L+2}(\frac{t}{T}\bar{N}_{T0}^{2}+\bar{N}_{T0}+L\ac^{2})\,dt
	\label{pfirstclickpc}\\
	\pPD{2}{(1,\pm)}&=\frac{e^{-\frac{\ssd{t}}{t\bar{N}_{T0}+T}L\ac^2}}{T}\left(\frac{T}{t\bar{N}_{T0}+T}\right)^{L+2}(\frac{t}{T}\bar{N}_{T0}^{2}+\bar{N}_{T0})\,dt
	\label{pfirstclicknpc}
\end{align}
\begin{align}
	P_{\text{Ken}_1}&(-|RM_{(1,\pm)};t)=\nonumber\\%&\pKen{1}{-}{(1,\pm)}=\nonumber\\
	&\quad\left(\frac{t\bar{N}_{T0}+T}{T}\right)^3(\frac{t}{T}\bar{N}_{T0}^{2}+\bar{N}_{T0}+L\ac^{2})^{-1}(1+\bar{N}_{T0})^{-3}\nonumber\\
	&\times e^{\frac{t\,L\ac^2}{t\bar{N}_{T0}+T}-\frac{L\ac^{2}\pm2\gmo\beta(t)\sqrt{\frac{T-t}{T}}+\beta(t)^2(1+\bar{N}_{T0}\frac{t}{T})}{1+\bar{N}_{T0}}}\nonumber\\
	&\times \left(\bar{N}_{T0}^2(1+\beta(t)^2\oneminuseta)+\bar{N}_{T0}(1\mp2\gmo\beta(t)\sqrt{\oneminuseta})+L\ac^2\;\right)\label{pken11-}\\\nonumber\\
	P_{\text{Ken}_1}&(+|RM_{(1,\pm)};t)=1-\pKen{1}{-}{(1,\pm)}\label{pken11+}\\\nonumber\\
	&\hspace{3em}P_{\text{erasure}}=(1+\bar{N}_{T0})^{-L}e^{-\frac{L\ac^2}{\ssd{\bar{N}_{T0}+1}}},
\end{align}
\begin{align}
    P_{\text{Ken}_2}(-|RM_{(1,\pm)};t)&=\frac{(t\bar{N}_{T0}+T)^2}{T^3(\bar{N}_{T0}+1)^3}\bigl((T-t)\bar{N}_{T0}\beta(t)^2+T(\bar{N}_{T0}+1)\bigr)e^{-\frac{\ssd{(t \bar{N}_{T0}+T)\beta(t)^2}}{\ssd{\bar{N}_{T0}+1}}}\\
    P_{\text{Ken}_2}(+|RM_{(1,\pm)};t)&=1-P_{\text{Ken}_2}(-|RM_{(1,\pm)};t)
\end{align}
\label{Xintegrandcomponents}
\end{subequations}
\end{widetext}
 where the nulling $\beta(t)=\gmo$ is taken to be exact. To evaluate the transition probabilities, we plug in expressions \ref{Xintegrandcomponents} into Eq.s~\eqref{tpentries} and numerically integrate over the pulse duration.

\end{document}